\pdfoutput=1
\documentclass[review,3p,times]{elsarticle}
\usepackage{amsmath,hyperref}
\usepackage{lineno}

\usepackage{subfigure}
\usepackage{epsfig}
\usepackage{caption,xcolor,float,graphicx}
\usepackage{graphics}
\usepackage{epstopdf}
\journal{XXX}

\begin{document}

	\begin{frontmatter}
		
		\title{Pore-scale investigation on natural convection melting in a square cavity with gradient porous media}
		\author{Jiangxu Huang}
		\author{Kun He}
		%\author{Yin Jiang}
		\author{Lei Wang\corref{mycorrespondingauthor}}
		\cortext[mycorrespondingauthor]{Corresponding author}
		\ead{leiwang@cug.edu.cn}
		%\author[mythirdaddress]{B}
		%\author[mysecondaddress,myfourthaddress]{Baochang Shi}
		%\author[mysecondaddress,myfourthaddress]{Zhenhua Chai\corref{mycorrespondingauthor}}
		
		\address{School of Mathematics and Physics, China University of Geosciences, Wuhan 430074, China}
		
		%\address[mysecondaddress]{ School of Mathematics and Statistics, Huazhong University of Science and Technology, Wuhan 430074, China}
		%\address[mythirdaddress]{School of Energy and Power Engineering, Huazhong University of Science and Technology, Wuhan 430074, China}
		
		%	\address[myfourthaddress]{Hubei Key Laboratory of Engineering Modeling and Scientific Computing, Huazhong University of Science and Technology, Wuhan 430074, China}
		
		\begin{abstract}
			In this paper, a numerical study on the melting behavior of phase change material (PCM) with gradient porous media has been carried out at the pore scales. In order to solve the governing equations, a pore-scale lattice Boltzmann method with the double distribution functions is used, in which a volumetric LB scheme is employed to handle the boundary. The Monte Carlo random sampling is adopted to generate microstructure of two-dimensional gradient foam metal which are then used to simulate the solid-liquid phase transition in cavity. The effect of several factors, such as gradient porosity structure, gradient direction, Rayleigh number and particle diameters on the liquid fraction of PCM are systematically investigated. It is observed that the presence of gradient media affect significantly the melting rate and shortens full melting time compared to that for constant porosity by enhancing natural convection. The melting time of positive and negative gradients will change with Rayleigh number, and there is a critical value for Rayleigh number. Specifically, when Rayleigh number is below the critical value, the positive gradient is more advantageous, and when Rayleigh number exceeds the critical value, the negative gradient is more advantageous. Moreover, smaller particle diameters would lead to lower permeability and larger internal surfaces for heat transfer.

		\end{abstract}
		
		
	\end{frontmatter}
	
	%\linenumbers
	\section{Introduction}
	
	Thermal energy storage (TES) \cite{TES1,TES2} technology has attracted wide attention due to its ability to solve the temporal and spatial imbalance between energy supply and energy demand. Among various TES methods, the latent heat storage (LHTES) \cite{LHTES,LHTES2} which could store thermal energy in latent heat through the solid-liquid phase change of phase change material (PCM) has been extensively developed in recent years. This is mainly due to LHTES technology has the advantages of high storage density, stable thermal energy storage temperature, and low cost. Unfortunately, most phase change materials in LHTES have a low thermal conductivity \cite{low thermal conductivity} which seriously limits the heat storage efficiency.

	In order to improve the performance of LHTES, scholars have proposed many heat transfer enhancement techniques, including adding highly conductive particles \cite{particles1,particles2}, adding highly conductive metallic fins \cite{fin1,fin2,fin3}, using multiple PCM methods \cite{multiple PCM1,multiple PCM2} , embedding PCMs in highly conductive porous media \cite{chenIJHMT2014,yangAE2016,zhaoIJHMT2016,taoATE2016,zhuATE2016,yaoIJTS2018,yangICHMT2021} and so on. Among these heat transfer enhancement methods, highly conductive porous media with high thermal conductivity, high heat penetration, high porosity and high specific surface area has been showing progressive potentials.  
	
	Metal foam is a typical porous media with high thermal conductivity. Metal foam perfectly maintains the excellent properties of base metals, such as high stability, light weight, high thermal conductivity, ductility, and improved heat transfer characteristics. Therefore, metal foam is a widely used thermal conductivity enhancers. Some studies have explored the problems of solid-liquid phase change heat transfer in PCM-filled metal foam \cite{chenIJHMT2014,yangAE2016,zhaoIJHMT2016,taoATE2016,zhuATE2016,yaoIJTS2018,yangICHMT2021}. Chen et al. \cite{chenIJHMT2014} used an infrared camera and microscope to experimentally study the heat transfer phenomenon of PCM in foamed metal and used the double-distributed lattice Boltzmann method to carry out a numerical simulation. It is found that the numerical value is consistent with the experimental results and the foam metal has a significant enhancement effect on the melting of PCM. Yang et al. \cite{yangAE2016} experimentally studied the effects of metal foam to melting and found that completely melting the PCM in metal foam takes over 1/3 less time than that of pure paraffin under the same conditions. Zhao et al. \cite{zhaoIJHMT2016} numerically studied the melting behavior of paraffin in foam metal, it is noted that the Rayleigh number, porosity and pore density have a significant influence on on the melting and solidification process. Tao et al. \cite{taoATE2016} used the lattice Boltzmann method to study the heat storage performance of paraffin foam metal composites. The author studied the influence of porosity and pore density on the melting rate, and proposed the best metal foam structure with a porosity of 0.94 and a PPI of 45. Zhu et al. \cite{zhuATE2016} used the finite volume method to explore the influence of the three strengthening methods of foam metal porosity, cold wall shape and discrete heat source on the thermal properties of foam metal/phase change materials.  Yao et al. \cite{yaoIJTS2018} conducted a visualized experiment to study the melting of paraffin in high porosity copper foam at pore scale. They concluded that the copper foam with a high porosity of 0.974 effectively extends the phase change interface and improves the heat storage of paraffin, while the reduction in the amount of latent heat is only 2.6\%. Yang et al. \cite{yangICHMT2021} numerically and experimentally explored the influence of the inclination angle of the inclined cavity containing foamed metal and the cavity aspect ratio on the melting of PCM. It is found that the tilt angle has little effect when the aspect ratio is given, and the smaller aspect ratio is better than the larger aspect ratio when the aspect ratio is not fixed.

	All the studies mentioned above only consider metal foams with fixed pore parameters. Recently, many investigations have found that gradient metal foams can further enhance melting heat transfer under the same conditions. Yang et al. \cite{yangIJHMT2015} numerically investigated the melting process of sodium nitrate inside porous copper foam with linearly changed porosity. The numerical re­sults show that  porosity linearly increased from bottom to top could improve the heat transfer performance and shorten the completely melted time compared to that for constant porosity by enhancing natural convection. Yang et al. \cite{yangIJHMT2016} numerically studied the solidification behavior of saturated distilled water in open-cell foam metal and compared it with ungraded foam metal. The authors found that the positive gradient porosity and negative gradient pore density structures have a faster solidification rate compared to the non-graded foams. Zhu et al. \cite{zhuATE2017} proposed an improved metal foam structure, which is composed of metal foam and finned metal foam with gradient pores. The finite volume method is also used to analyze the influence of structural parameters on energy storage performance. It is found that this structure can shorten the melting time by changing the melting sequence of the phase change material. Zhang et al. \cite{zhangATE2017} numerically studied the melting behavior in gradient foam metal which was consist of three different homogeneous porosity slices, and results show that the gradient porosity structure can overcome the corner phenomenon at the bottom to increase the heat storage rate. Yang et al. \cite{yangAE2020} experimentally and numerically studied the effect of gradient porosity and gradient density in tube latent heat thermal energy storage. Results indicated that the positive gradient design of porosity can significantly reduce the melting time of PCMs filled in the pore space and obtain simultaneously a better temperature uniformity. Ghahremannezhad et al. \cite{ghahremannezhadATE2020} used a finite volume approach to numerically simulate the melting behavior of PCMs in gradient foam metal under different heating modes. They found that the direction of gradient porosity and PPI can affect the heat transfer rate. Hu et al. \cite{huATE2020} use a three-dimensional model to numerically simulate the melting behavior in a gradient metal foam saturated with PCM and  quantitatively explored the effect of gradient size and gradient difference on the heat storage characteristics of PCMs. It found that gradient metal foam effectively improves and accelerates the heat storage efficiency and there is an optimal gradient difference under a fixed average porosity. Marri et al. \cite{marriIJHMT2021} experimentally numerically studied the thermal function of cylindrical foam metal/PCM composite heat sinks with gradients of porosity and density. The results show that the three-level gradient and the two-level gradient have comparable thermal performance, which are 4.4 times and 4 times stronger than the thermal performance of the uniform structure, respectively.

	From the above experiments and numerical studies, it can be seen that the gradient pore structure has a significant impact on the heat transfer characteristics of PCM, and it is an effective means to enhance phase change heat transfer. However, most of the research on the influence mechanism of the gradient metal foam on the phase change process of the phase change material (PCM) is based on the representative elementary volume (REV) models, and seldom is studied by the pore-scale models. Compared with the REV model, the pore-scale model can provide detailed local information of the fluid flow and heat transfer inside the gradient metal foam. As far as the author knows, the effect of some key parameters such as Reynolds number on the solid-liquid phase transition in gradient metal foam has not been systematically studied. Therefore, we studied the melting behavior of PCM under a gradient porosity structure at the pore scale.

	In this work, a pore-scale LB model was used to study the melting of composite metal foam PCM with porosity gradient. Lattice Boltzmann method (LBM) has been developed as an effective numerical method to solve the problem of solid-liquid phase transition \cite{boltzmann1,boltzmann2,boltzmann3}, and has solved the problem of solid-liquid phase transition of PCM in porous media \cite{chenIJHMT2014,taoATE2016,LBM_media1,LBM_media2}. The remainder of this paper is structured as follows. The governing equations for the melting problem and physical problem are presented in Section 2, followed by the LBM for the enthalpy equation and the fluid flow in Section 3.1 and 3.2. Subsequently, the melting model is verified and boundary treatment in Section 3.3 and Section 3.4, respectively. Numerical results are presented and discussed in Section 4, and finally, some conclusions from the present investigation are summarized in Section 5.

	\section{Problem statement and governing equations}
	\label{section2}
\begin{figure} [H]
		\centering 
		\subfigure[Case A: horizontal porosity gradient]{
			\includegraphics[width=0.4\linewidth]{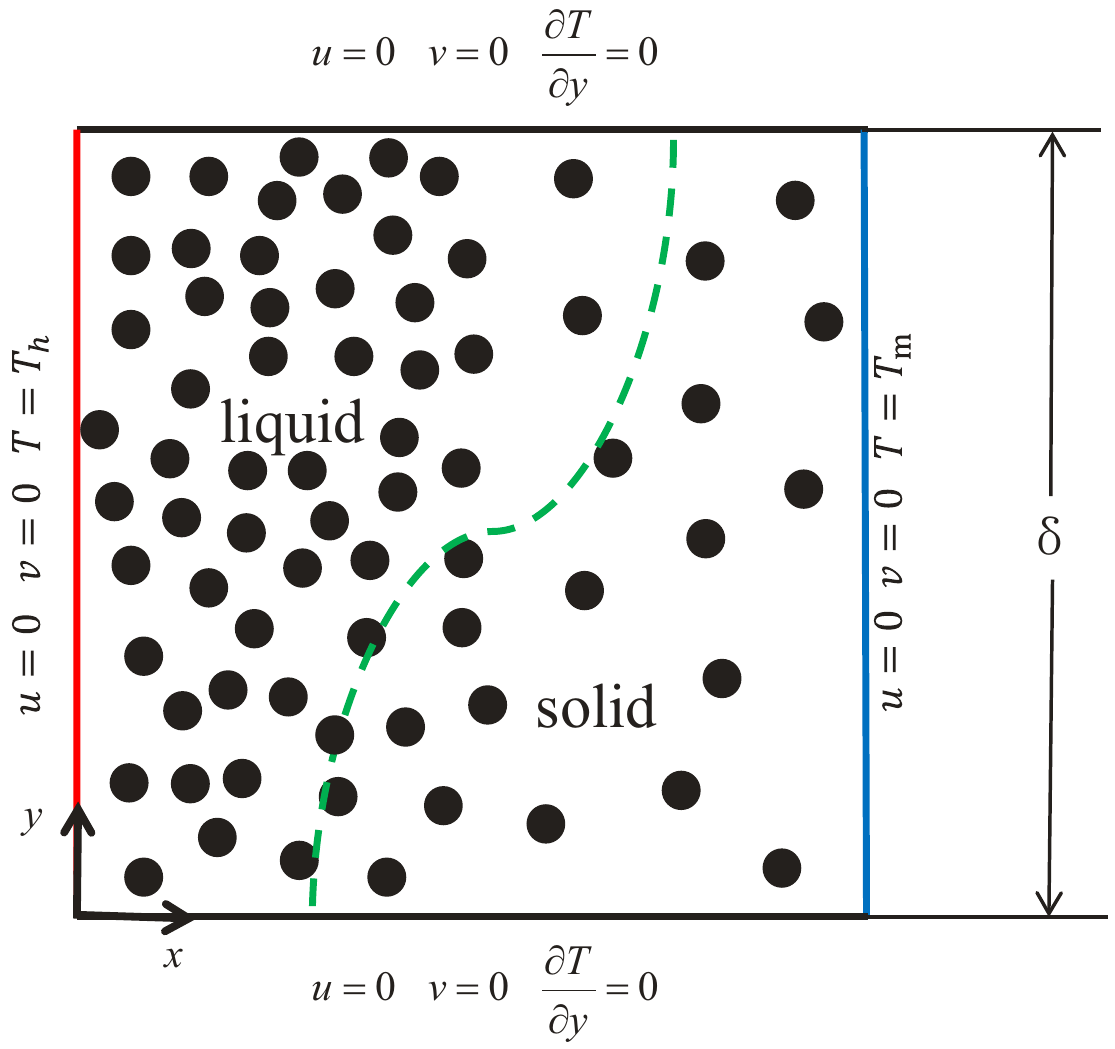}}

		%\vfill
		\subfigure[Case B: vertical porosity gradient]{
			\includegraphics[width=0.4\linewidth]{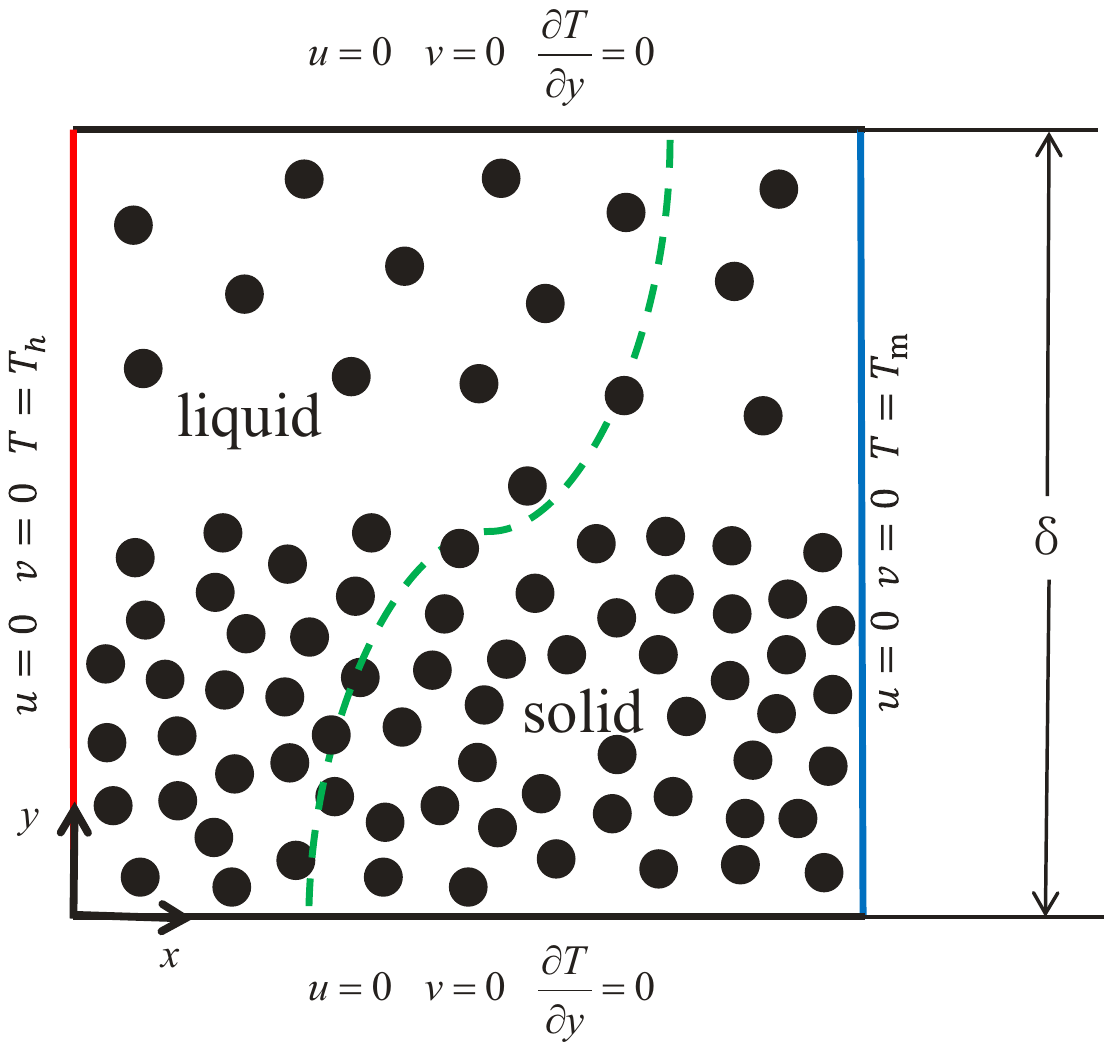}}
		\caption{Schematic of phase change in gradient structure.}	
		\label{fig1}
	\end{figure}

	In this work, we consider the numerical simulation of solid-liquid phase change in a square cavity equipped filled with a gradient porous media. The schematic diagram of the two-dimensional physical model considered in this study is depicted in Fig.\ref{fig1}. A square cavity with side length $ \delta $ encloses the gradient foam metal, which is filled with solid PCMs at the melting temperature $ T_{m} $. In the solid-liquid phase change, the left wall is raised to a constant temperature $ T_{h} $, higher than $ T_{m} $, the temperature of the right wall is kept at the
	
	\noindent  melting temperature $ T_{m} $, while the other two walls are assumed to be adiabatic boundary. Additionally, the detailed description of the microstructure of porous media is a necessary prerequisite for the pore scale model. In the present work, the stochastic simulation method of porous media proposed by Chen et al. \cite{zhouCMS2014} is employed to generate porous structure. Since the gradient porosity structure created with tri-layer media have almost the same performance as a bi-layer media \cite{marriIJHMT2021}, so we adopt a two-layer structure for simplicity. Three structures of porous media have been defined with fixed average porosity as follows: positive gradient structure (porosity increases from 0.8 to 0.95), negative gradient structure (porosity decreases from 0.8 to 0.95), and uniform gradient structure (porosity keep at 0.875). The positive porosity gradient is obtained by increasing the porosity of porous media  along the positive x-axis or y-axis direction, while the negative one meant the decreased porosity along the positive x-axis or y-axis direction. Case A and Case B study horizontal gradient porosity and vertical gradient porosity, respectively. Furthermore, It is worth noting that the particle diameter with 7.5 $ l.u. $ (lattice unit) and the volume of PCM that can be filled in the gradient structure is the same, which lays the foundation for comparisons of melting behaviors during phase change process. 
	
	Moreover, The following simplifying assumptions are made to establish the mathematical model: (1) the fluid in the cavity is incompressible and the Boussinesq approximation is applied, (2) the volume change during the phase transition is ignored, (3) all materials meet the assumptions of uniformity and isotropy, (4) the physical properties of PCM are different from porous media, (5) the physical parameters of phase change materials and porous media are almost constant. Based on the above assumptions, a mathematical model of the solid-liquid phase transition in the pore-sale model can be expressed by the following equations.
	
	\begin{equation}
		\nabla \cdot \boldsymbol{u}=0,
	\end{equation}
	\begin{equation}
		\frac{\partial \mathbf{u}}{\partial t}+\nabla \cdot \mathbf{u} \mathbf{u}=-\frac{1}{\rho_{f}} \nabla p+\nabla \cdot\left[v_{f}\left(\nabla \mathbf{u}+(\nabla \mathbf{u})^{\mathrm{T}}\right)\right]-\mathbf{g}\left[1-\beta_{f}\left(T-T_{m}\right)\right],
	\end{equation}
	\begin{equation}
		\frac{\partial\left[\left(\rho C_{p}\right)_{f} T\right]}{\partial t}+\nabla \cdot\left[\left(\rho C_{p}\right)_{f} \mathbf{u} T\right]=\nabla \cdot \lambda_{f} \nabla T
	\end{equation}
	and the energy equation for the solid phase is:
	\begin{equation}
		\frac{\partial\left[\left(\rho C_{p}\right)_{s} T\right]}{\partial t}=\nabla \cdot \lambda_{s} \nabla T
	\end{equation}
	where the subscript $l$ and $s$ denote the liquid and solid phases, respectively.$ \boldsymbol{g},\boldsymbol{u},T,p$ are the gravity acceleration vector, velocity, temperature,  and pressure, respectively. $ \beta_{f} $ is the thermal volumetric expansion coefficient of the fluid. $ \rho,C_{p},\lambda $ represent the density, heat capacity, and thermal conductivity coefficient, respectively. 
	
	Moreover, thermal conductivity coefficient of PCM is:
	\begin{equation}
		\lambda_{P C M}=(1-f l) \lambda_{s}+f_{l} \lambda_{l}
	\end{equation}
	In two phase regions, the enthalpy H can be solved by the enthalpy-based method and is given by
	\begin{equation}
		H=C_{p} T+f l L
	\end{equation}
	here $ f_{l} $ denotes the liquid fraction and values within 0 and 1. $ L $ represent the latent heat of PCM.
	
	This problem can be characterized by the following three main dimensionless parameters, i.e. Rayleigh number Ra, Prandtl number Pr, Stefan number Ste, which are defined by
	\begin{equation}
		R a=\frac{|\boldsymbol{g}| \beta (T_{h}-T_{m}) H^{3}}{\nu \alpha} \quad \operatorname{Pr}=\frac{\nu}{\alpha} \quad \text { Ste }=\frac{C_{p} (T_{h}-T_{m})}{L}
		\label{equation_Ra}
	\end{equation}
	where $ \nu $ and $ \alpha $ are the kinetic viscosity and thermal diffusivity, respectively

	%Finally, in order to illustrate the heat transfer characteristics of the solid-liquid phase transition in a closed cavity, the local Nusselt number $Nu$ and average Nusselt number $Nu_{ave}$ of the left wall is considered and expressed as \cite{Nusselt}
	%\begin{equation}
	%	\begin{aligned}
	%		Nu&=\frac{\partial {T} }{\partial {y}}\bigg|_{y=0} \quad  Nu_{ave}=\frac{1}{L} \int_{0}^{L}{Nu}dx,\\
	%	\end{aligned}
	%\end{equation}
	
	\section{The lattice Boltzann model and model validation}
	\label{section3}

	\subsection{Lattice Boltzmann method for velocity field}
	\label{section3_1}
	In the present study, the incompressible lattice Bhatnagar–Gross–Krook (LBGK) model proposed by Guo et al. \cite{LBGK} is used to simulate the fluid flow, the evolution equation of the particle velocity field is
	\begin{equation}
		{f_i}\left( {{\bf{x}} + {{\bf{c}}_i}\Delta t,t + \Delta t} \right) -
		{f_i}\left( {{\bf{x}},t} \right) =  - \frac{1}{{{\tau _f}}}\left[
		{{f_i}\left( {{\bf{x}},t} \right) - f_i^{(eq)}\left( {{\bf{x}},t}
			\right)} \right] + \Delta t{F_i},
		\label{F}
	\end{equation}
	where $ f_{i}(x,t) $ is the probability density distribution functions with velocity eiat position x and time t, $ \Delta t $ is the time increment. $ \tau _f $ is the dimensionless relaxation time determined by $\nu=\rho_{0} c_{s}^{2}\left(\tau_{l}-0.5\right) \Delta t $. In addition, $f_i^{eq}$ is the equilibrium distribution function given by \cite{LBGK}
	\begin{equation}
		f_{i}^{(e q)}(\mathbf{x}, t)=\eta_{i} p+\omega_{i}\left[\frac{\mathbf{c}_{i} \cdot \mathbf{u}}{c_{s}^{2}}+\frac{\mathbf{u u}:\left(\mathbf{c}_{i} \mathbf{c}_{i}-c_{s}^{2} \mathbf{I}\right)}{2 c_{s}^{4}}\right],
	\end{equation}
	where $ \eta_{i} $ is the model parameter satisfying $\eta_{0}=\left(\omega_{0}-1\right) / c_{s}^{2}+\rho_{0},  \eta_{i}=\omega_{i} / c_{s}^{2}(i \neq 0)$ with the constant $ \rho_{0} $ being the fluid average density. $ \omega_{i} $ is the weight coefficient, represented as
	\begin{equation}
		\omega_{i}=\left\{\begin{array}{ll}
			4 / 9, & i=0 \\
			1 / 9, & i=1,2,3,4 \\
			1 / 36, & i=5,6,7,8
		\end{array}\right.
	\end{equation}
	The direction of the discrete velocity $ c_{i} $ of the model in $ i $ direction are given by
	\begin{equation}
		\mathbf{c}_{i}=\left\{\begin{array}{ll}
			(0,0) & i=0 \\
			c(\cos [(i-1) \pi / 2], \sin [(i-1) \pi / 2]) & i=1,2,3,4 \\
			\sqrt{2} \mathrm{c}(\cos [(2 i-9) \pi / 4], \sin [(2 i-9) \pi / 4]) & i=5,6,7,8
		\end{array}\right.
	\end{equation}
	where $ i $ is velocity direction, and c is the lattice speed satisfying $ c=\Delta x/\Delta t $, $ \Delta x $ is the lattice spacing. The body force term $ F_{i} $ is defined as \cite{Guo_F}
	\begin{equation}
		F_{i}(\mathbf{x}, t)=\omega_{i}\left(1-\frac{1}{2 \tau_{f}}\right)\left[\frac{\mathbf{c}_{i} \cdot \mathbf{F}}{c_{s}^{2}}+\frac{(\mathbf{F u}+\mathbf{u F}):\left(\mathbf{c}_{i} \mathbf{c}_{i}-c_{s}^{2} \mathbf{I}\right)}{2 c_{s}^{4}}\right],
	\end{equation}
	where $ c_{s} = c/ \sqrt{3} $ is the sound speed of the model, $ \mathbf{F} $ is the buoyancy force, and can be calculated according to Boussinesq assumption:
	\begin{equation}
		\mathbf{F}=\rho \mathbf{g} \beta\left(T-T_{ref}\right)
	\end{equation}After a serious of iterative computations, the macroscopic pressure and velocity can be obtained by
	\begin{equation}
		\mathbf{u}=\sum_{i} \mathbf{c}_{i} f_{i}+\frac{\Delta t}{2} \mathbf{F},
	\end{equation}
	\begin{equation}
		p=\frac{c_{s}^{2}}{1-\omega_{0}}\left(\sum_{i \neq 0} f_{i}-\omega_{0} \frac{|\mathbf{u}|^{2}}{2 c_{s}^{2}}\right).
	\end{equation}

	\subsection{Lattice Boltzmann method for temperature field}
	\label{section3_2}
	For the heat transfer in the fluid and solid domains, the double relaxation time LB model proposed by Lu et al. \cite{luIJTS2019} is adopted to simulate the temperature. It can effectively suppress the non-physical diffusion phenomenon at the solid-liquid interface during the phase transition simulation process. The evolution equation of the temperature distribution function can be described as 
	\begin{equation}
		\begin{aligned}
			g_{i}\left(\mathbf{x}+\mathbf{c}_{i} \Delta t, t+\Delta t\right)= g_{i}(\mathbf{x}, t)
			-\frac{1}{\tau_{g}^{s}}\left[g_{i}^{s}(\mathbf{x}, t)-g_{i}^{seq}(\mathbf{x}, t)\right] 
			-\frac{1}{\tau_{g}^{a}}\left[g_{i}^{a}(\mathbf{x}, t)-g_{i}^{aeq}(\mathbf{x}, t)\right],
		\end{aligned}
	\end{equation}
	where $ g_{i}^{s}(\mathbf{x}, t) $ and $ g_{i}^{a}(\mathbf{x}, t) $ are the symmetric and anti-symmetric parts of the particle distribution function
	where the superscript $ s $ and $ a $ represent the symmetric and anti-symmetric parts of distribution function, respectively. $ {\tau_{g}}^s $ and $ {\tau_{g}}^a $ are the symmetric relaxation time and anti-symmetric relaxation time, respectively. The
	expressions of $ g_{i}^{s}(\mathbf{x}, t), g_{i}^{a}(\mathbf{x}, t), g_{i}^{seq}(\mathbf{x}, t), g_{i}^{aeq}(\mathbf{x}, t)  $ are defined by
	\begin{equation}
		g_{i}^{s}=\frac{g_{i}+{g}_{\bar{i}}}{2} , g_{i}^{a}=\frac{g_{i}-{g}_{\bar{i}}}{2}, g_{i}^{seq}=\frac{g_{i}^{eq}+{g}_ {\bar{i}}^{\mathrm{eq}}}{2}, g_{i}^{\mathrm{aeq}}=\frac{g_{i}^{\mathrm{eq}}-{g}_{\bar{i}}^{\mathrm{eq}}}{2}.
	\end{equation}
	in which $ i $ represents the opposite direction of $ i $. The equilibrium distribution function can be expressed as
	\begin{equation}
		g_{i}^{e q}=\left\{\begin{array}{ll}
			H-C_{p} \theta+\omega_{i} C_{p} \theta\left(1-\frac{|\mathbf{u}|^{2}}{2 c_{s}^{2}}\right), & i=0 \\
			\omega_{i} C_{p} \theta\left[1+\frac{c_{i} \cdot \mathbf{u}}{c_{s}^{2}}+\frac{\left(\mathbf{c}_{1} \cdot \mathbf{u}\right)^{2}}{2 c_{s}^{4}}-\frac{|\mathbf{u}|^{2}}{2 c_{s}^{2}}\right], & i \neq 0
		\end{array}\right.
	\end{equation}
	The dimensionless relaxation time $ \tau_{g}^{s} $ and $ \tau_{g}^{a} $ can be obtained by
	\begin{equation}
		\frac{\lambda}{\rho_{0} c_{p}}=c_{s}^{2}\left(\tau_{g}^{a}-0.5\right) \Delta t ,
	\end{equation}
	\begin{equation}
		\frac{1}{\tau_{g}^{s}}+\frac{1}{\tau_{g}^{a}}=2.
	\end{equation}
	And the total enthalpy can be obtained as follows
	\begin{equation}
		H=\sum_{i} g_{i}.
	\end{equation}
	Finally, the liquid fraction $ f_{l} $ and the temperature $ T $ can be calculated from total enthalpy,
	\begin{equation}
		f_{l}=\left\{\begin{array}{ll}
			0 & H \leq H_{s} \\
			\frac{H-H_{s}}{H_{l}-H_{s}} & H_{s}<H<H_{l} \\
			1 & H \geq H_{l}
		\end{array}\right.
	\end{equation}
	\begin{equation}
		T=\left\{\begin{array}{ll}
			\frac{H}{C_{p}} & H \leq H_{s} \\
			\frac{H_{l}-H}{H_{l}-H_{s}} T_{s}+\frac{H-H_{s}}{H_{l}-H_{s}} T_{l} & H_{s}<H<H_{l} \\
			T_{l}+\frac{H-H_{l}}{C_{p}} & H \geq H_{l}
		\end{array}\right.
	\end{equation}
	in which $ H_{s}=C_{p}T_{s} $ and $ H_{l}=C_{p}T_{s}+L $ are total enthalpy of the solid phase and the liquid phase, respectively. $ T_{s} $ and $ T_{l} $ denote the solidus and liquidus temperatures, respectively.

	\subsection{LBM boundary conditions}
	In this context, a volumetric LB scheme proposed by Huang et al. is adopted, which can be expressed as
	\begin{equation}
		f_{i}=f_{l} f_{i}^{*}+\left(1-f_{l}\right) f_{i}^{\mathrm{eq}}\left(\rho, \mathbf{u}_{s}\right)
	\end{equation}
	where the $ f_{i}^{*}=f_{i}\left(\mathbf{x}+\mathbf{c}_{i} \Delta t, t+\Delta t\right) $ is given by Eq. \ref{F} and $\mathbf{u}_{s}=\mathbf{0}$ is the the velocity of solid phase. What's more, the non­equilibrium
	extrapolation scheme (NEES) proposed by Guo et al. \cite{GuoPf2002} is applied to deal with all boundary of walls for its second­ order accuracy.

	\subsection{Validation of phase change model}
	\label{section3_3}
	
	\begin{figure}[h]
		\centering 
		\subfigure[]{ \label{fig3}
			\includegraphics[width=0.475\textwidth]{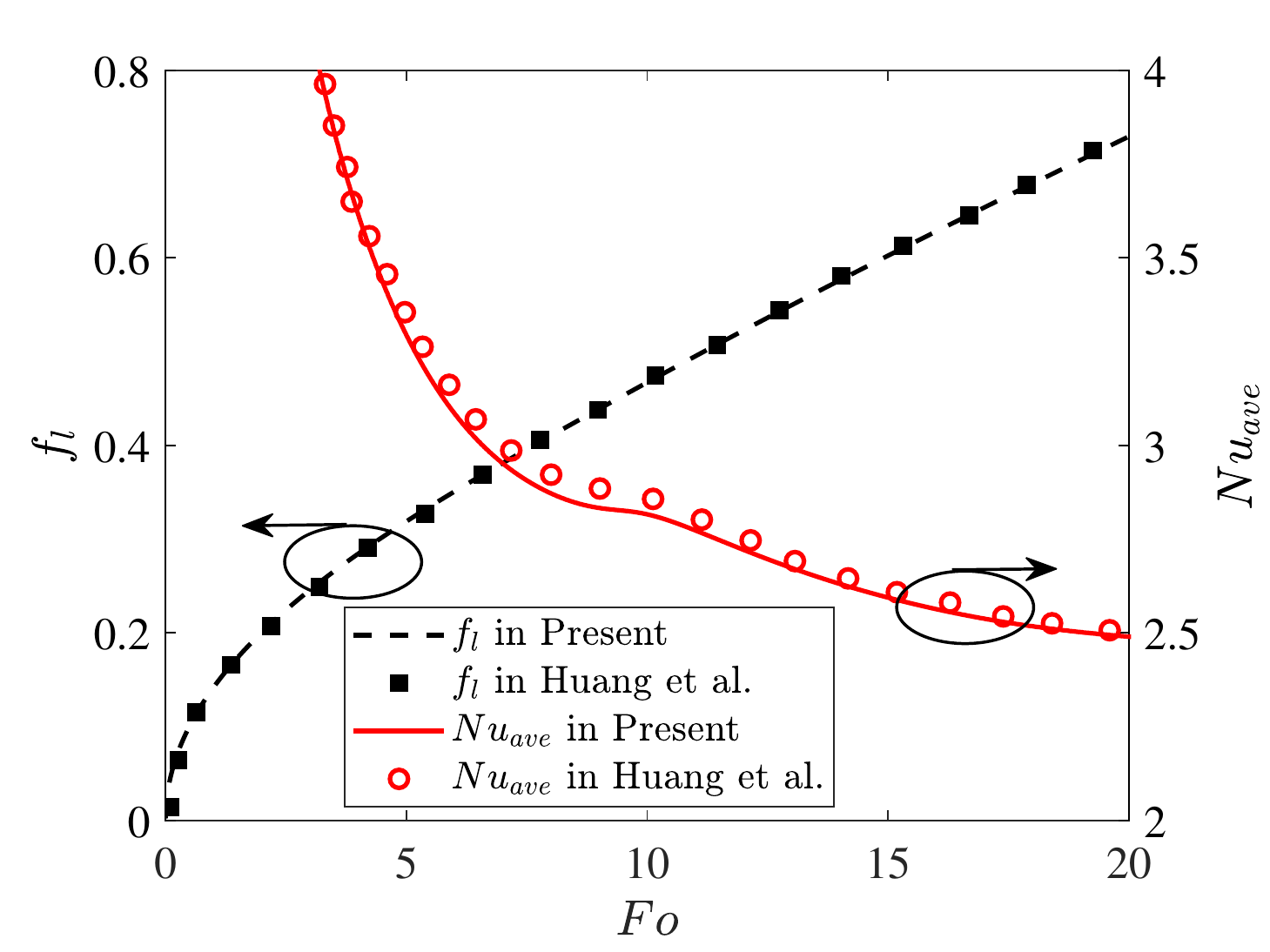}} 
		\subfigure[]{ \label{fig4}
			\includegraphics[width=0.475\textwidth]{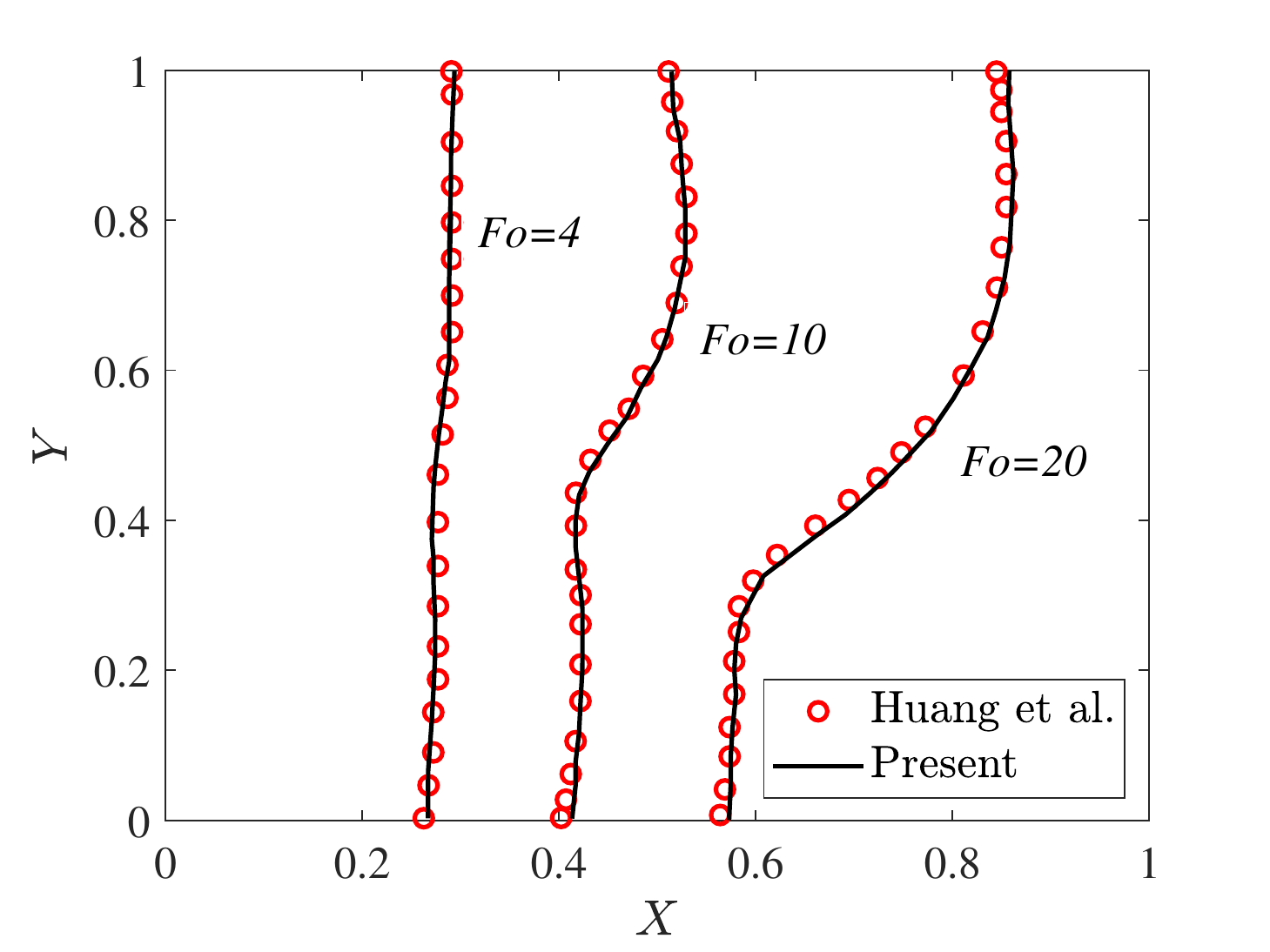}}
		\caption{Comparing results between present and Huang et.al of (a) $ Nu_{ave} $, $ f_{l} $ and (b) phase interface position at Fo= 4, 10, 20.}
	\end{figure}
	
	\noindent In order to demonstrate the ability of the present code in simulation of fluid flow and heat transfer, we simulated the pure PCM melting at $ Ra=25000.0 $, $ Ste = 0.01 $, and $ Pr= 0.02 $ in a square cavity. Additionally, a grid resolution of 512 × 512 is employed in the simulations, which is fine enough to give the grid-independent solution. We compare the average Nusselt number of left wall, total liquid fraction and melting interface with Huang et.al \cite{huangIJHMT2013}. Fig.\ref{fig3} shows $ Nu_{ave} $ and $ f_{l} $ at different Fourier numbers ($ Fo $). It can be clearly seen that the heat transfer performance matches the Huang’s results well at different $ Fo $, a good agreement is obtained. Moreover, the comparison of the position of the solid-liquid interface at $ Fo=4, Fo=10 $  and  $Fo=20 $ is shown in Fig.\ref{fig4}, and shows good agreement. The present numerical results are in good agreement with the results of Huang's results, indicating that the present LBM model and simulation code are accurate and reliable.

	\section{Results and discussion}
	\label{section4}
	In this section, numerical simulations are carried out for the solid-liquid phase transition in gradient porous media via the LBM. And we mainly investigate the influence of the geometrical structure and arrangement of gradient porous media on the melting improvement of PCM, which is measured by the nondimensional Fourier numaber ($ Fo $), liquid phase fraction ($ f_{l} $). Unless otherwise stated, the Stefan numbe and the Prandtl number are selected as 0.1 and 0.2, which have been studied [38]. In order to show the main physical difference between porous media and PCM, we set the thermal conductivity of porous media to 500 times that of PCM.

\begin{figure}[H]
		\centering 
		\subfigure[Case A]{ \label{fig5}
			\includegraphics[width=0.4\textwidth]{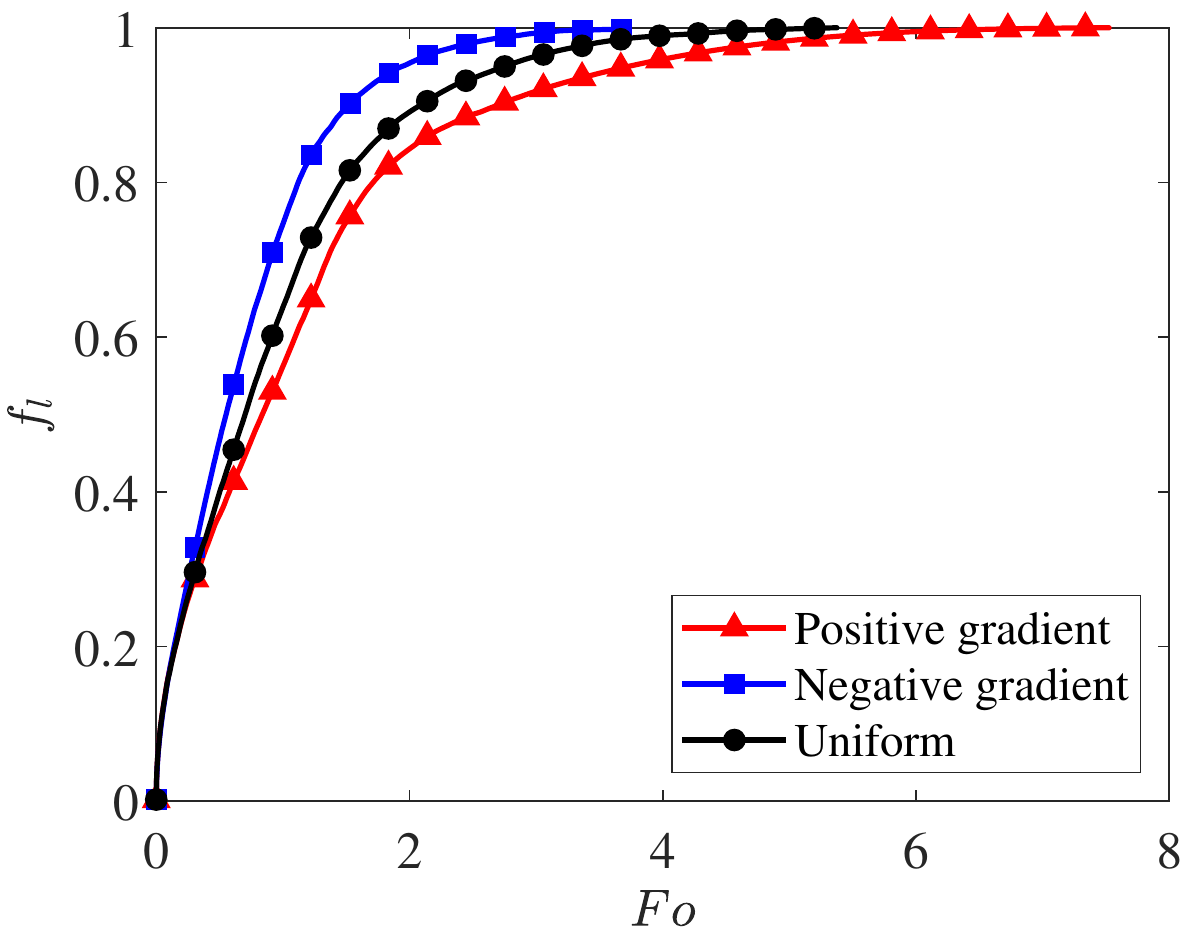}} 
		\subfigure[Case B]{ \label{fig6}
			\includegraphics[width=0.4\textwidth]{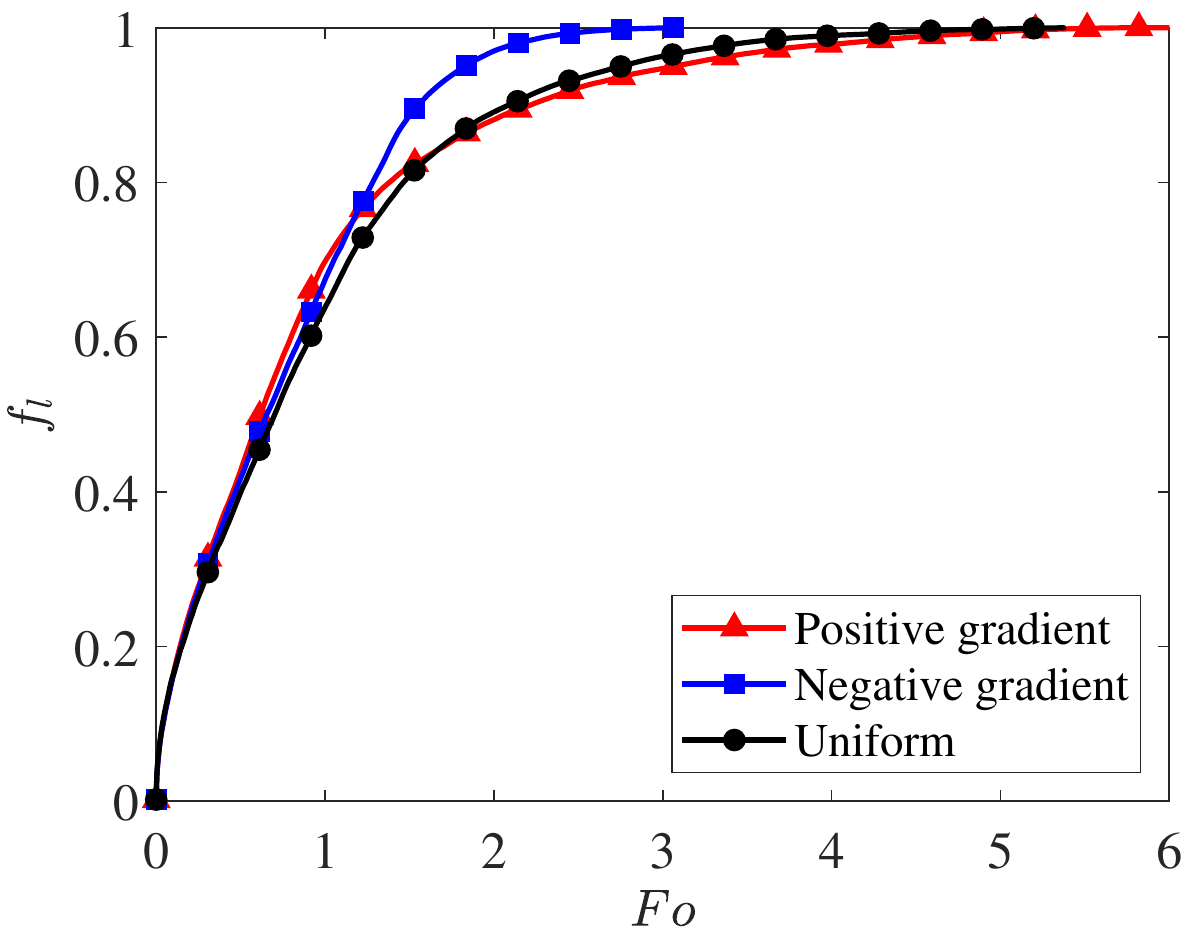}}
		\caption{Effects of gradient porosity on (a) total liquid fraction  and (b) average Nusselt number along the left wall at $ Ra=10^{6} $.}
		\label{fig7}
	\end{figure}

	%\subsection{The effect of gradient porosity}
	%\label{section4_1}
	We first examine the effect of the gradient porosity on melting behavior at $ Ra=10^{6} $. Fig. \ref{fig7} illustrates the liquid fraction $ f_{l} $ as a function of dimensionless time $ Fo $ for Case A and Case B with different gradient structures: uniform, positive and negative gradients. Based on the difference in the slope of the melting curve, the melting process can be divided into three stages. At the early stage, heat is mainly transferred through conduction so that the curves of the three gradient porosity structure completely overlap. With time elapsed, $ f_{l} $ incresaes as $ Fo $ incresses, and the trend is firstly steep and then gradually weaker, which indicates thermal conductivity  performance degradation for melting interface. Specifically, the liquid fraction of the negative gradient in Case A is higher than the other two structures, and this trend is maintained until completely melted, which may due to the 
	
	\noindent fact that the left part with relatively low porosity of negative gradient leads to a growth in natural convection strength. However, the situation of Case B can get a little complicated, a turning point is observed and the liquid fraction of negative gradient increases with a higher rate compared to the other cases while the melting rate of other cases decreases. Eventually, since the fact that the main heat conduction mechanism has changed from the natural convection to the heat conduction, the melting rate of the three gradients structure is reduced. The negative gradient ends up with a relatively short period of time in Case A and Case B, followed by the uniform and the negative cases.

\begin{figure}[H]
		\centering
		
		\begin{minipage}[c]{0.2\textwidth}
			\centering
			\caption*{Case A: negative gradient}
			%\label{fig:side:caption}
		\end{minipage}
		\begin{minipage}[c]{0.2\textwidth}
			\includegraphics[width=\textwidth]{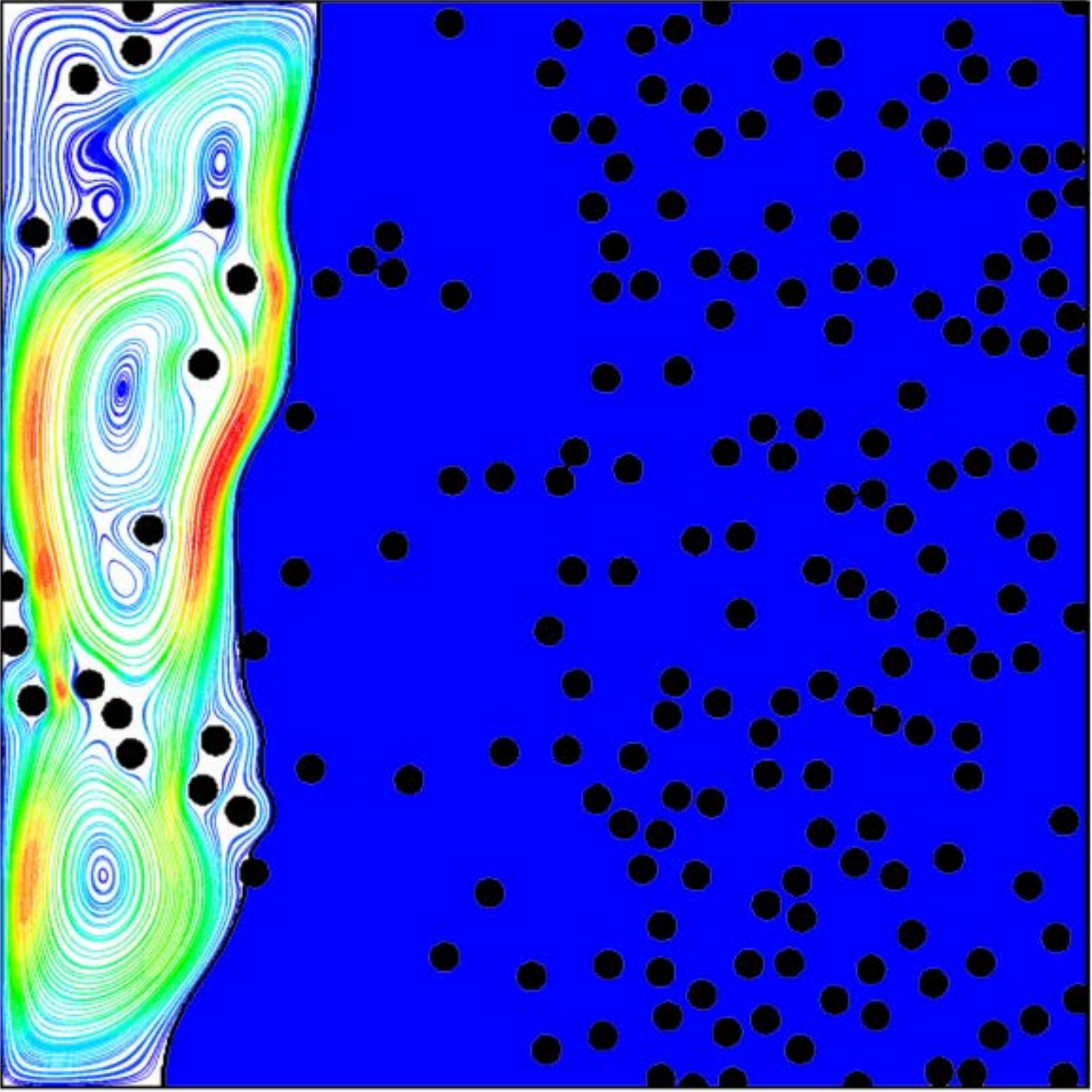}
		\end{minipage}
		\begin{minipage}[c]{0.2\textwidth}
			\includegraphics[width=\textwidth]{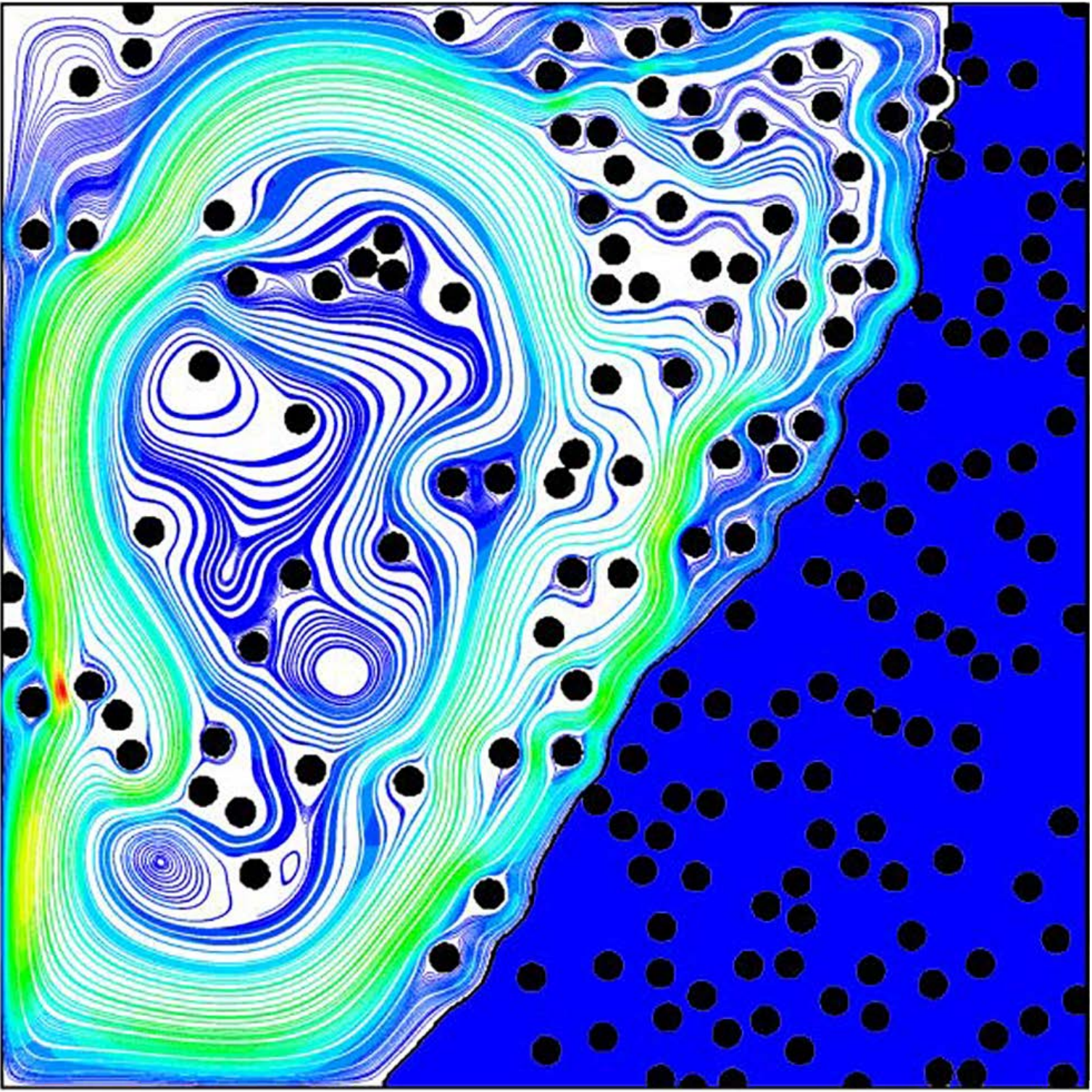}
		\end{minipage}
		\begin{minipage}[c]{0.2\textwidth}
			\includegraphics[width=\textwidth]{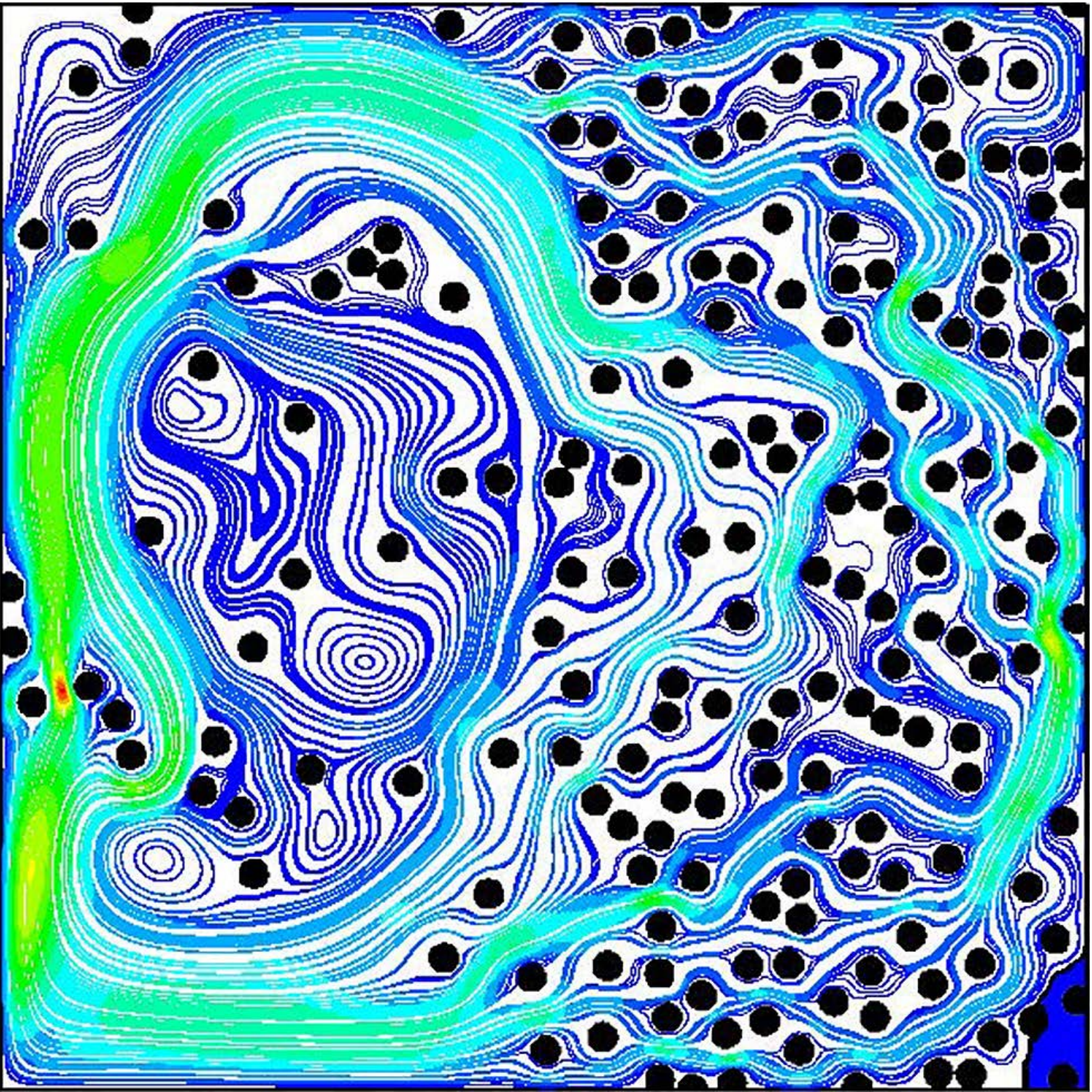}
		\end{minipage}

		\begin{minipage}[c]{0.2\textwidth}
			\centering
			\caption*{Case A: positive gradient}
			%\label{fig:side:caption}
		\end{minipage}
		\begin{minipage}[c]{0.2\textwidth}
			\includegraphics[width=\textwidth]{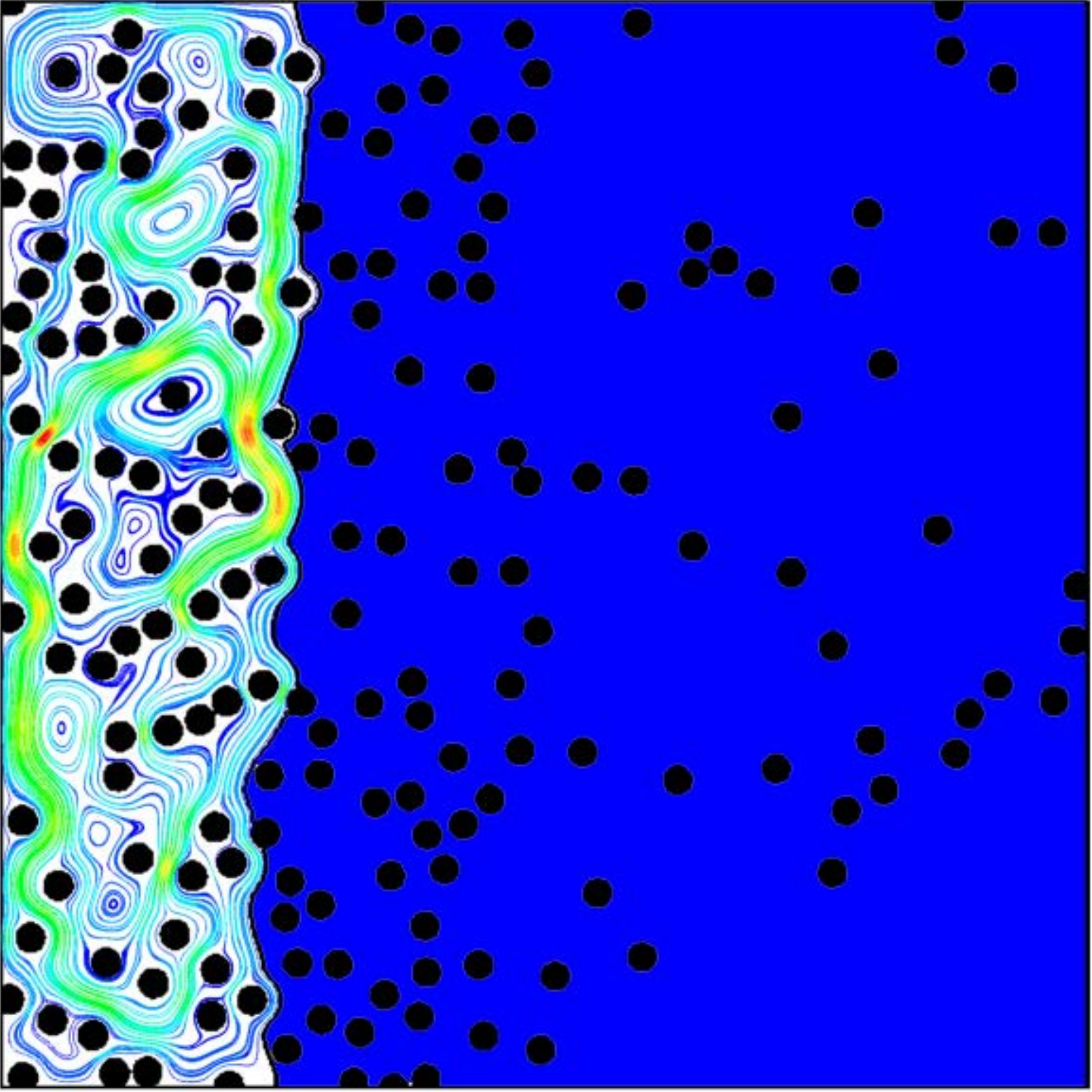}
		\end{minipage}
		\begin{minipage}[c]{0.2\textwidth}
			\includegraphics[width=\textwidth]{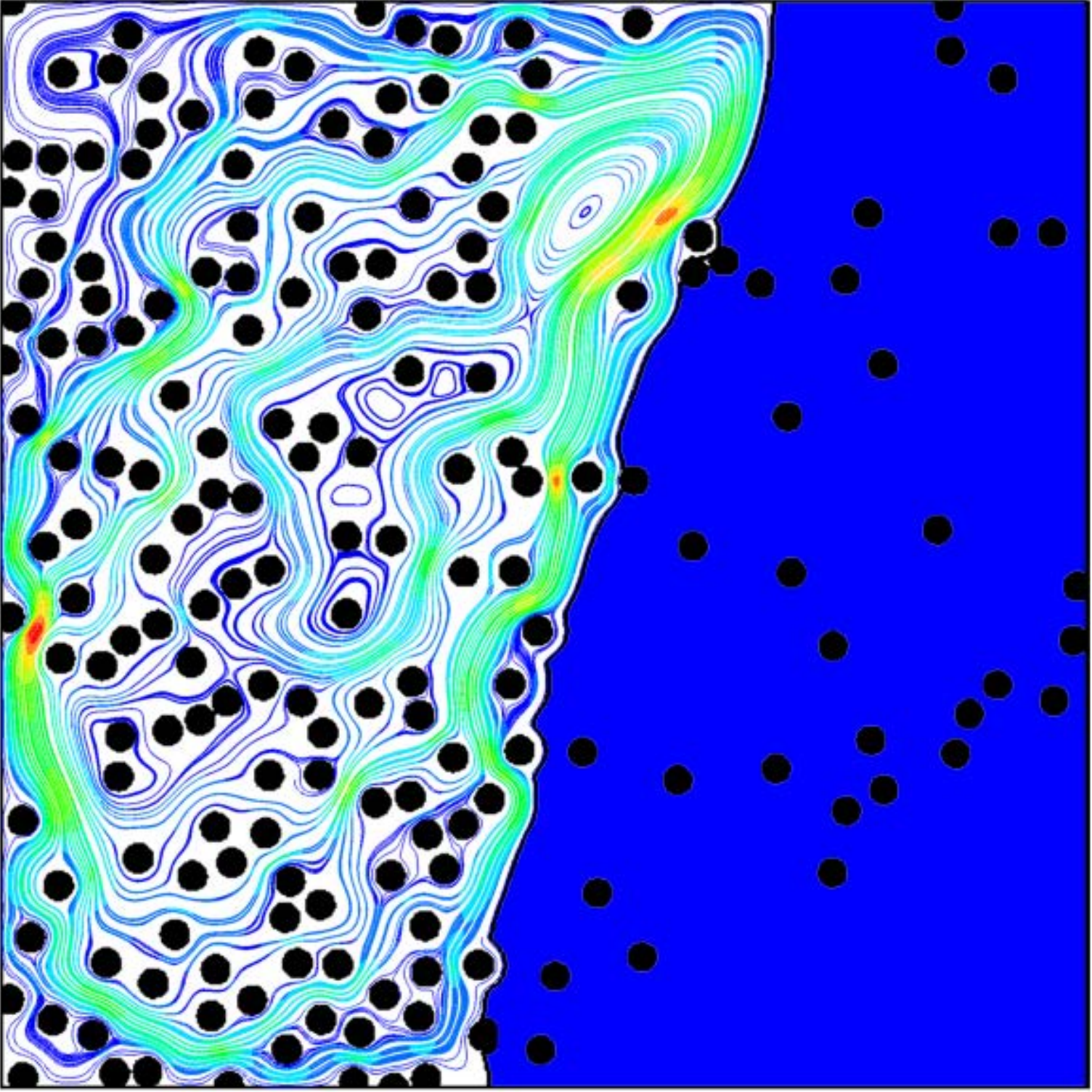}
		\end{minipage}
		\begin{minipage}[c]{0.2\textwidth}
			\includegraphics[width=\textwidth]{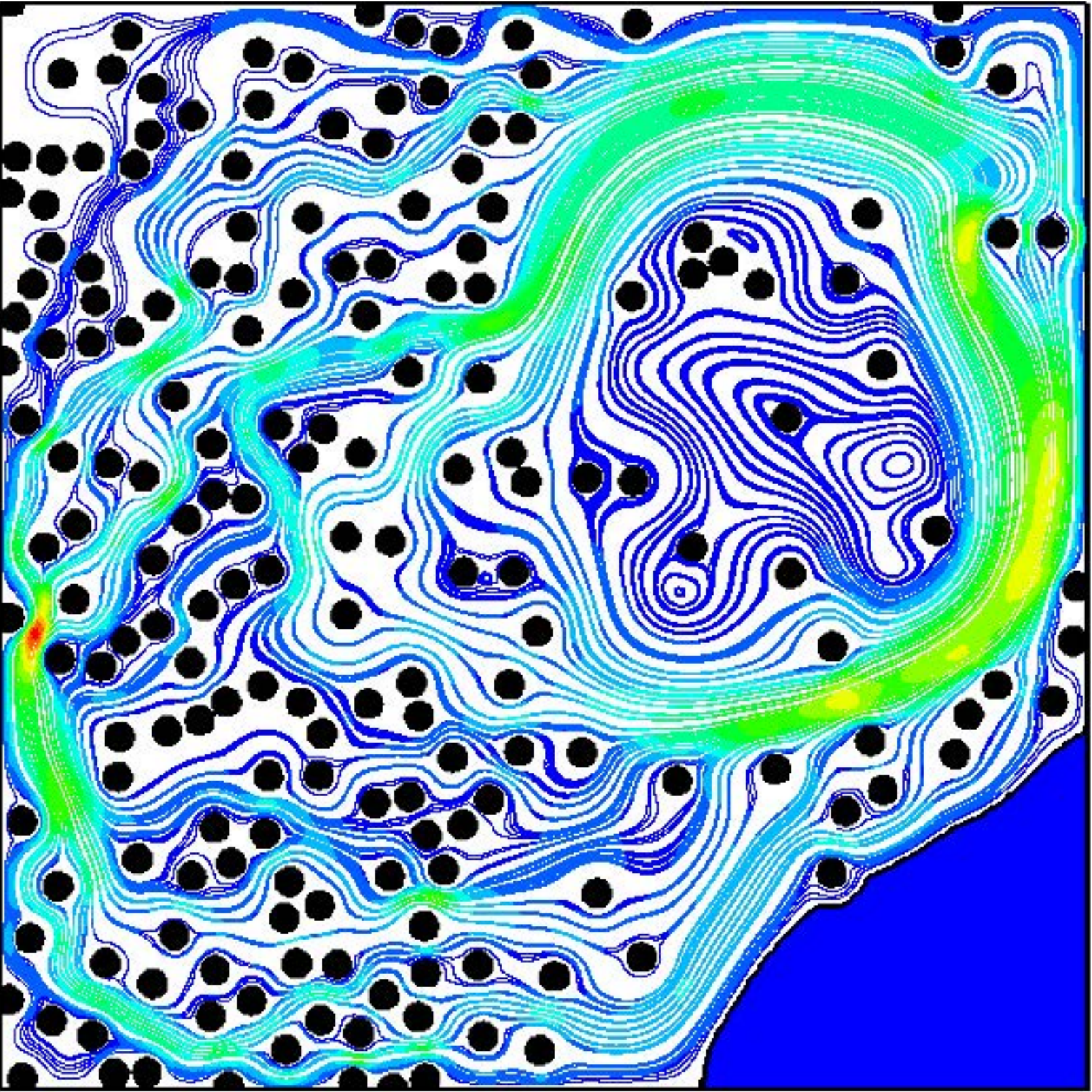}
		\end{minipage}
		
		\begin{minipage}[c]{0.2\textwidth}
			\centering
			\caption*{Uniform gradient}
			%\label{fig:side:caption}
		\end{minipage}
		\begin{minipage}[c]{0.2\textwidth}
			\includegraphics[width=\textwidth]{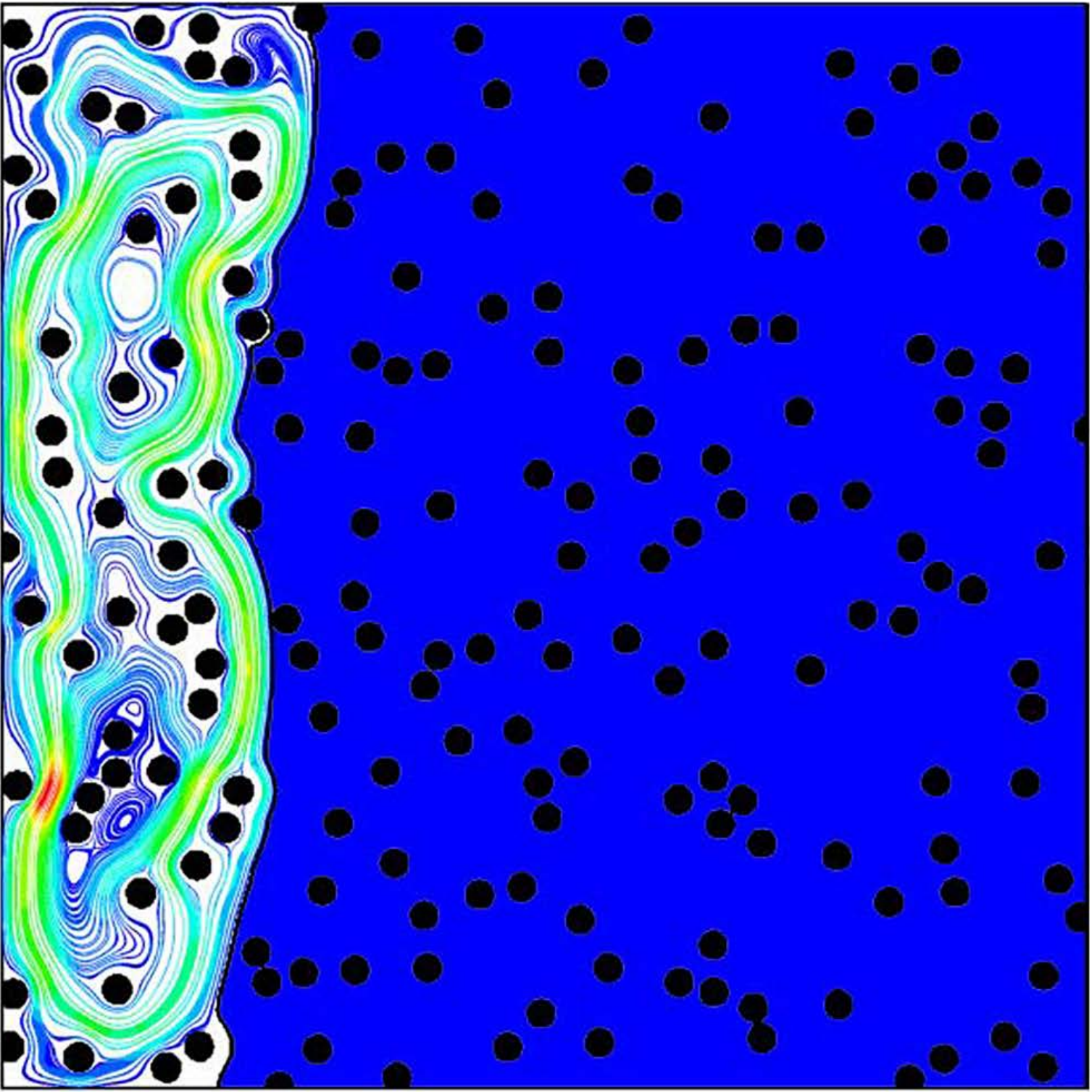}
		\end{minipage}
		\begin{minipage}[c]{0.2\textwidth}
			\includegraphics[width=\textwidth]{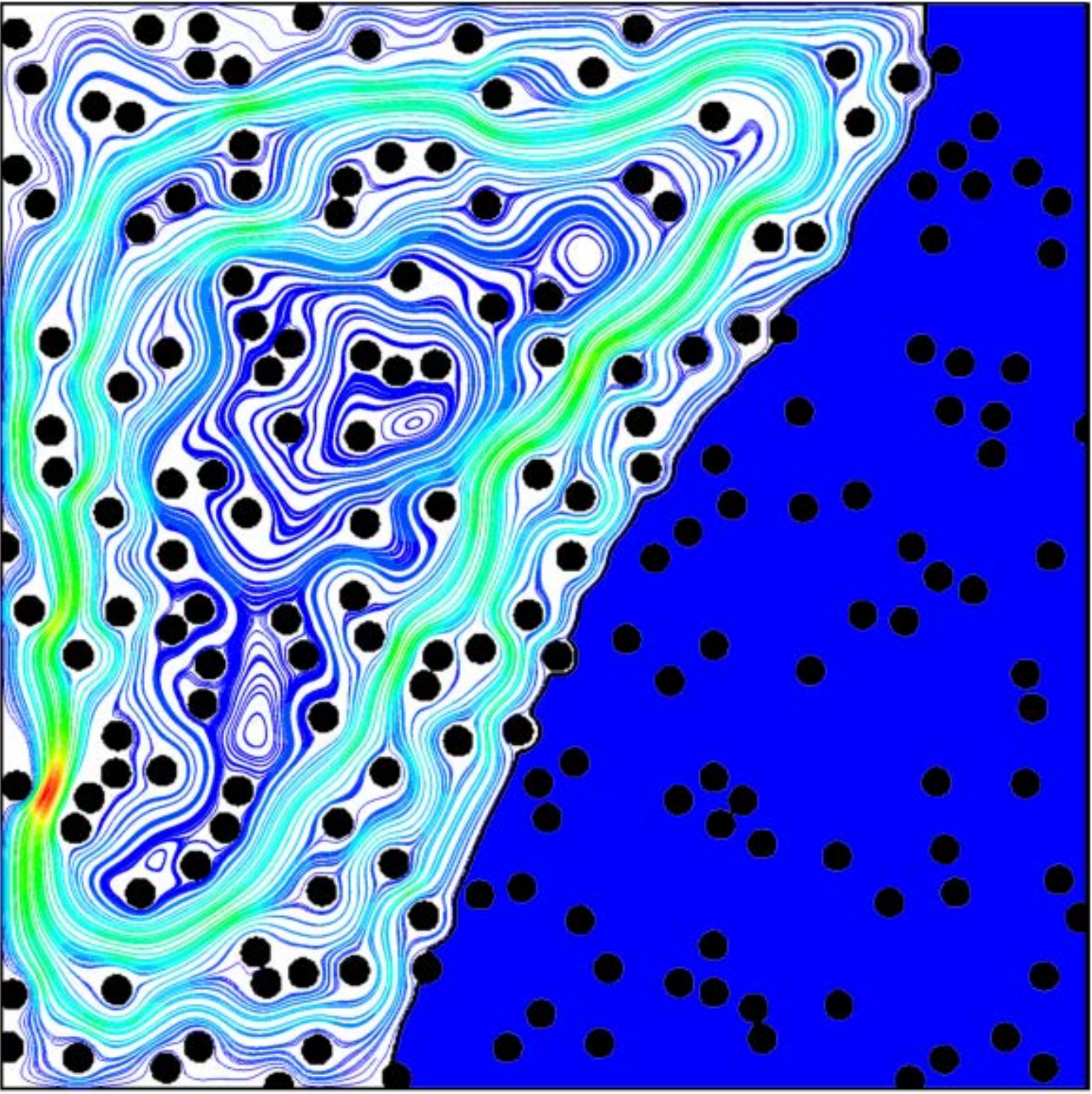}
		\end{minipage}
		\begin{minipage}[c]{0.2\textwidth}
			\includegraphics[width=\textwidth]{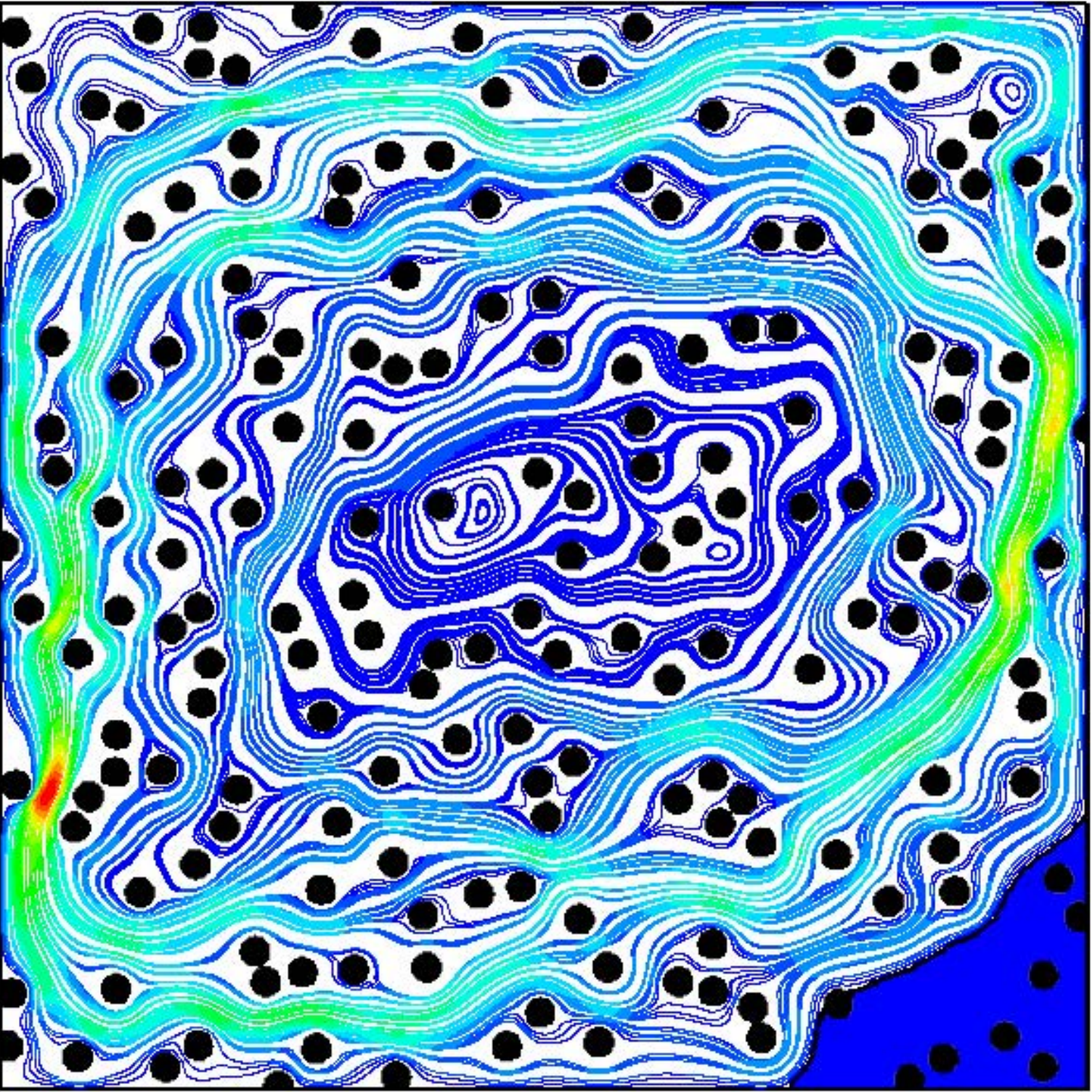}
		\end{minipage}
		
		\begin{minipage}[c]{0.2\textwidth}
			\centering
			\caption*{Case B: negative gradient}
			%\label{fig:side:caption}
		\end{minipage}
		\begin{minipage}[c]{0.2\textwidth}
			\includegraphics[width=\textwidth]{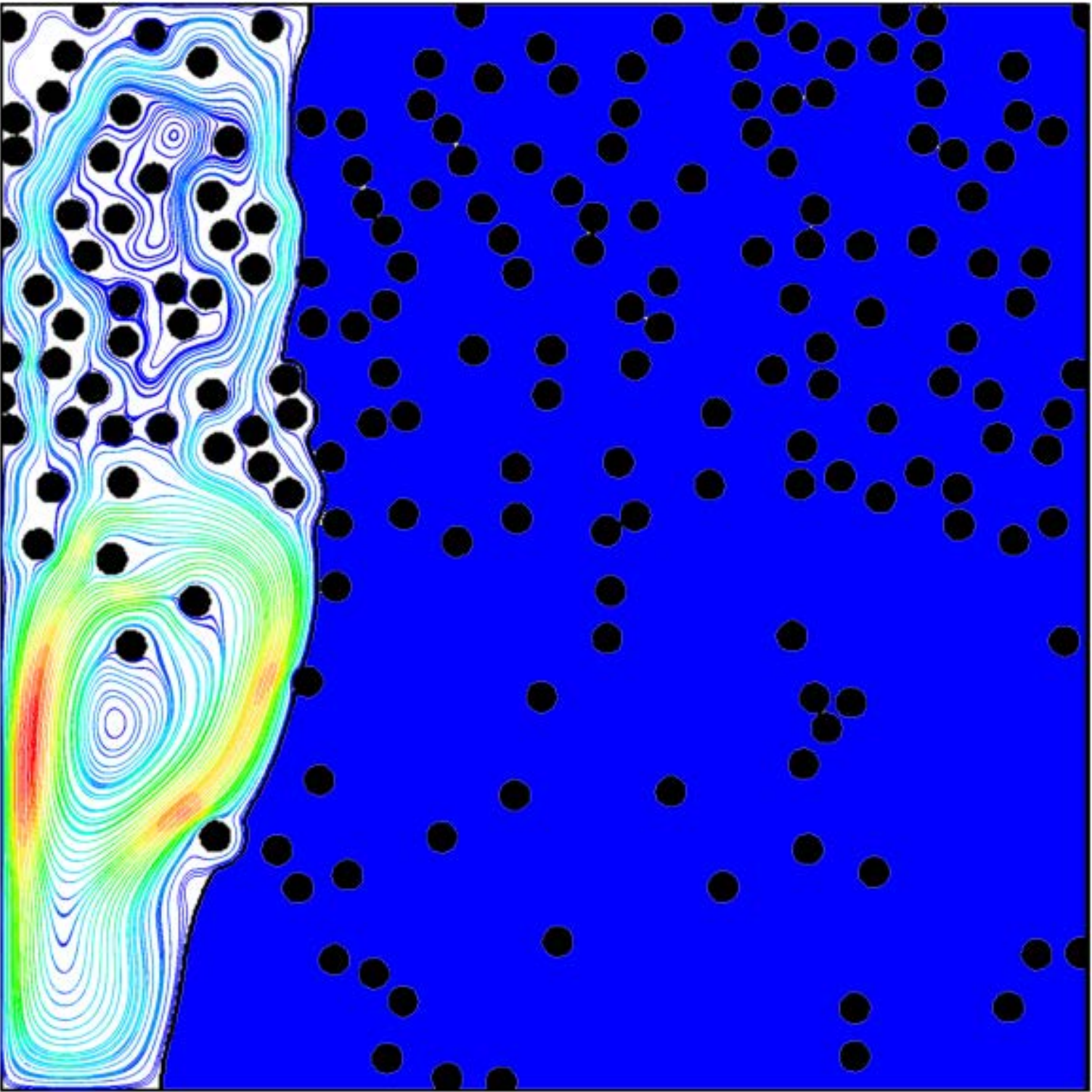}
		\end{minipage}
		\begin{minipage}[c]{0.2\textwidth}
			\includegraphics[width=\textwidth]{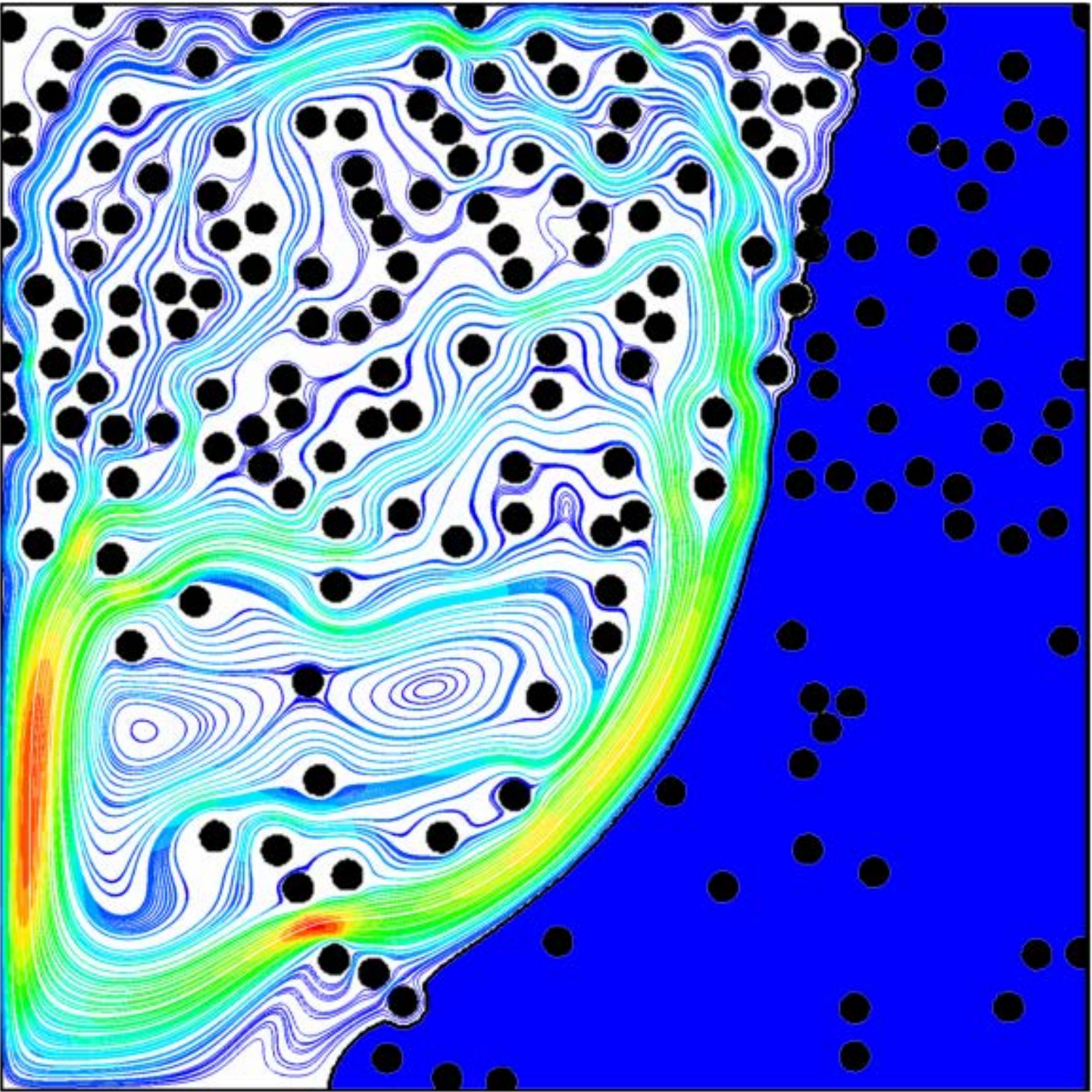}
		\end{minipage}
		\begin{minipage}[c]{0.2\textwidth}
			\includegraphics[width=\textwidth]{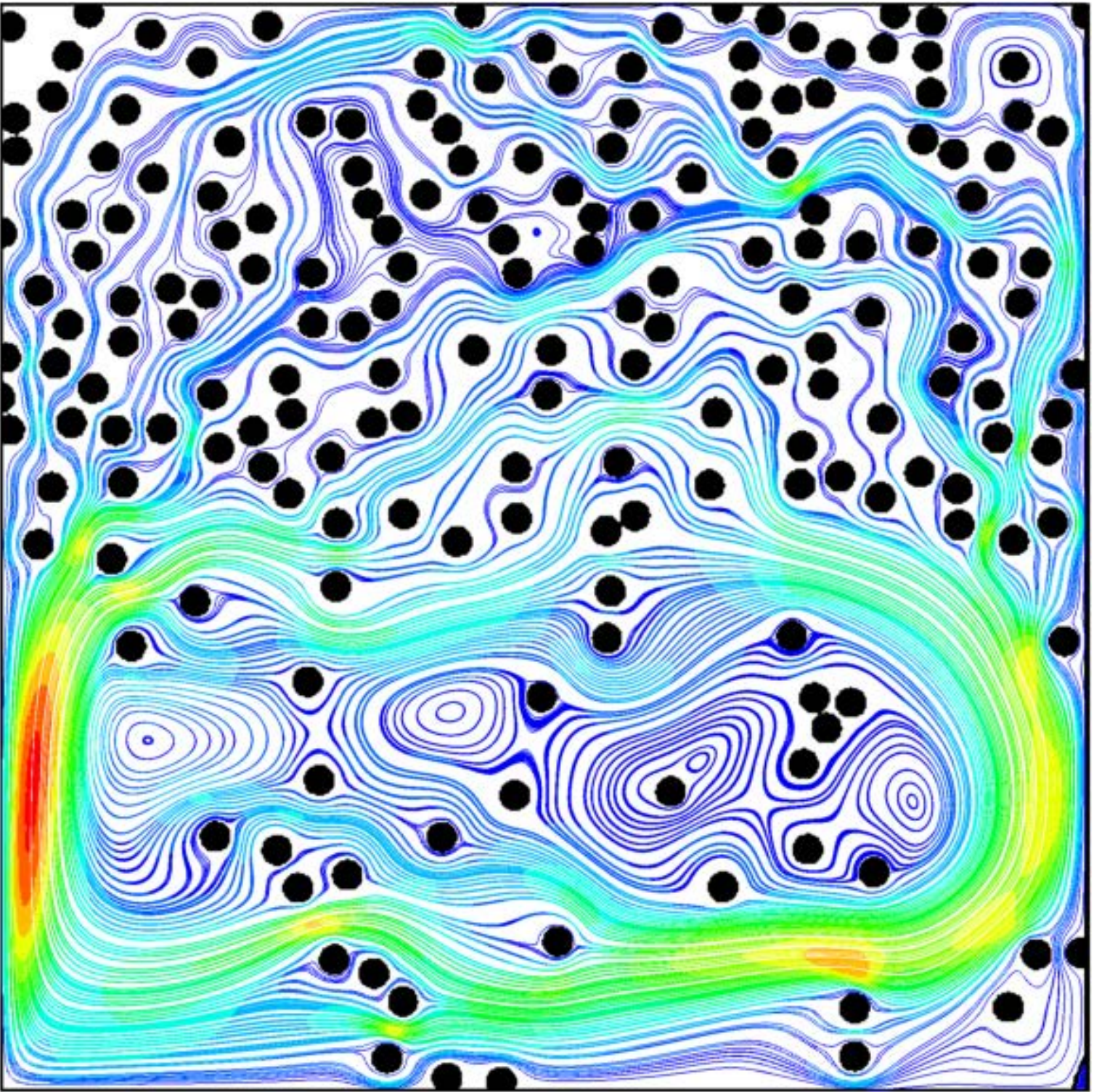}
		\end{minipage}
		
		\begin{minipage}[c]{0.2\textwidth}
			\centering
			\caption*{Case B: positive gradient}
			%\label{fig:side:caption}
		\end{minipage}
		\begin{minipage}[c]{0.2\textwidth}
			\includegraphics[width=\textwidth]{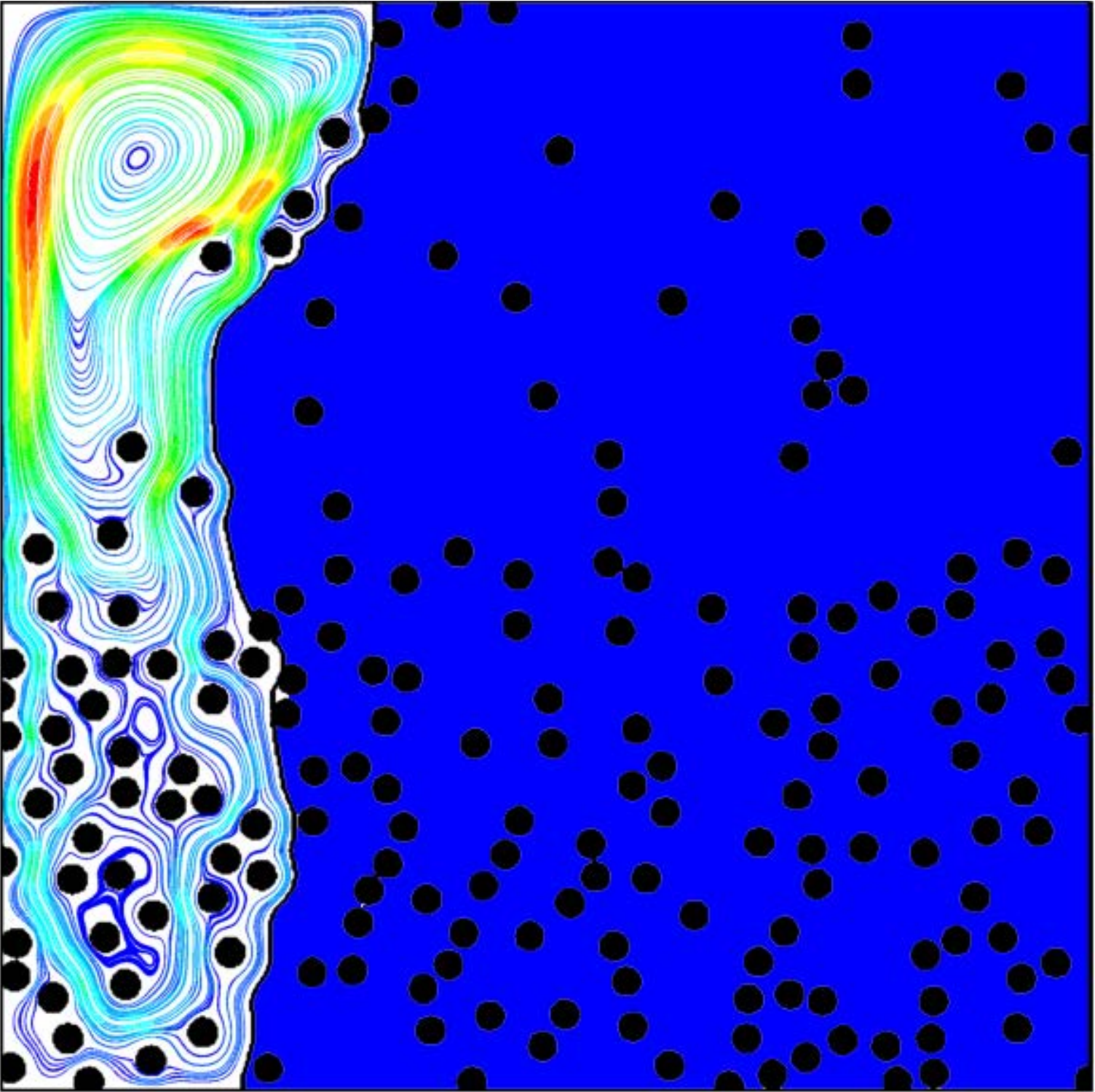}
			\caption*{$ Fo=0.2 $}
		\end{minipage}
		\begin{minipage}[c]{0.2\textwidth}
			\includegraphics[width=\textwidth]{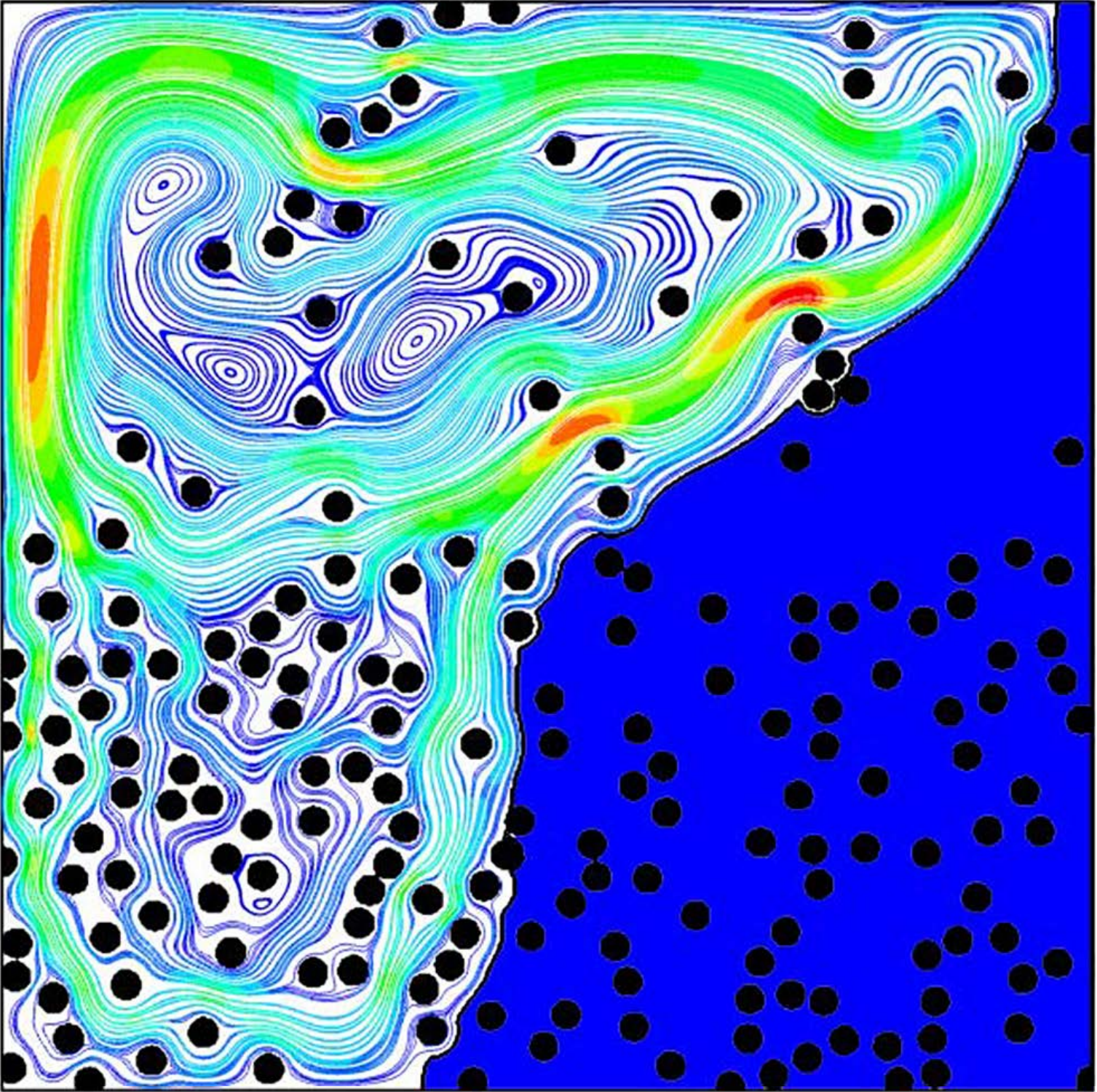}
			\caption*{$ Fo=0.9 $}
		\end{minipage}
		\begin{minipage}[c]{0.2\textwidth}
			\includegraphics[width=\textwidth]{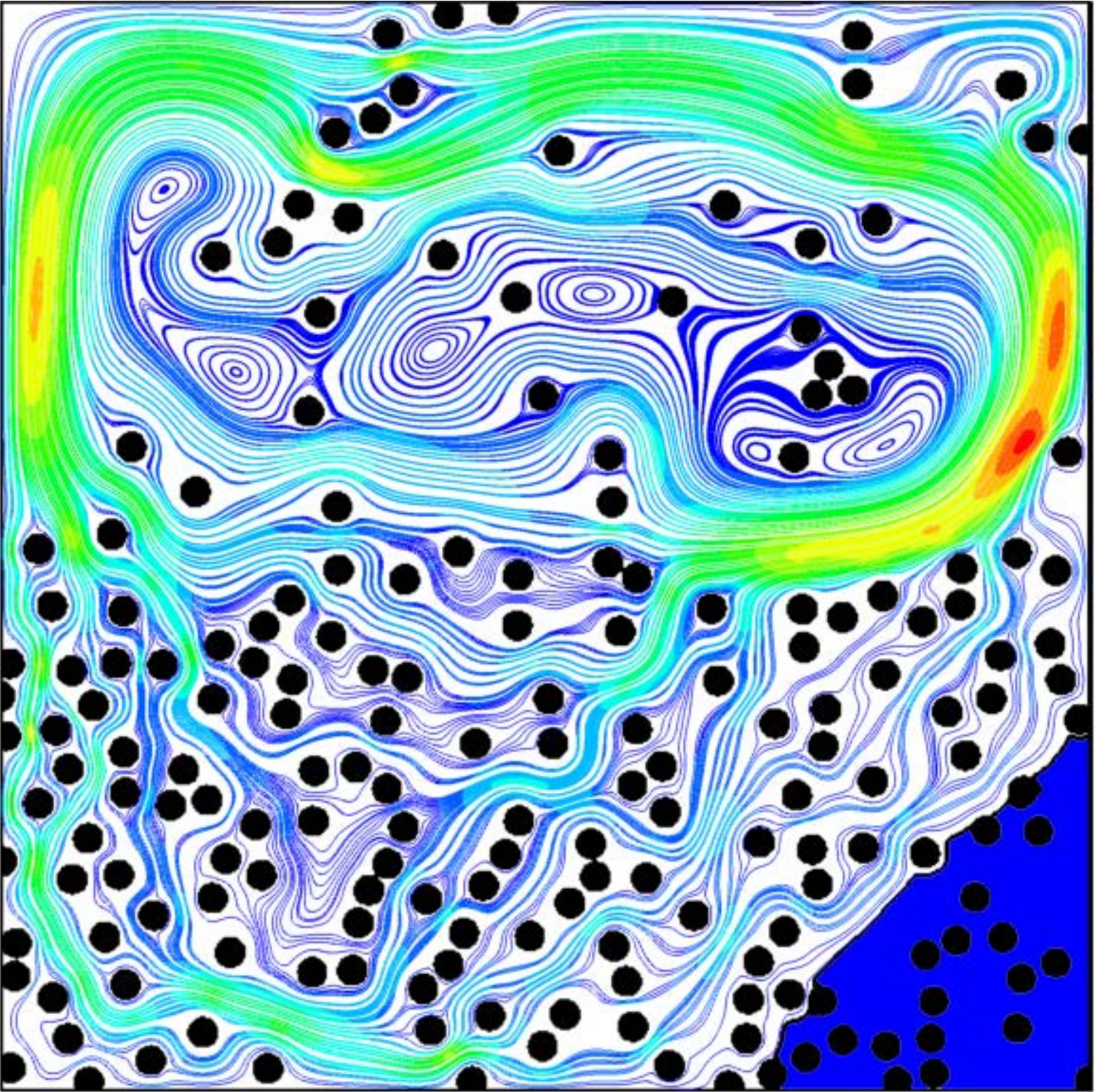}
			\caption*{$ Fo=3.0 $}
		\end{minipage}
		\caption{The effect of gradient porosity on the total liquid fraction at $ Ra=10^{6} $.} 
		\label{fig8}
	\end{figure}

	In order to intuitively understand the flow transition process in different gradient structures. Fig. \ref{fig8} shows streamline and total liquid fraction distributions with different gradient structure at $ Ra=10^{6} $ condition, where

	\noindent  the liquid and solid phase zone are represented by the white and blue region, porous metal foams are denoted by black region. In the first place, the phenomenon in Fig. \ref{fig8} is caused by the combined effect of natural convection and conduction during the phase change. The final performance depends on which of natural convection and heat conduction is dominant during the melting process. In the initial stage of melting, there is not enough liquid PCM that can support natural convection, and natural convection has little effect on heat transfer resulting in heat conduction as the main heat conduction mechanism. As a consequence, the general trend of the solid-liquid interface is roughly perpendicular to the upper and lower wall for three different gradient structures. As the $ Fo $ increase ($ Fo=0.9 $), the gradual increase in melted PCM provides a larger carrier for natural convection resulting that the main heat transfer mechanism gradually changed from heat conduction to natural convection. As the results showed, the velocity fields were consistent with the phase-change interface. The molten PCM carrying energy flowed upward driven by the buoyancy force. It then changes the flow direction under the influence of the temperature gradient when it reaches the top of the wall, transfers energy to the vicinity of the solid-liquid interface. Consequently, the phase change interface moves forward and gradually tilts, the melting zone expands, and the liquid zone gradually thickens until the top reaches the right side. In the final stage of the phase transition process, the driving force of natural convection is reduced in the final stage, the PCM at the bottom right will slowly melt. This phenomenon is called "bottom corner phenomenon" \cite{guai}, which worsens the conduction heat transfer at the bottom region and reduce energy storage efficiency and should be avoided. Furthermore, the PCM near the foamed metal always preferentially melt due to the conductivity of porous metal foams is higher that it of PCM. It eventually lead to tilt and fluctuation of the solid-liquid interface. However, at the bottom of the PCM, natural convection is weaker due to weaker buoyancy and increasing thermal resistance. Therefore, the heat conduction plays a major role in the heat transfer process. 
	
	On the other hand, the difference in gradient structure is also one of the influencing factors. The increase of porosity contributes to the decrease of the intensity of natural convection, resulting in a strong suppression of the natural convection of the molten PCM. Compared to uniform case, negative gradient in Case A provides the lowest thermal conductivity and the lowest convection resistance in the left area because it has the highest porosity on the left region. As a result, stronger natural convection is formed in the left area, bringing more heat near the solid-liquid interface, so that the negative gradient has the fastest melting speed in the mid-term. The negative gradient has the lowest porosity and the highest thermal conductivity on the right side. Therefore, heat is transferred more effectively to the right side of the cavity during the final melting phase resulting in a smaller total melting time. Compared with the negative gradient, the positive gradient porosity has the completely opposite porosity gradient direction, which provides the highest resistance to natural convection at left area and also lower thermal conductivity at the bottom right section of the enclosure. Therefore, this gradient structure has a lower melting rate. Moreover, although the highest thermal conductivity on the left side of the positive gradient makes the melting area larger in the early stage, its advantages are not shown on the above melting curve due to the area of the PCM in this area is minimized. For Case B, negative gradient provides higher conductivity at the bottom at the top and lower suppression of circulations. As a result, the two main heat transfer mechanisms are effectively present at the heat source. And this structure further strengthened the scouring action of local natural convection at the bottom. The advantage of lower suppression of circulations at the bottom becomes more effective in the final melting stages especially at the bottom right section of the unit where other porous structures have slower melting rate. Additionally, the combined effect of natural convection and heat conduction leads to differences in local energy storage, which are shown in Fig. \ref{fig5} as local protrusions at the phase interface. Compared with Case A, the negative gradient of Case B has a shorter melting time mainly because the structure can eliminate the corner phenomenon of the bottom through the continuous scouring action of the local natural convection at the bottom to shorten the melting time. This shows the importance of providing lower suppression of circulations at the bottom for faster melting of PCM in energy storage units.

\begin{figure}[H]
		\centering 
		\subfigure[Case A]{ \label{fig9}
			\includegraphics[width=0.4\textwidth]{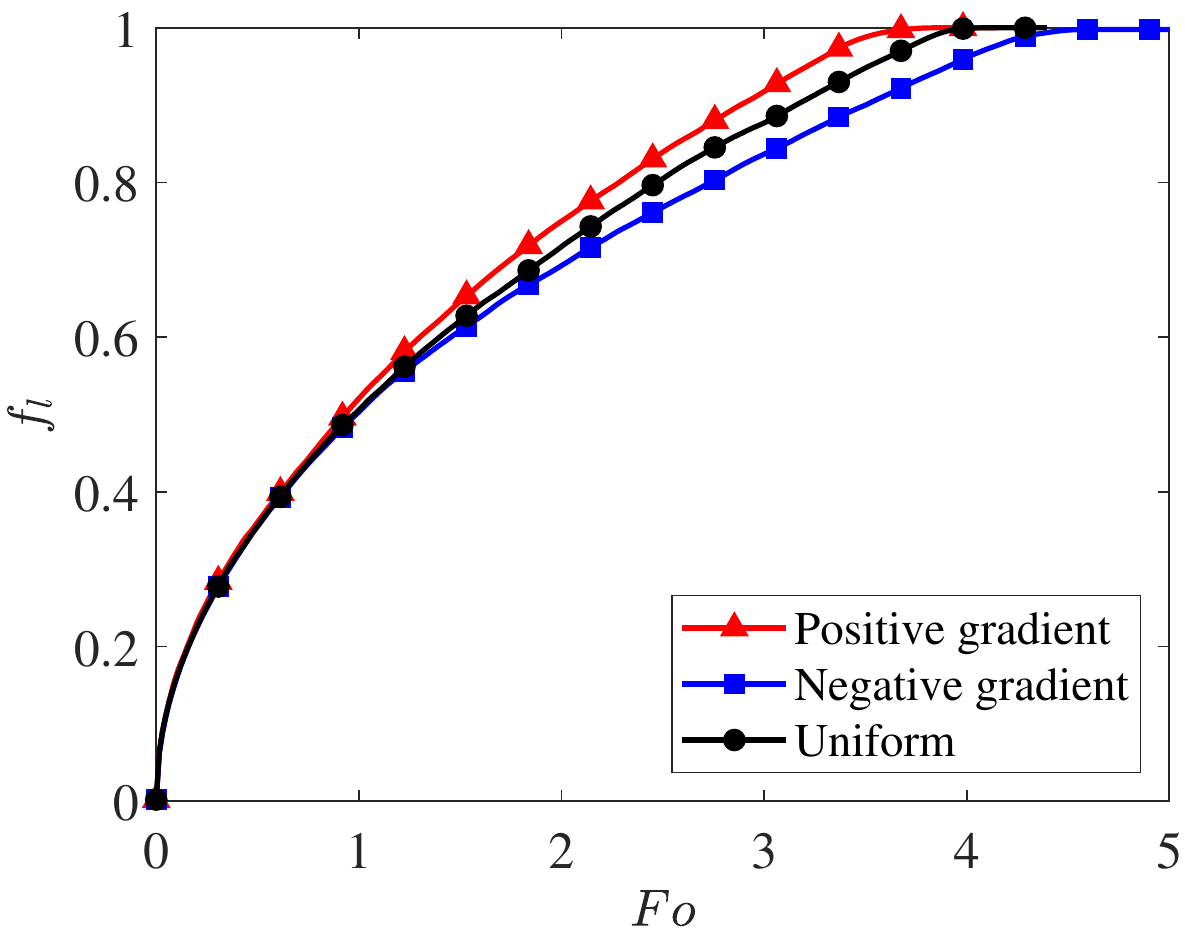}} 
		\subfigure[Case B]{ \label{fig10}
			\includegraphics[width=0.4\textwidth]{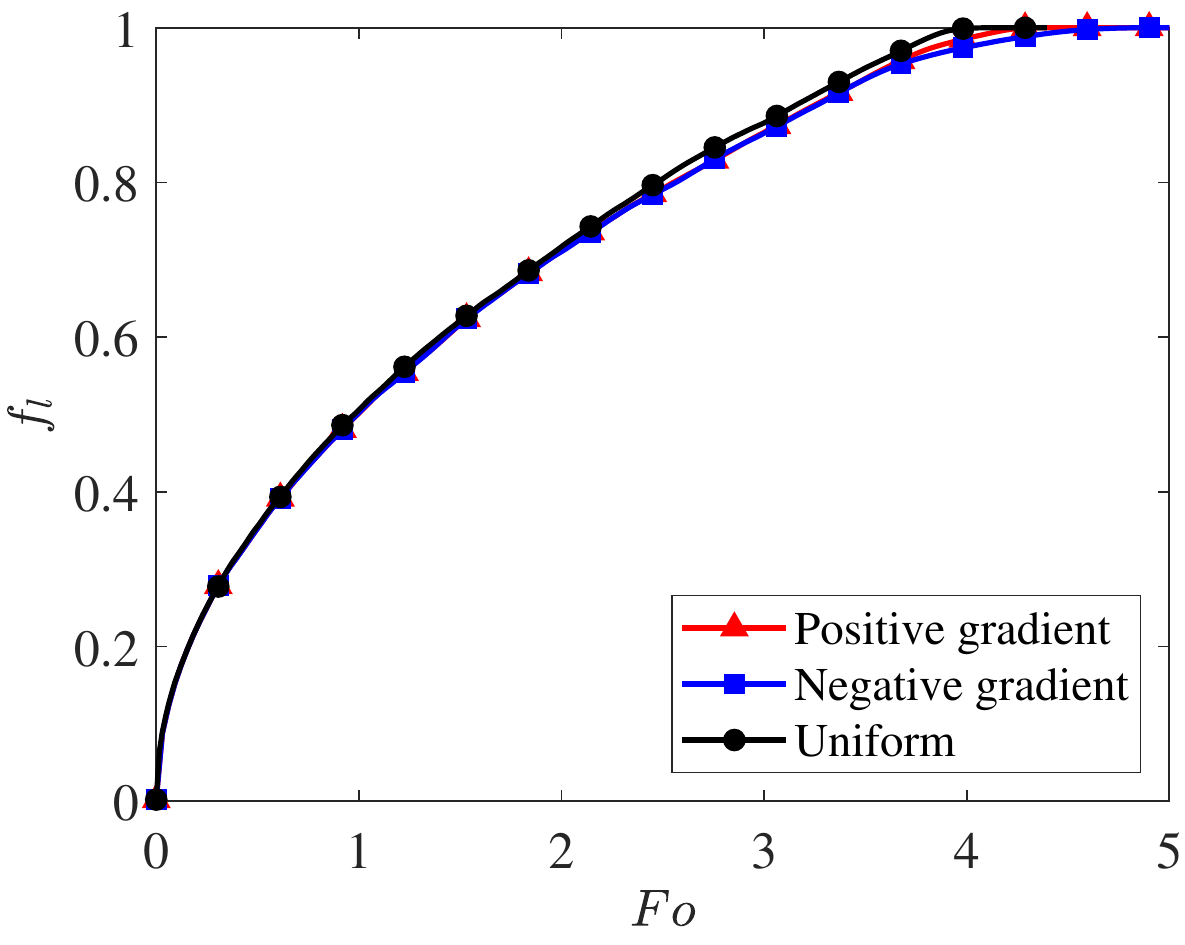}}
		\caption{Effects of gradient porosity on (a) total liquid fraction  and (b) average Nusselt number along the left wall at $ Ra=10^{4} $.}
		\label{fig11}
	\end{figure}

\begin{figure}[H]
		\centering
		
		\begin{minipage}[c]{0.2\textwidth}
			\centering
			\caption*{Case A: negative gradient}
			%\label{fig:side:caption}
		\end{minipage}
		\begin{minipage}[c]{0.2\textwidth}
			\includegraphics[width=\textwidth]{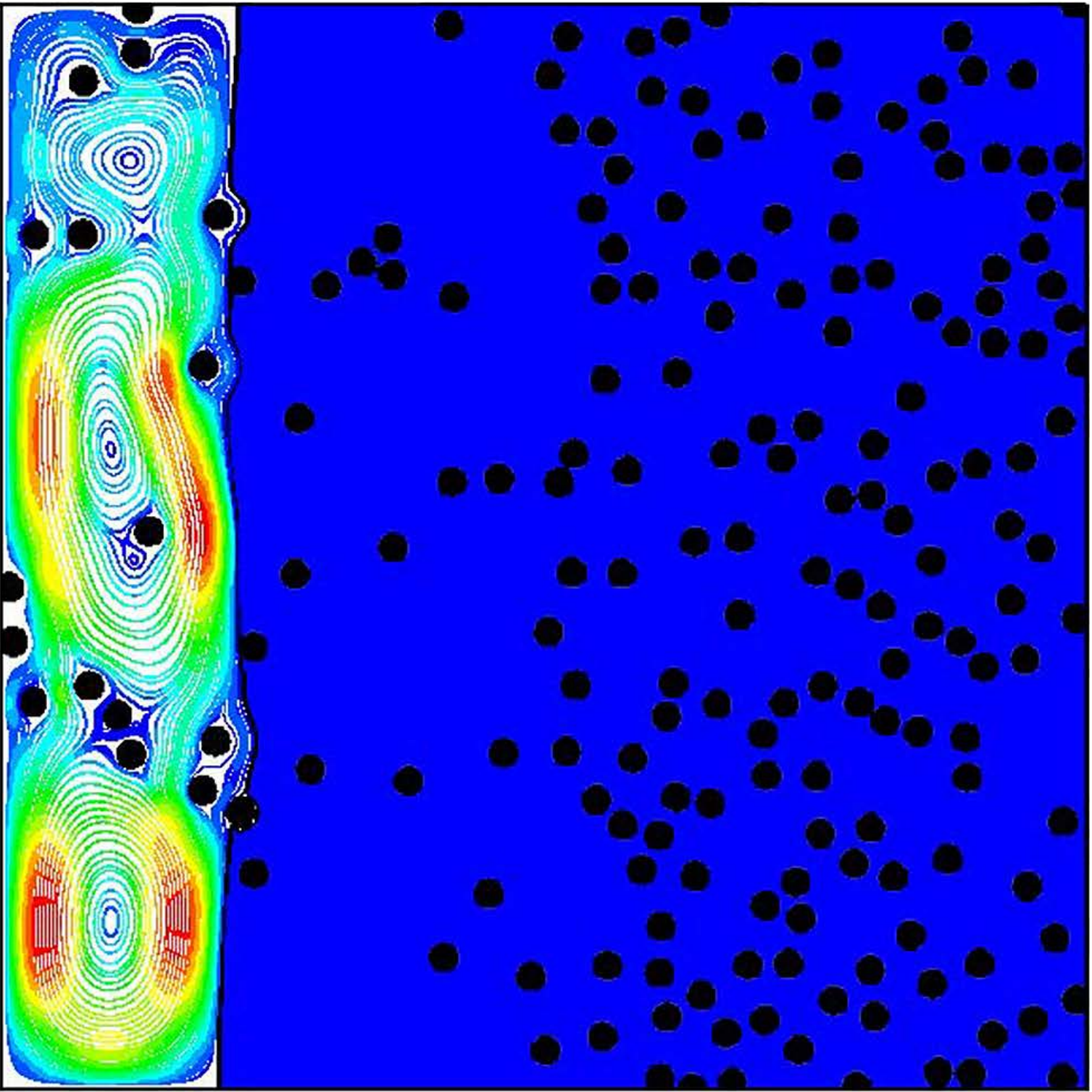}
		\end{minipage}
		\begin{minipage}[c]{0.2\textwidth}
			\includegraphics[width=\textwidth]{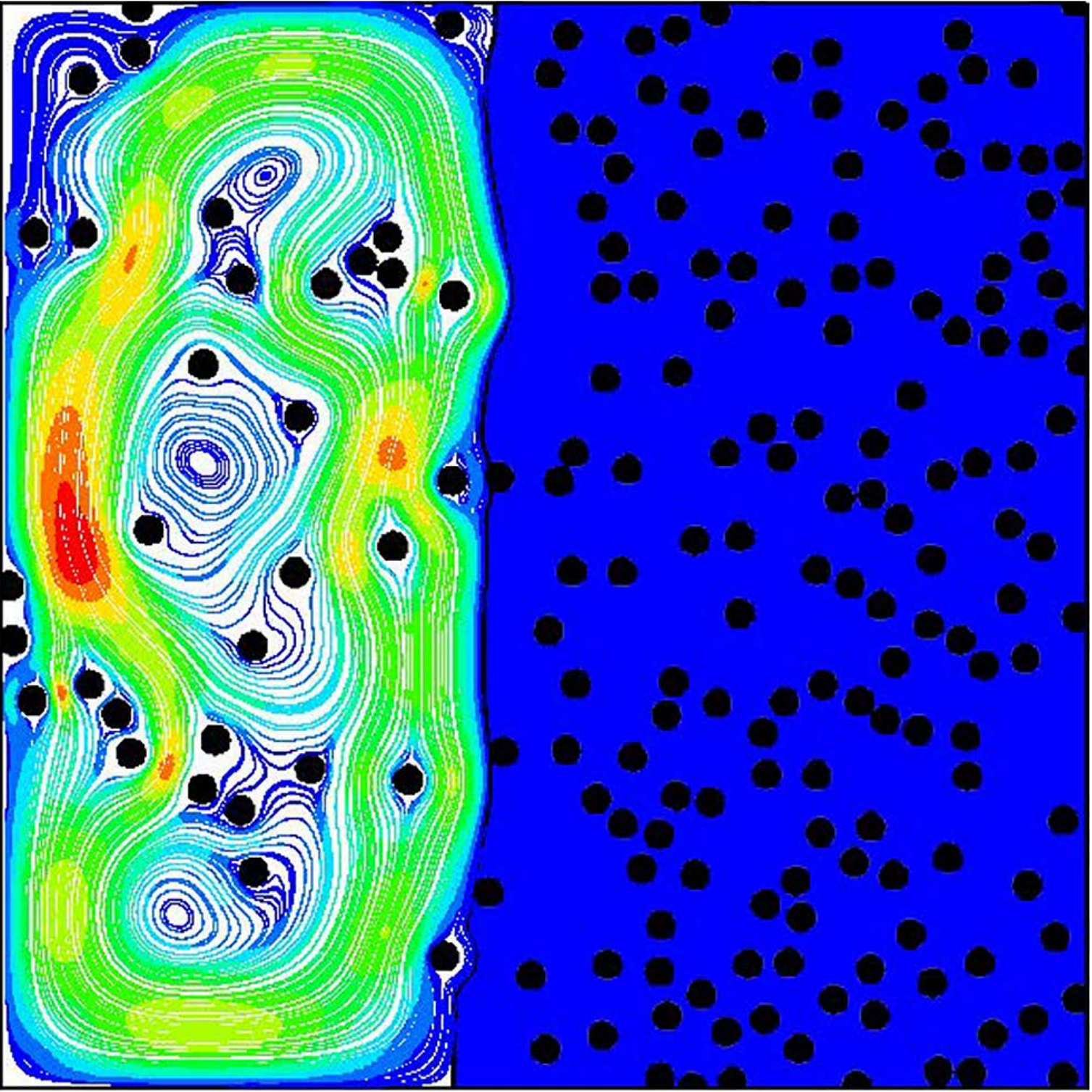}
		\end{minipage}
		\begin{minipage}[c]{0.2\textwidth}
			\includegraphics[width=\textwidth]{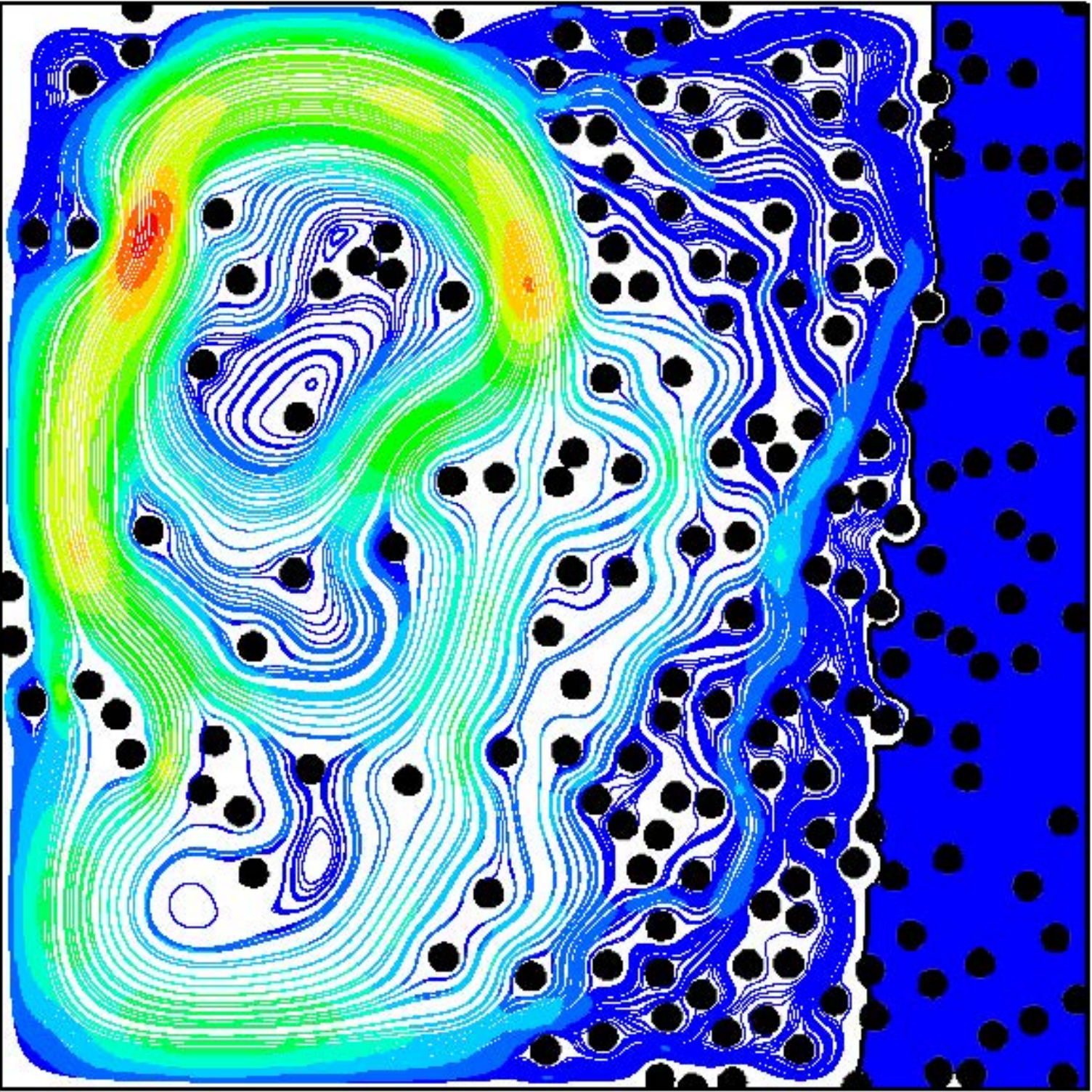}
		\end{minipage}

		\begin{minipage}[c]{0.2\textwidth}
			\centering
			\caption*{Case A: positive gradient}
			%\label{fig:side:caption}
		\end{minipage}
		\begin{minipage}[c]{0.2\textwidth}
			\includegraphics[width=\textwidth]{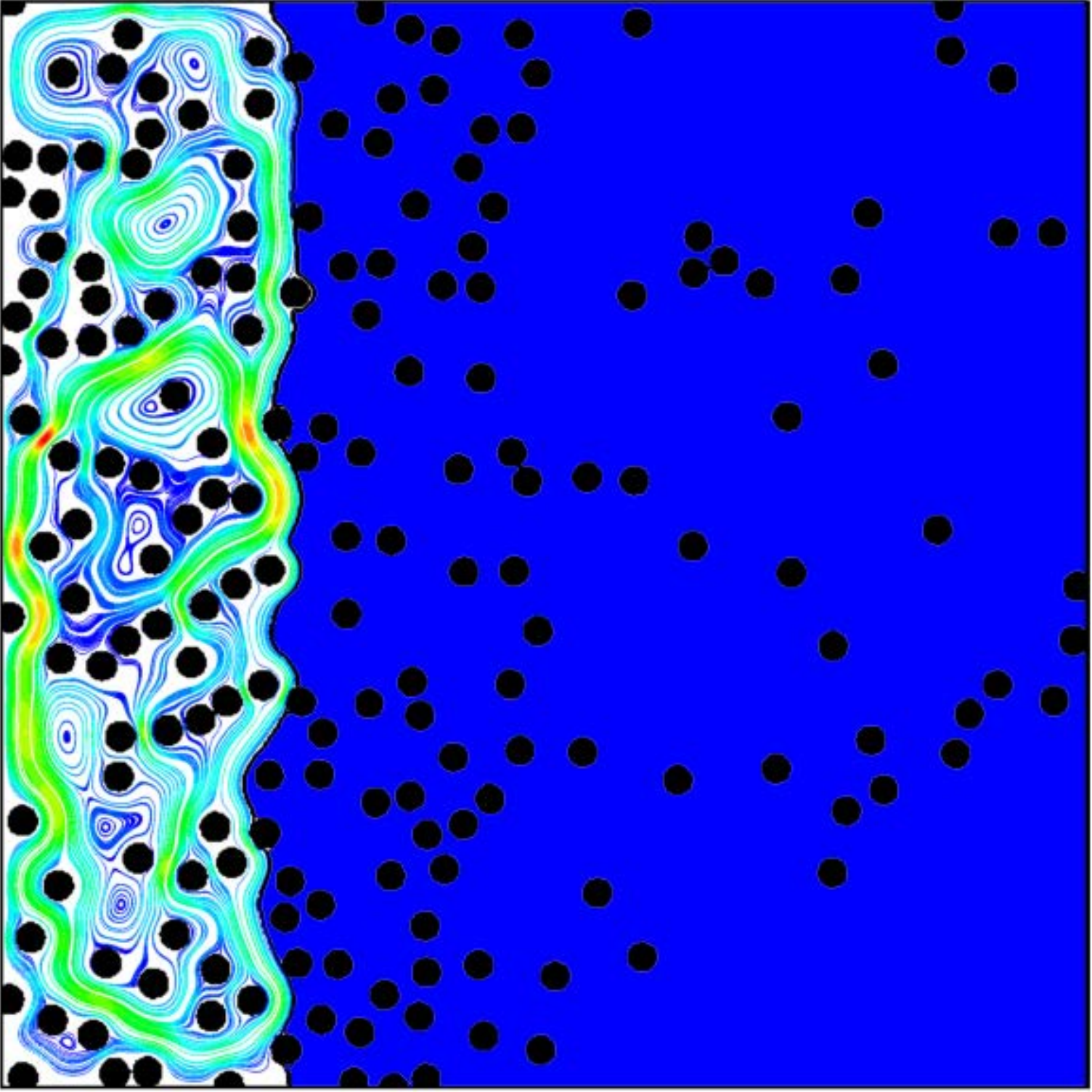}
		\end{minipage}
		\begin{minipage}[c]{0.2\textwidth}
			\includegraphics[width=\textwidth]{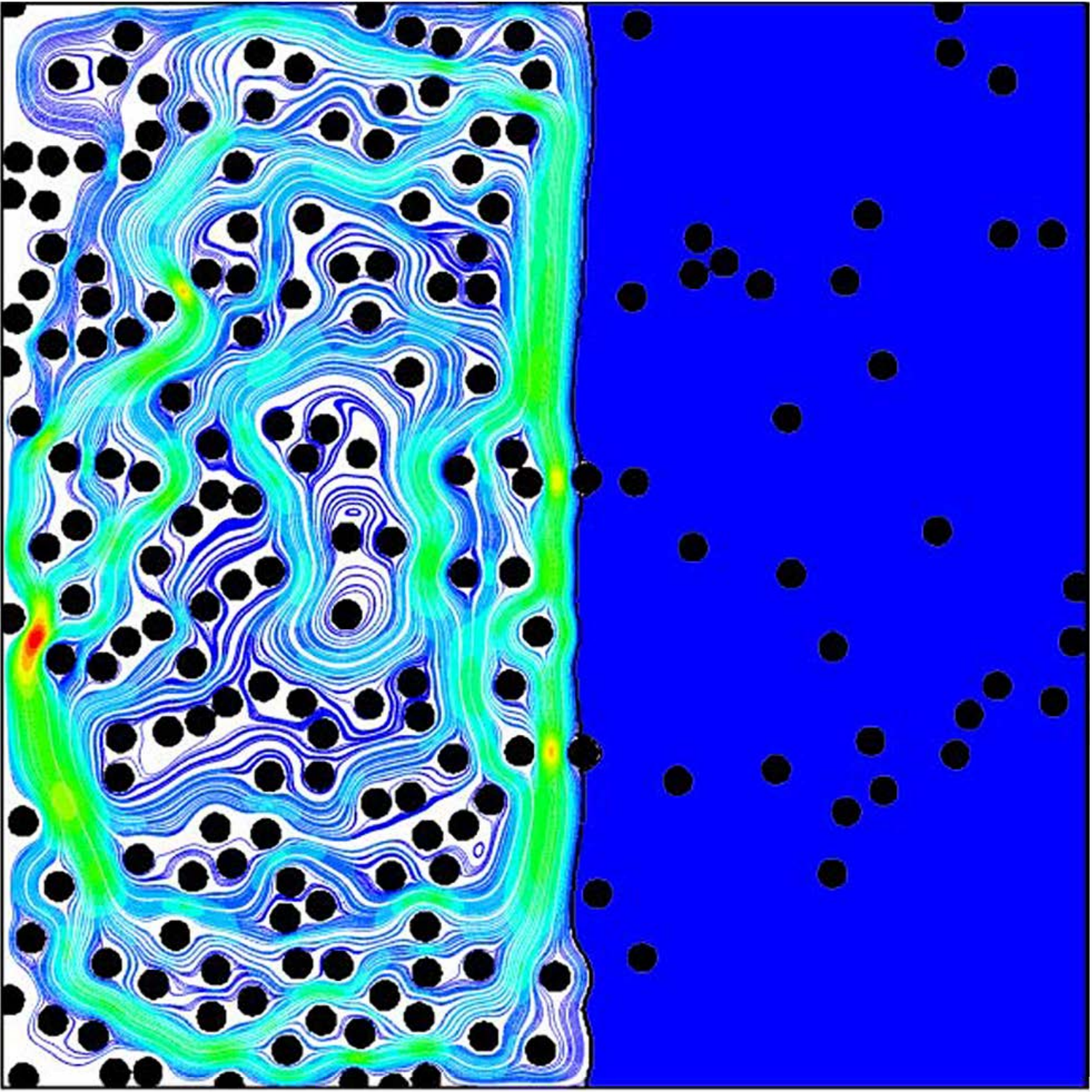}
		\end{minipage}
		\begin{minipage}[c]{0.2\textwidth}
			\includegraphics[width=\textwidth]{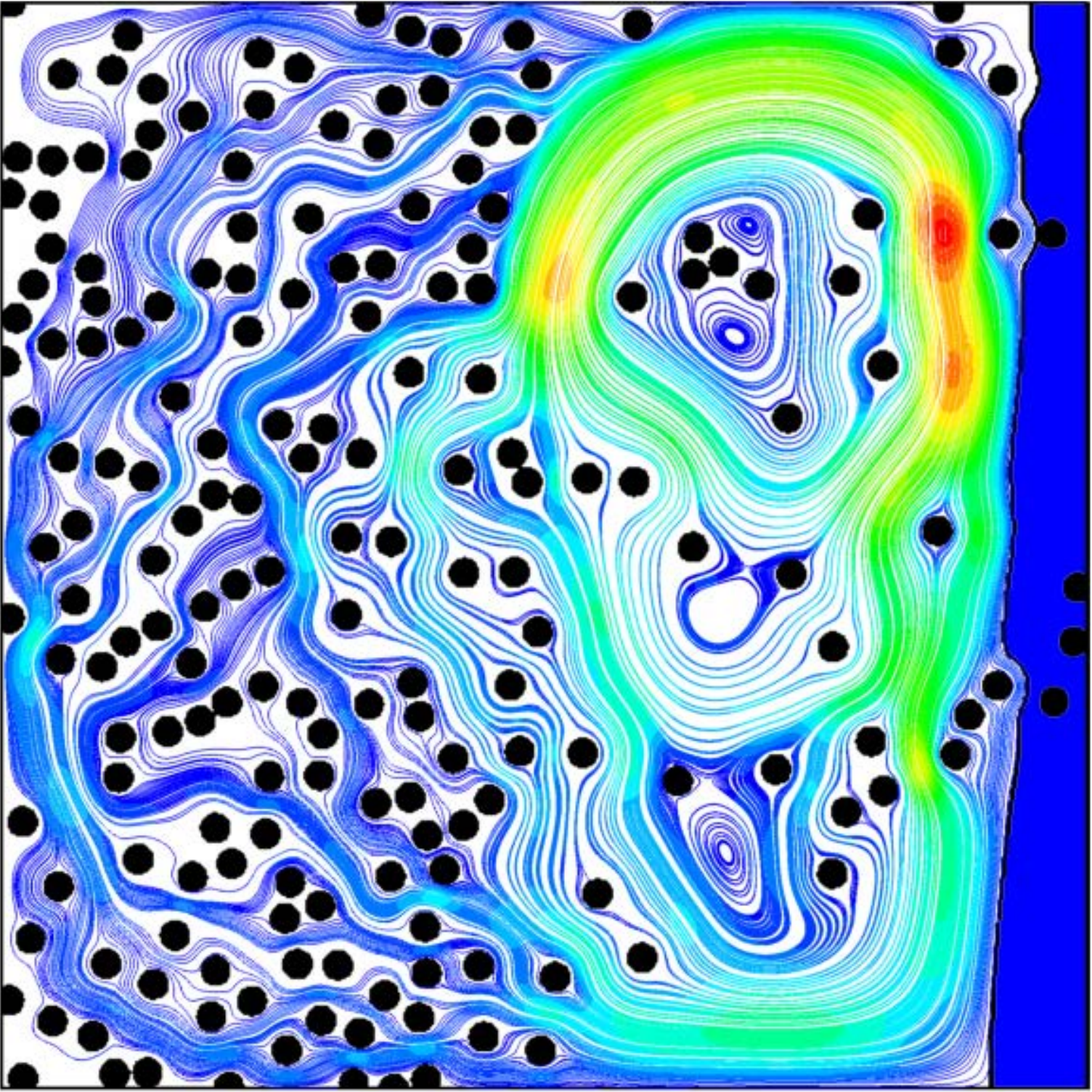}
		\end{minipage}

		\begin{minipage}[c]{0.2\textwidth}
			\centering
			\caption*{Uniform gradient}
			%\label{fig:side:caption}
		\end{minipage}
		\begin{minipage}[c]{0.2\textwidth}
			\includegraphics[width=\textwidth]{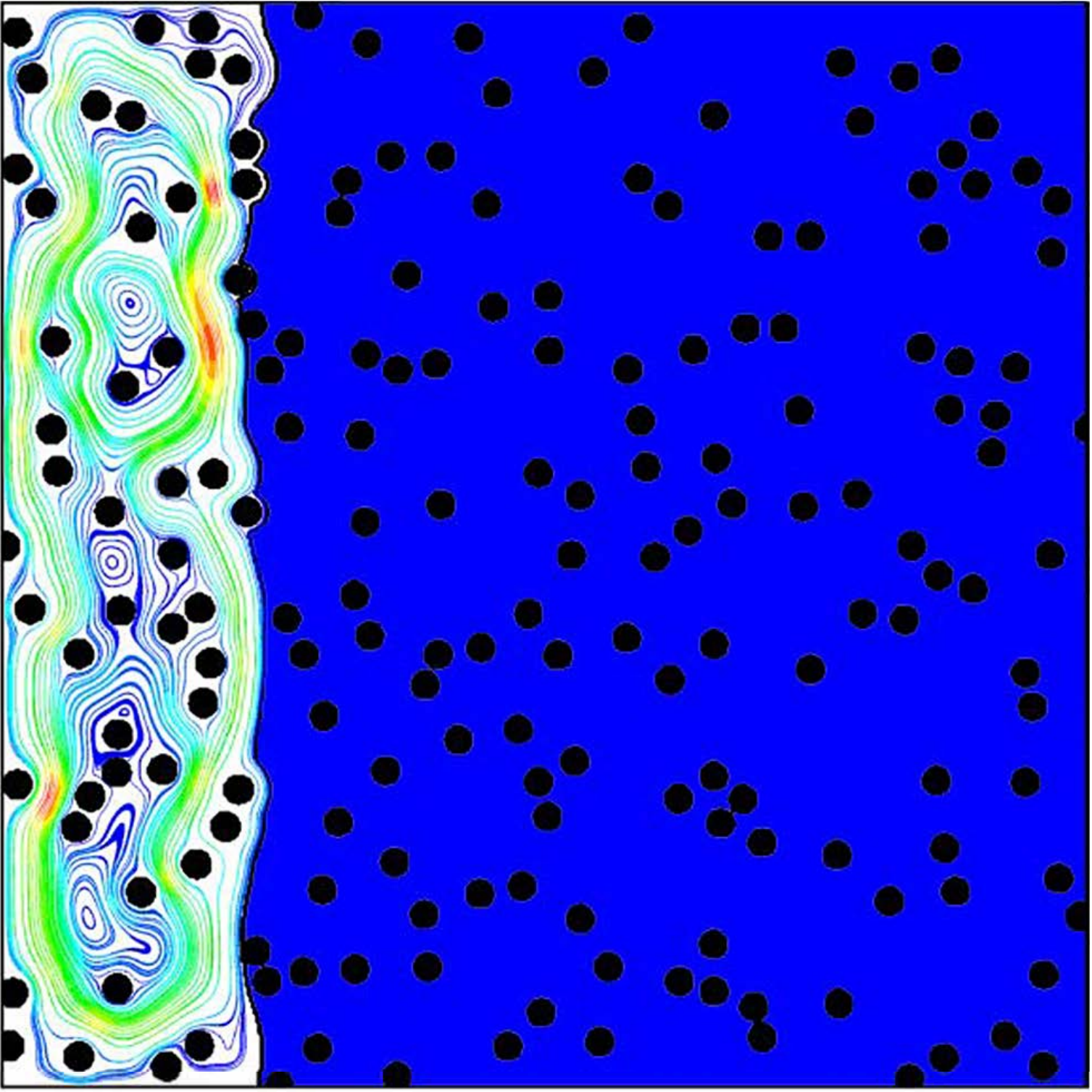}
		\end{minipage}
		\begin{minipage}[c]{0.2\textwidth}
			\includegraphics[width=\textwidth]{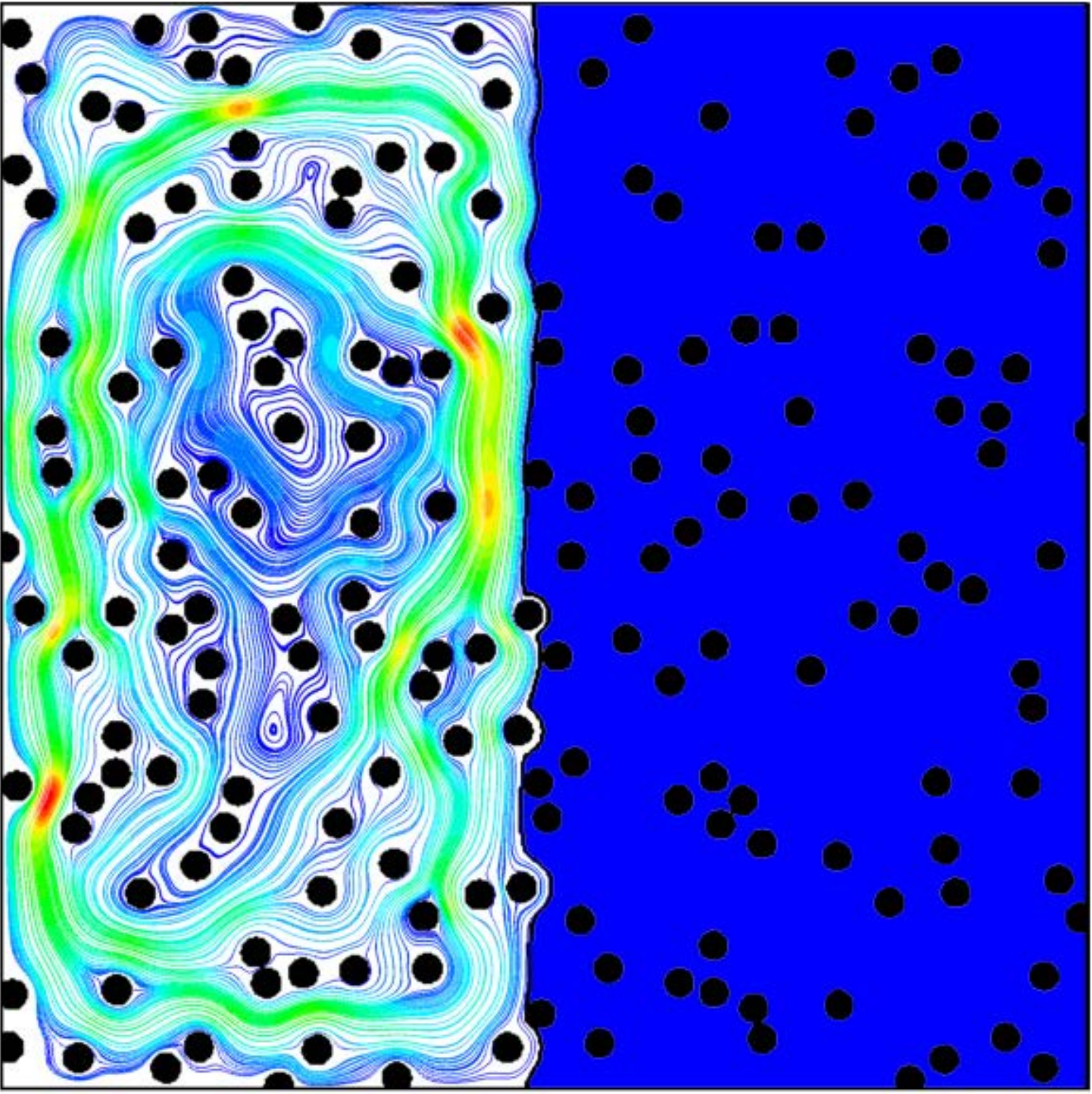}
		\end{minipage}
		\begin{minipage}[c]{0.2\textwidth}
			\includegraphics[width=\textwidth]{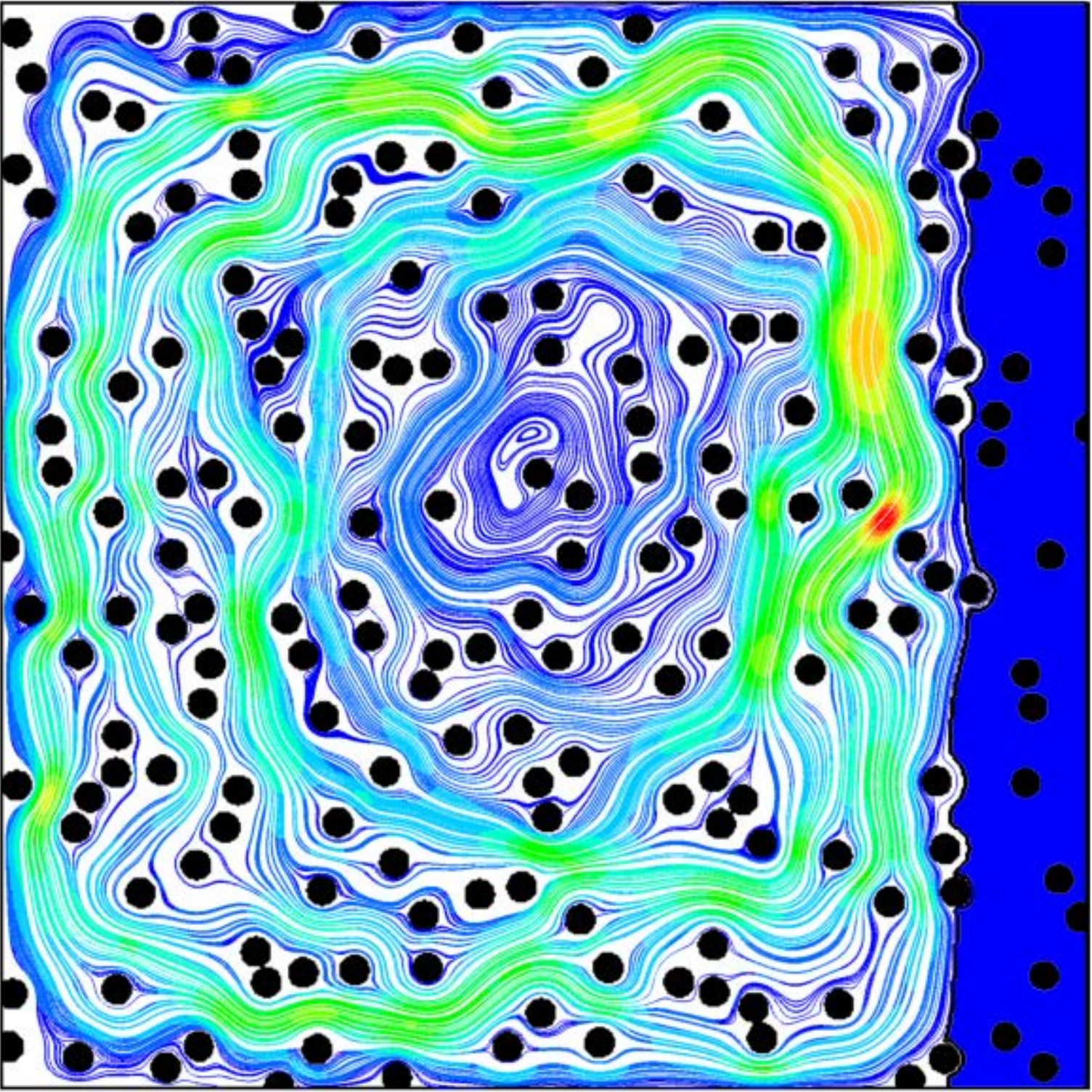}
		\end{minipage}

		\begin{minipage}[c]{0.2\textwidth}
			\centering
			\caption*{Case B: negative gradient}
			%\label{fig:side:caption}
		\end{minipage}
		\begin{minipage}[c]{0.2\textwidth}
			\includegraphics[width=\textwidth]{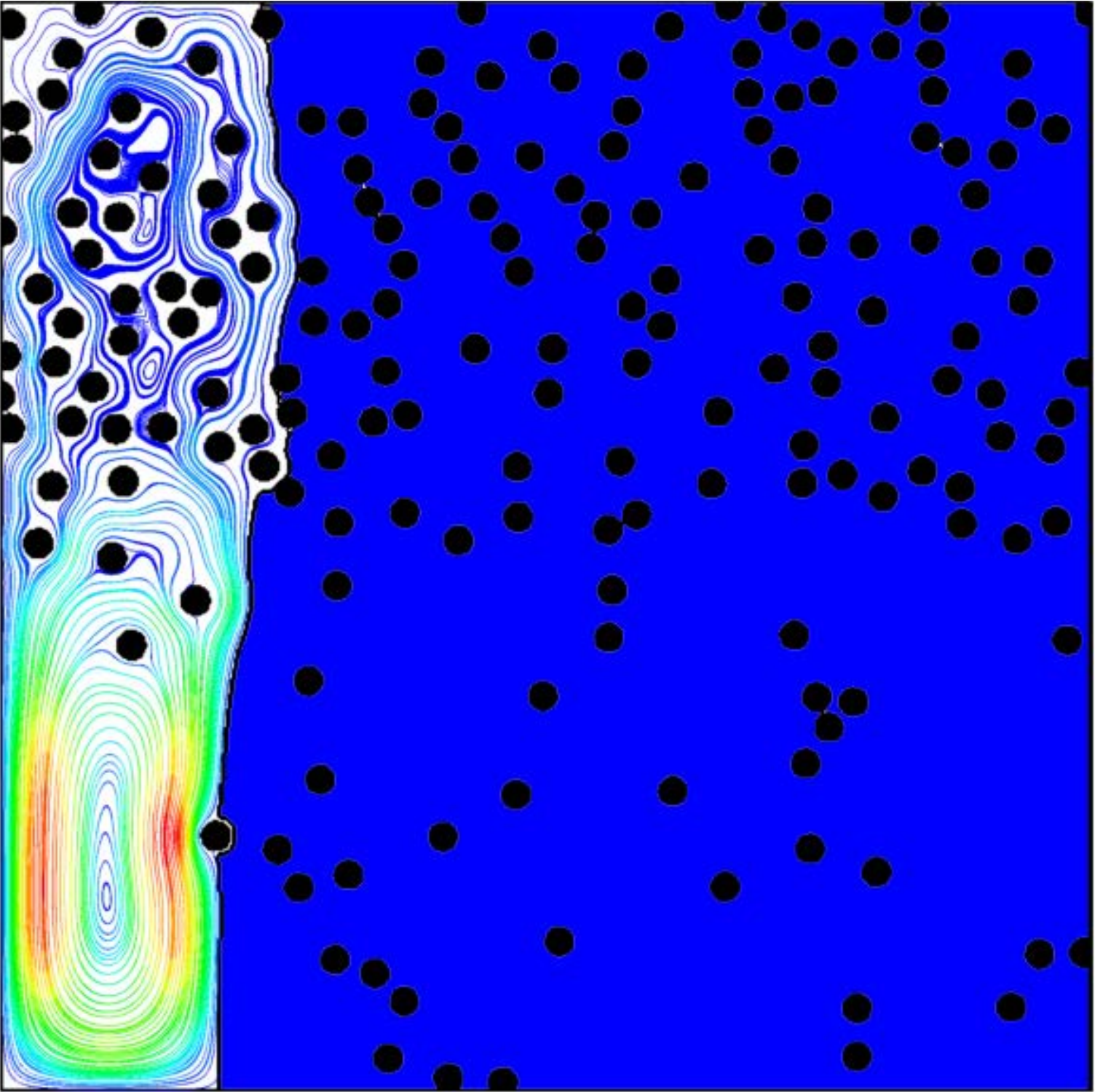}
		\end{minipage}
		\begin{minipage}[c]{0.2\textwidth}
			\includegraphics[width=\textwidth]{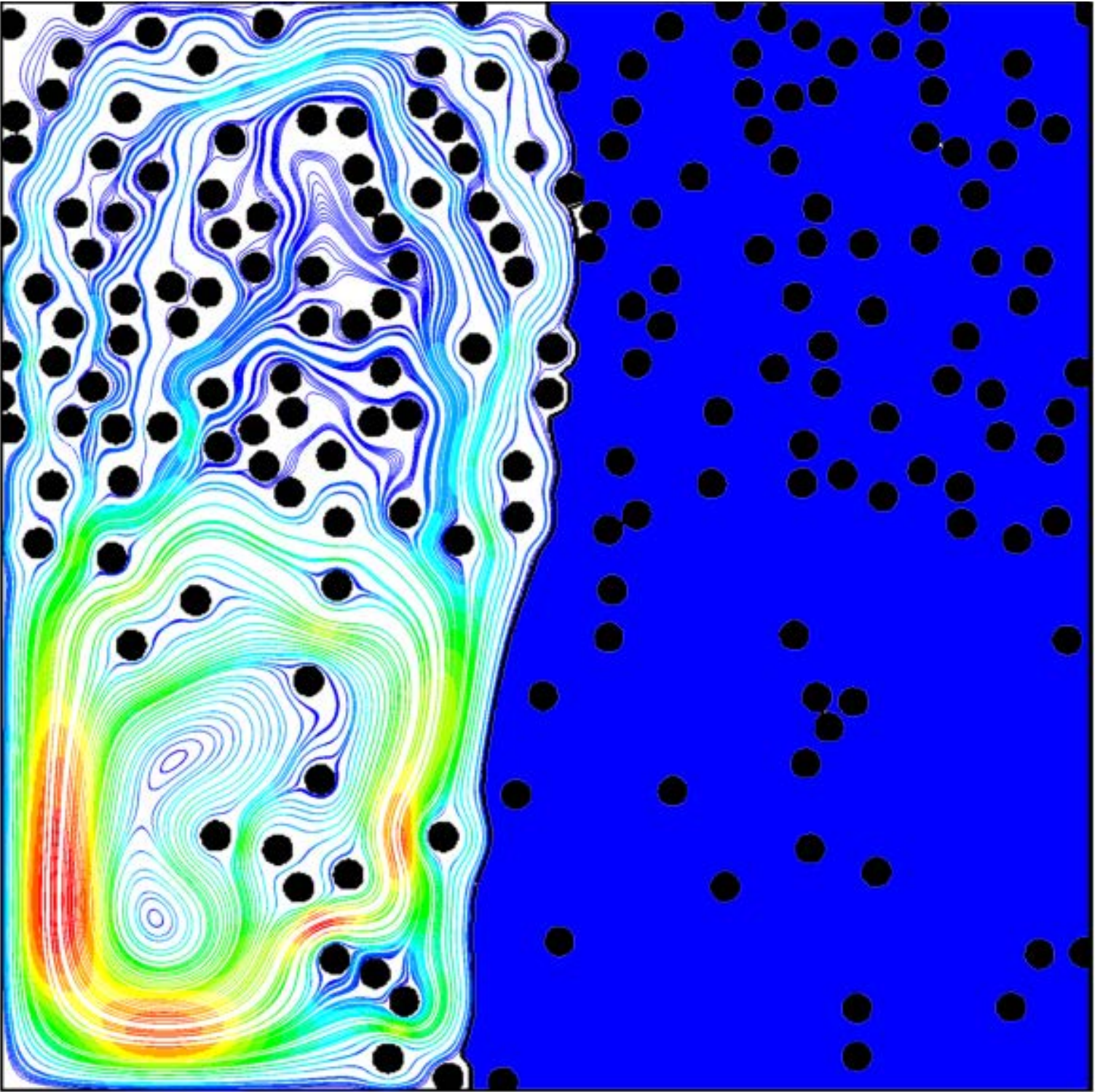}
		\end{minipage}
		\begin{minipage}[c]{0.2\textwidth}
			\includegraphics[width=\textwidth]{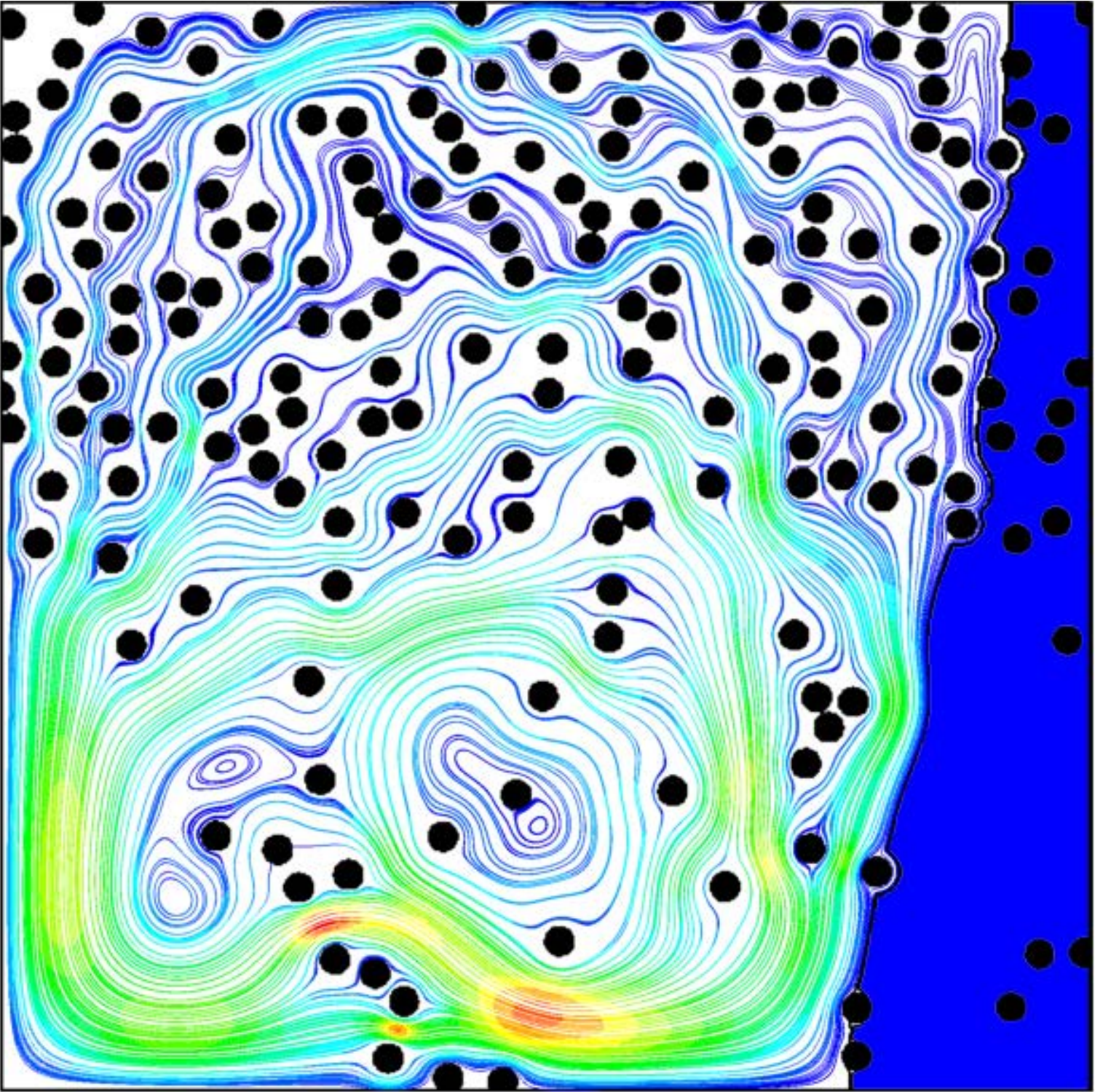}
		\end{minipage}

		\begin{minipage}[c]{0.2\textwidth}
			\centering
			\caption*{Case B: positive gradient}
			%\label{fig:side:caption}
		\end{minipage}
		\begin{minipage}[c]{0.2\textwidth}
			\includegraphics[width=\textwidth]{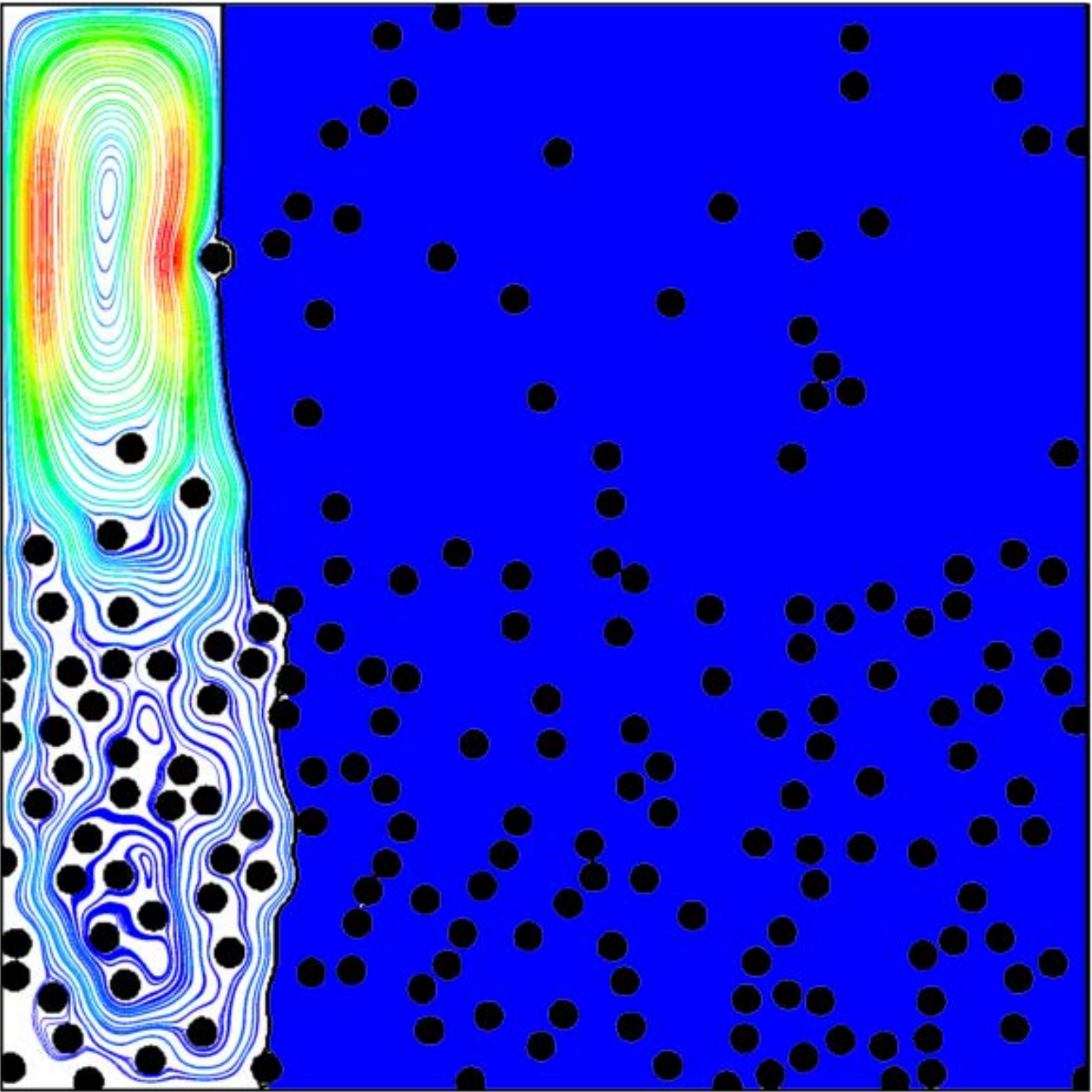}
			\caption*{$ Fo=0.2 $}
		\end{minipage}
		\begin{minipage}[c]{0.2\textwidth}
			\includegraphics[width=\textwidth]{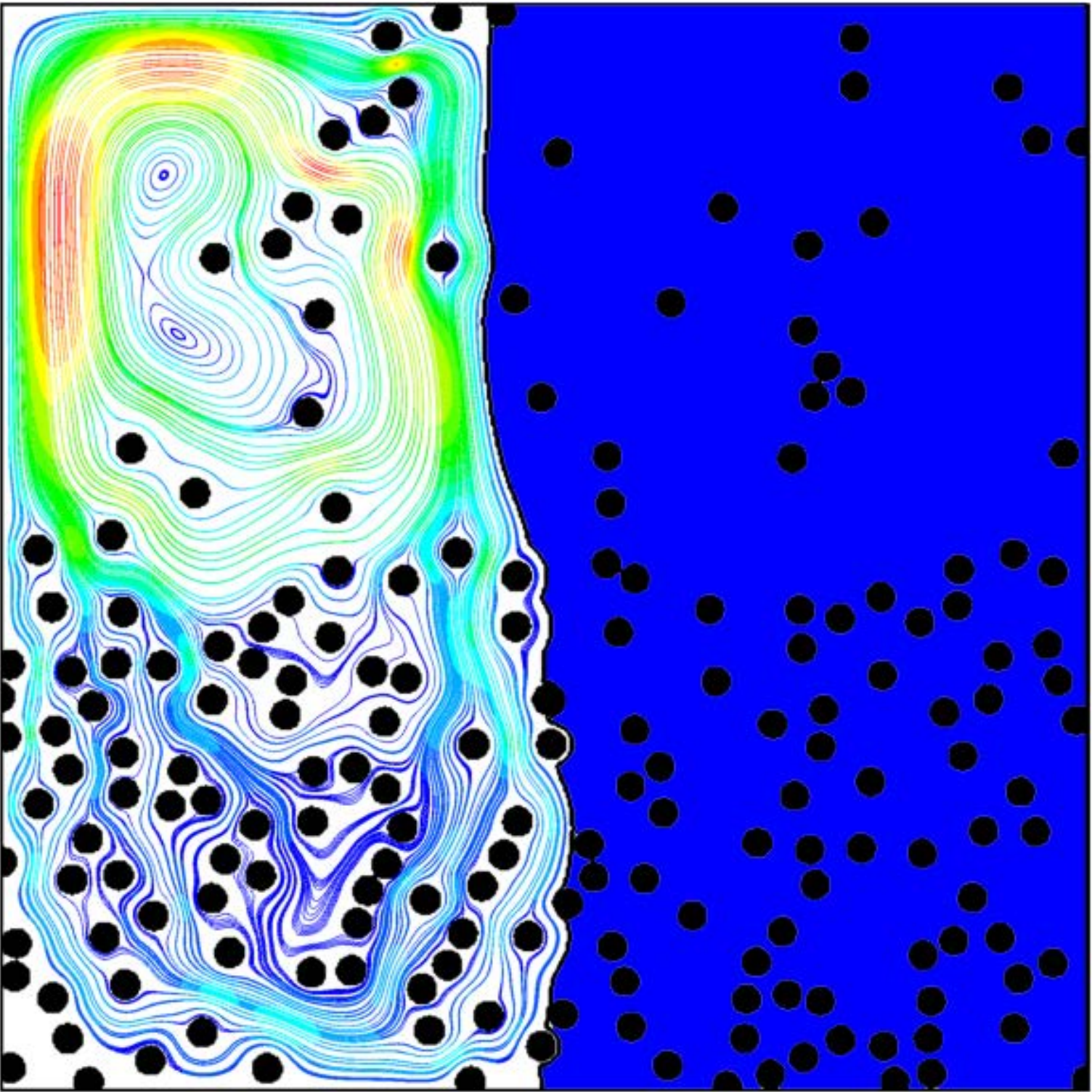}
			\caption*{$ Fo=0.9 $}
		\end{minipage}
		\begin{minipage}[c]{0.2\textwidth}
			\includegraphics[width=\textwidth]{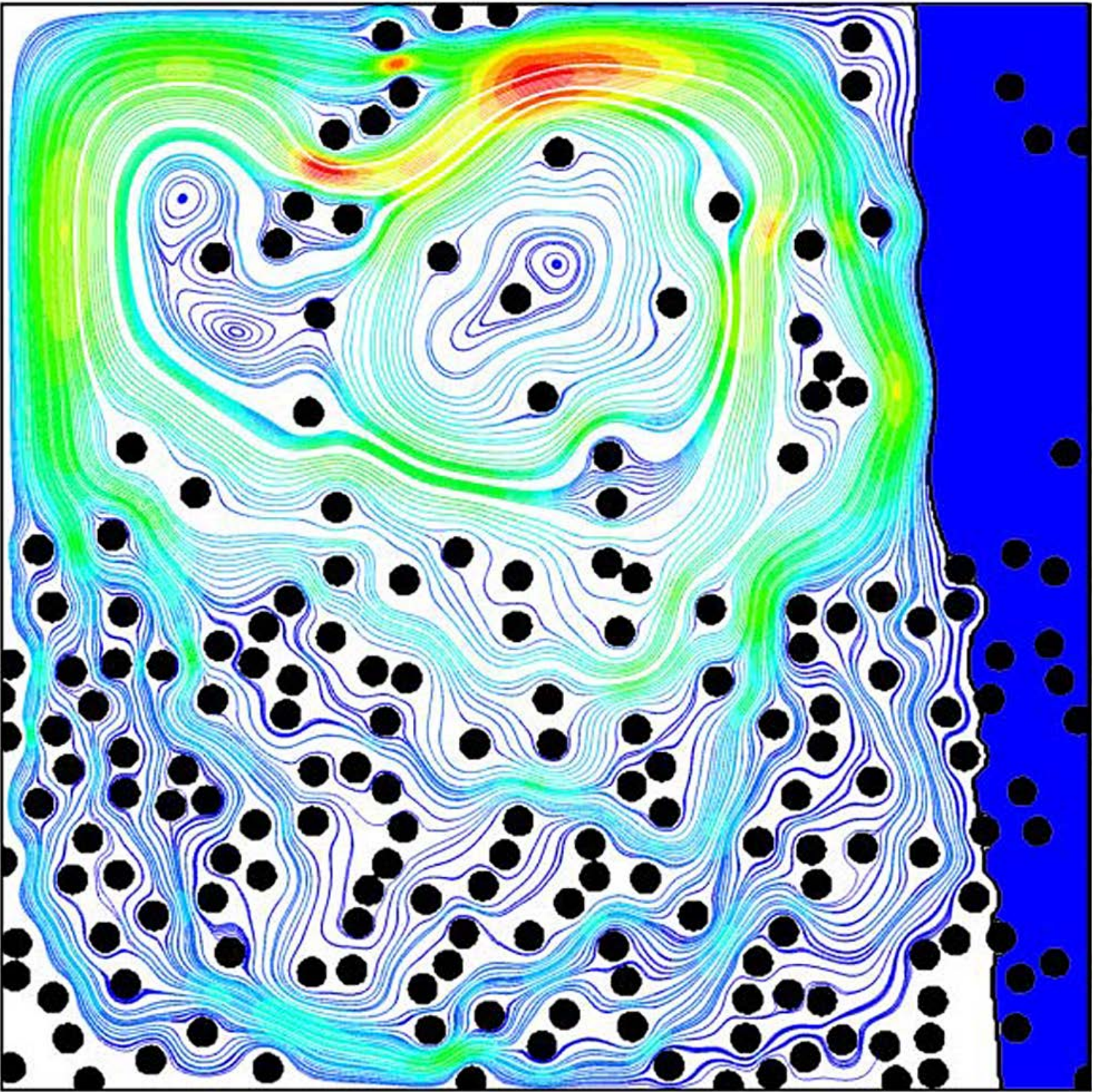}
			\caption*{$ Fo=3.0 $}
		\end{minipage}
		\caption{The effect of gradient porosity on the total liquid fraction at $ Ra=10^{4} $.} 
		\label{fig12}
	\end{figure}

	In addition, it is found that the results of different gradients have changed under low Rayleigh number conditions through numerical simulation results. Fig. \ref{fig11} illustrates the $ Fo $ dependent $ f_{_{l}} $ with three different gradient porosity conditions under $ Ra=10^{4} $ condition. For Case A, the variation tendency of the liquid fraction are extremely similar to that under high Rayleigh number conditions. The main difference is that the order in which the melting of the positive and negative gradients is completed has changed, and the positive gradient structure is firstly melted with a higher liquid fraction. It can be observed from the liquid fraction distribution in Fig \ref{fig12} that the positive gradient structure relies on the high thermal conductivity brought by the lower porosity on the left to melt faster in the left area, and the positive gradient structure is the first to melt due to the strong natural convection erosion caused by the lower natural convection resistance when melting to the right area. For Case B, there is little difference between the melting curves for the three gradient in the whole melting process. It can be seen from the liquid fraction distribution in Fig \ref{fig12} that the area with higher porosity melts more quickly, for this reason, the last melted area is always the corner with lower porosity. There are obvious steps in the solid-liquid interface, and the steps are located exactly at the interface where the porosity changes. Compare Case A, it is found that positive gradient of Case A is firstly melt due to the continuous scouring action of the local natural convection. The evolution of melting front further illustrated that the superiority of natural convection in eliminating corners.
	
\begin{figure}[H]
		\centering 
		\subfigure[Case A]{ \label{fig13}
			\includegraphics[width=0.4\textwidth]{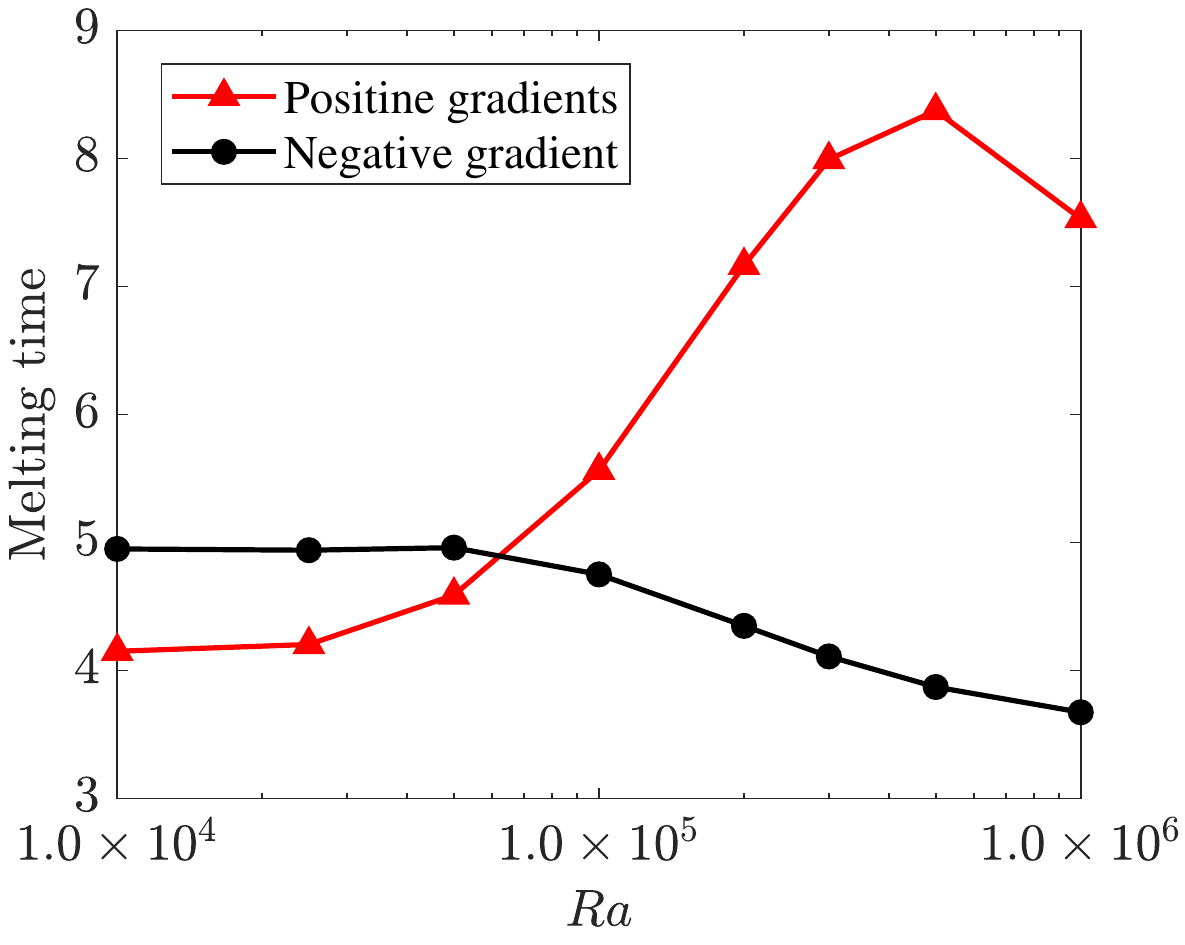}} 
		\subfigure[Case B]{ \label{fig14}
			\includegraphics[width=0.4\textwidth]{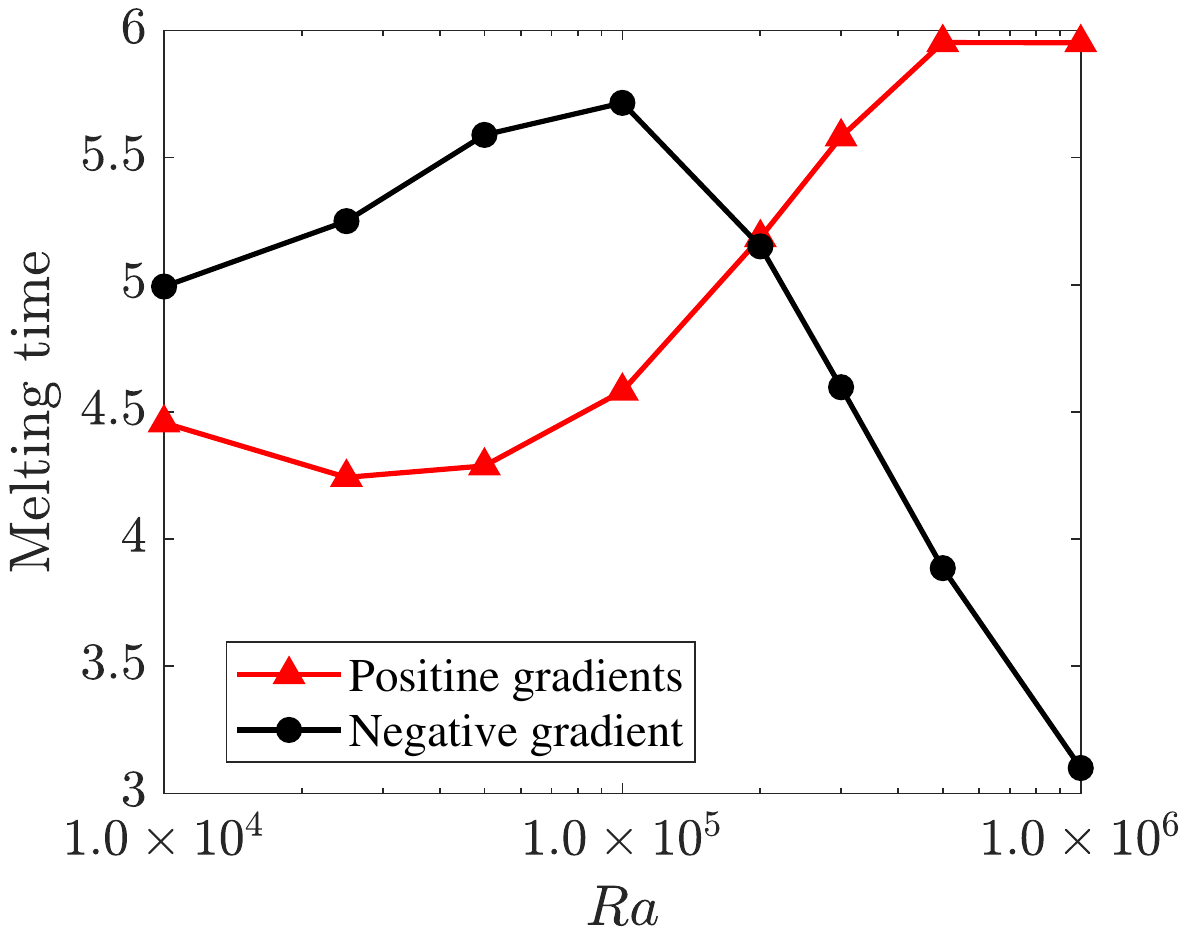}}
		\caption{Effects of Ra on (a) Case A  and (b) Case B.}
		\label{fig15}
	\end{figure}

	The simulation results have changed dramatically under different Rayleigh number, we next turn to investigate the influence of $ Ra $ on melting. Rayleigh number is an important parameter reflecting the intensity of natural convection. It can be seen from Eqs. \ref{equation_Ra} that the buoyancy force, as the driving force of natural convection, gradually increases with the Ra increasing. Consequently, the intensity of natural convection became increasing with the increase of the $ Ra $. Fig. \ref{fig15} depicts the influence of the Rayleigh number on positive and negative gradient

\begin{figure}[H]
		\centering 
		\subfigure[Case A]{ \label{fig16}
			\includegraphics[width=0.4\textwidth]{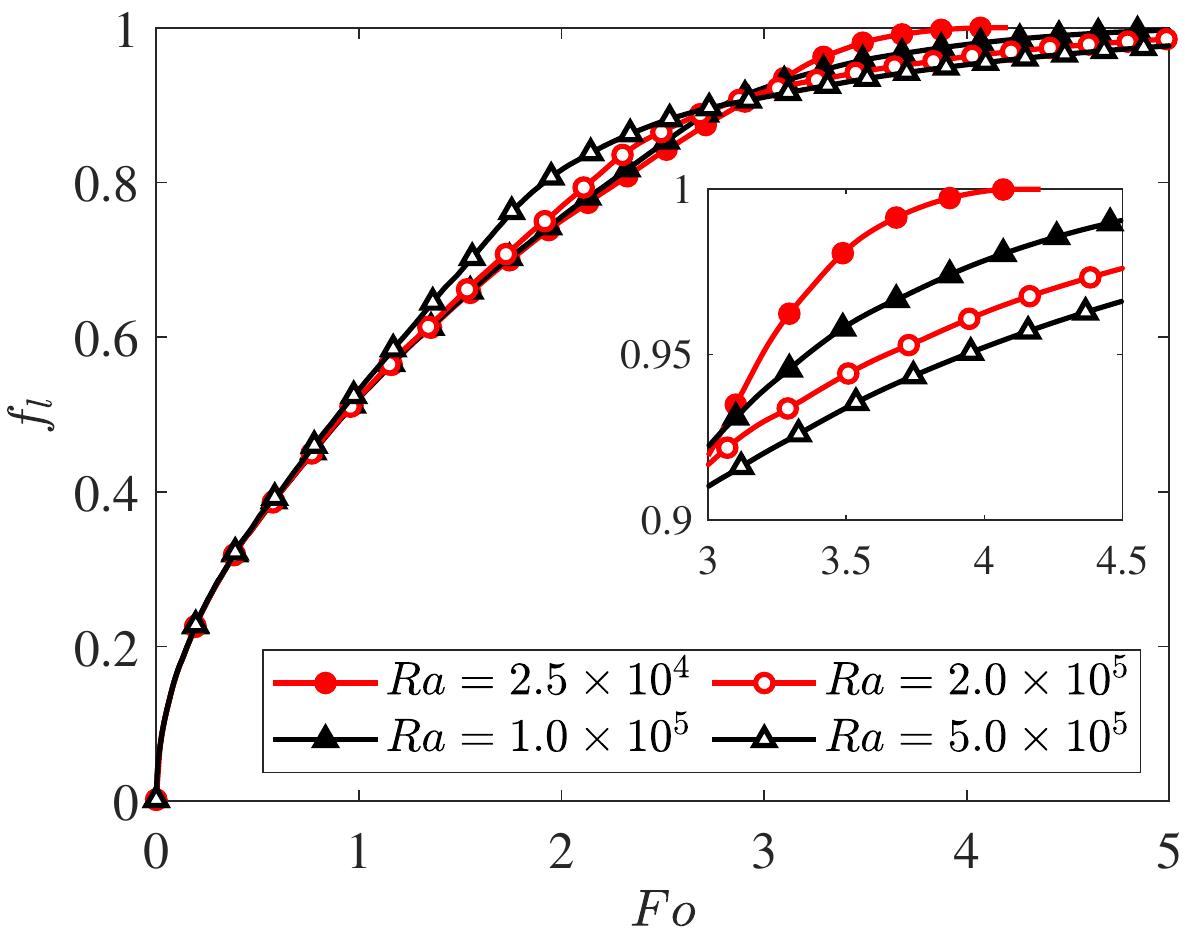}} 
		\subfigure[Case B]{ \label{fig17}
			\includegraphics[width=0.4\textwidth]{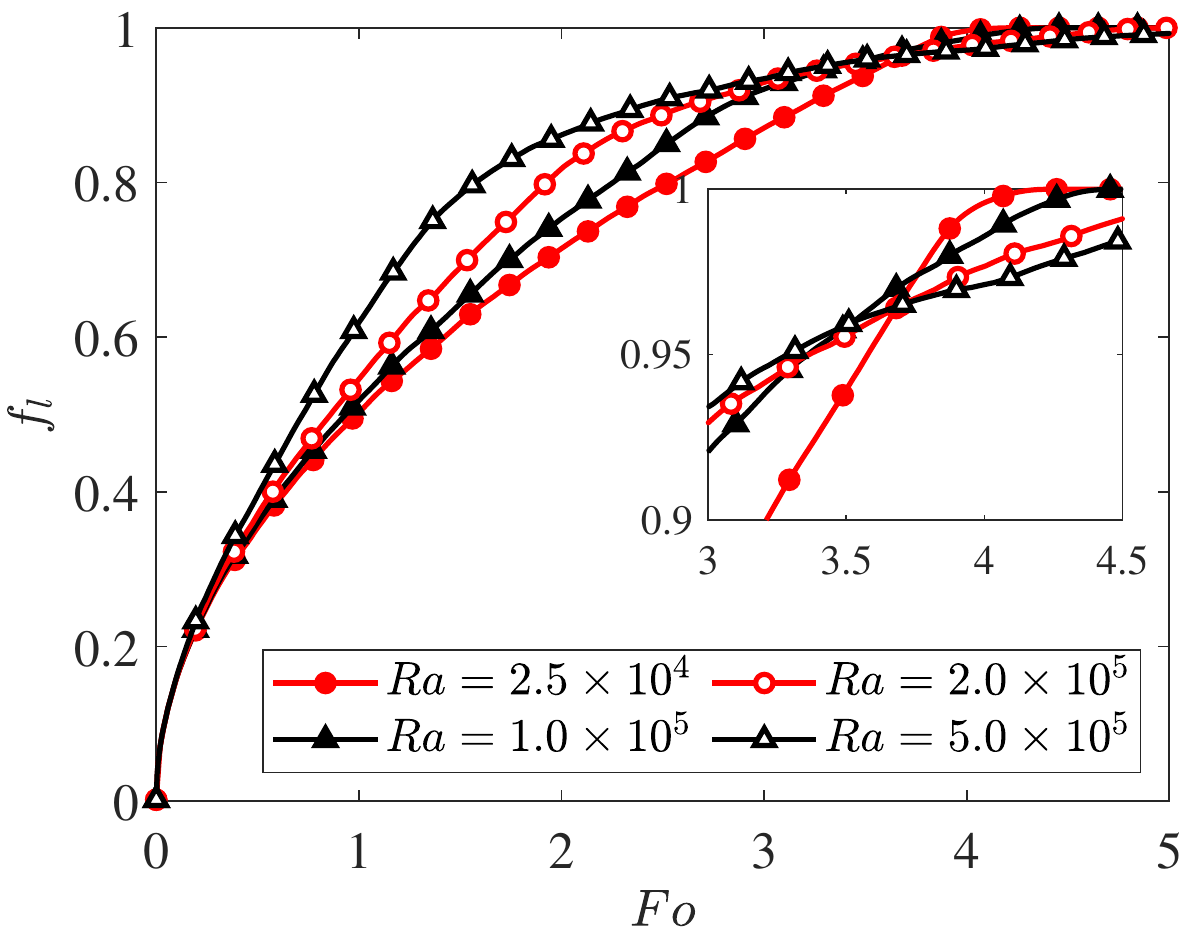}}
		\caption{The effect of Rayleigh number on melting time for (a) Case A and (b) Case B.}
		\label{fig18}
	\end{figure}

\begin{figure}[H]
		\centering 
		\subfigure[]{ \label{fig19}
			\includegraphics[width=0.4\textwidth]{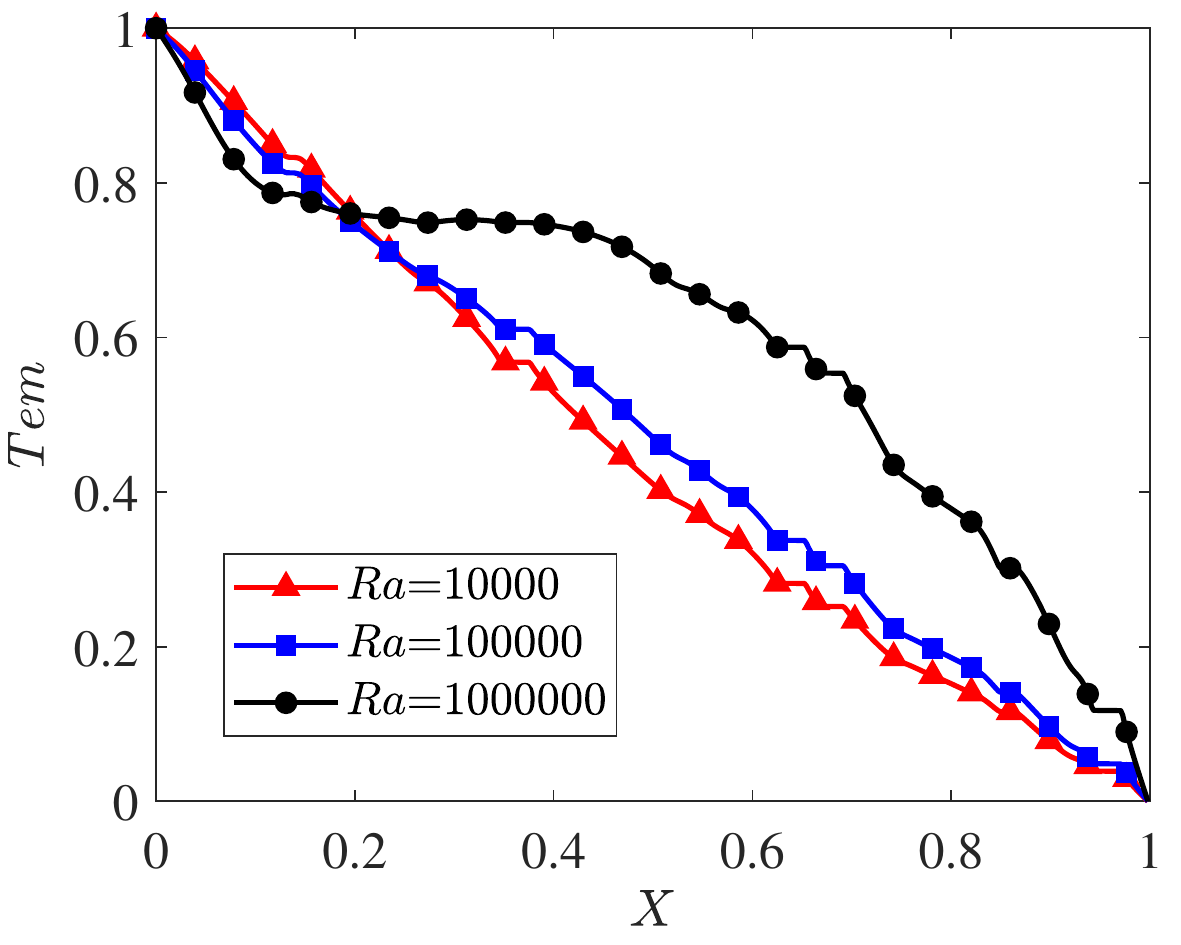}} 
		\subfigure[]{ \label{fig20}
			\includegraphics[width=0.4\textwidth]{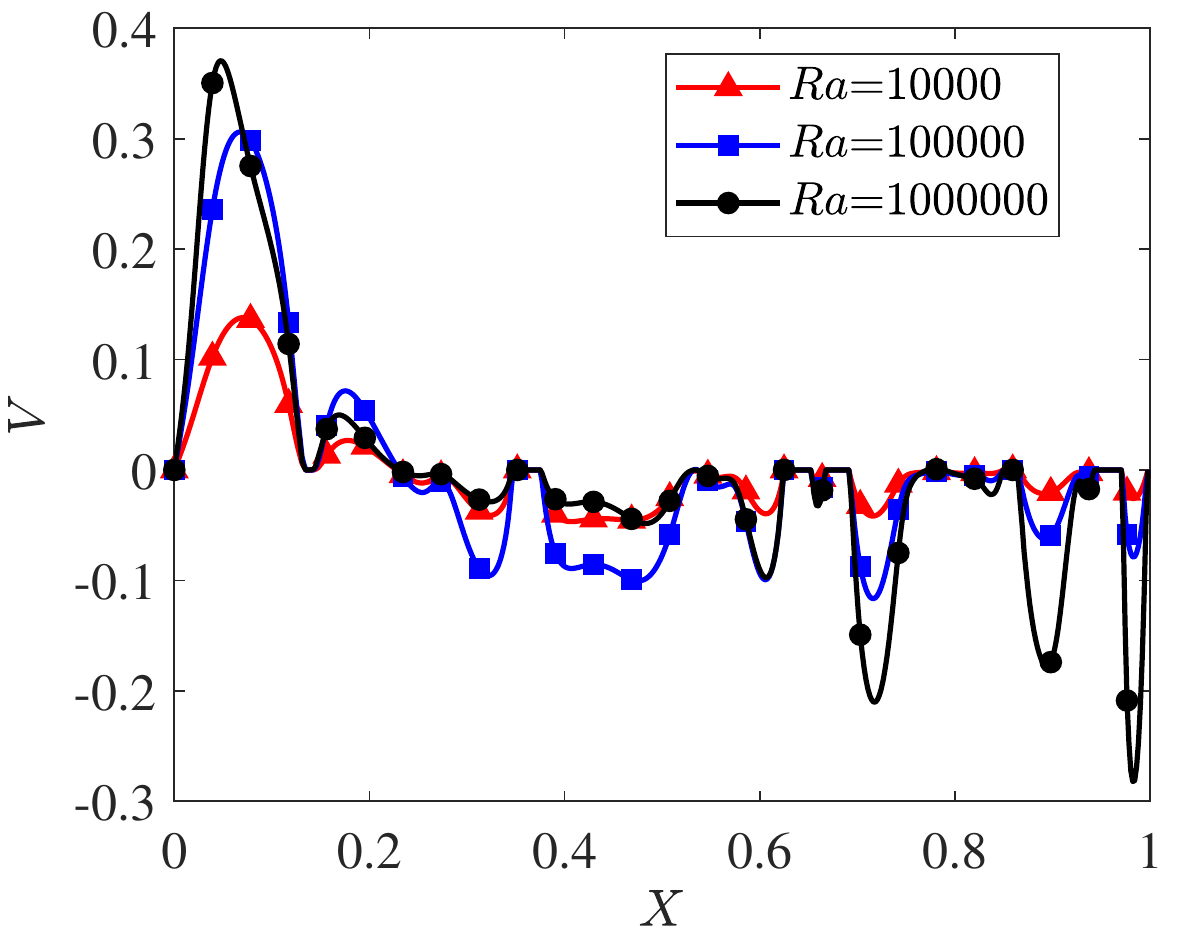}}
		\caption{(a) Temperature distributions and (b) vertical velocity distributions along with Y = 0.5 cross-section (fully melted) for Case A: positive gradient.}
		\label{fig21}
	\end{figure}

\begin{figure*}[htbp]%加*的作用是跨栏（双栏和单栏latex的区别）
		\centering
		
		%每列占整个文档的文本宽度的0.2，每列两张图像，两列图像，这些可以自行设置
		
		\subfigure{
			\begin{minipage}[b]{0.2\textwidth}
				\includegraphics[width=\textwidth]{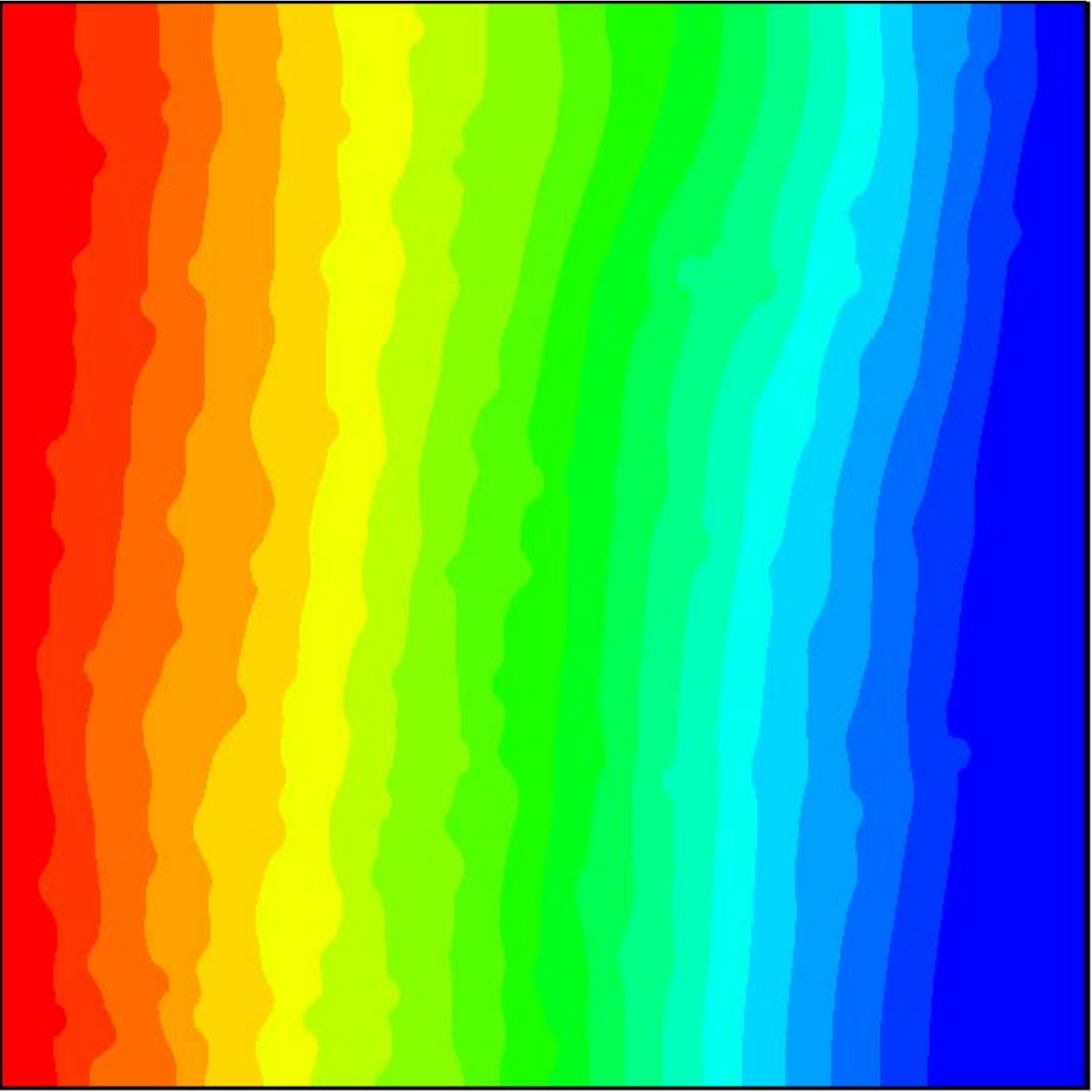}
				\includegraphics[width=\textwidth]{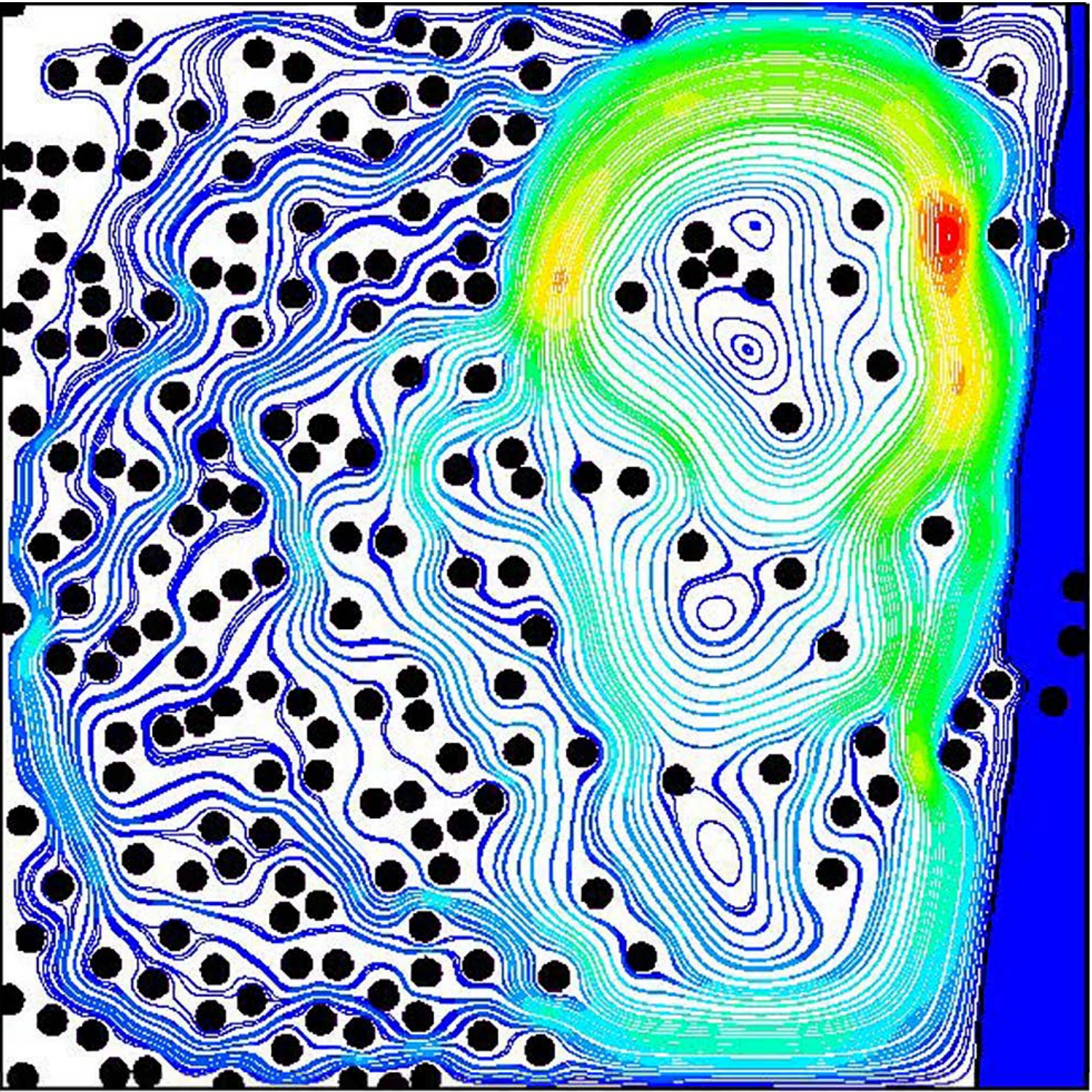}
				\caption*{$ Ra=2.5 \times 10^{4} $} 
				
			\end{minipage}
		}
		\subfigure{
			\begin{minipage}[b]{0.2\textwidth}
				\includegraphics[width=\textwidth]{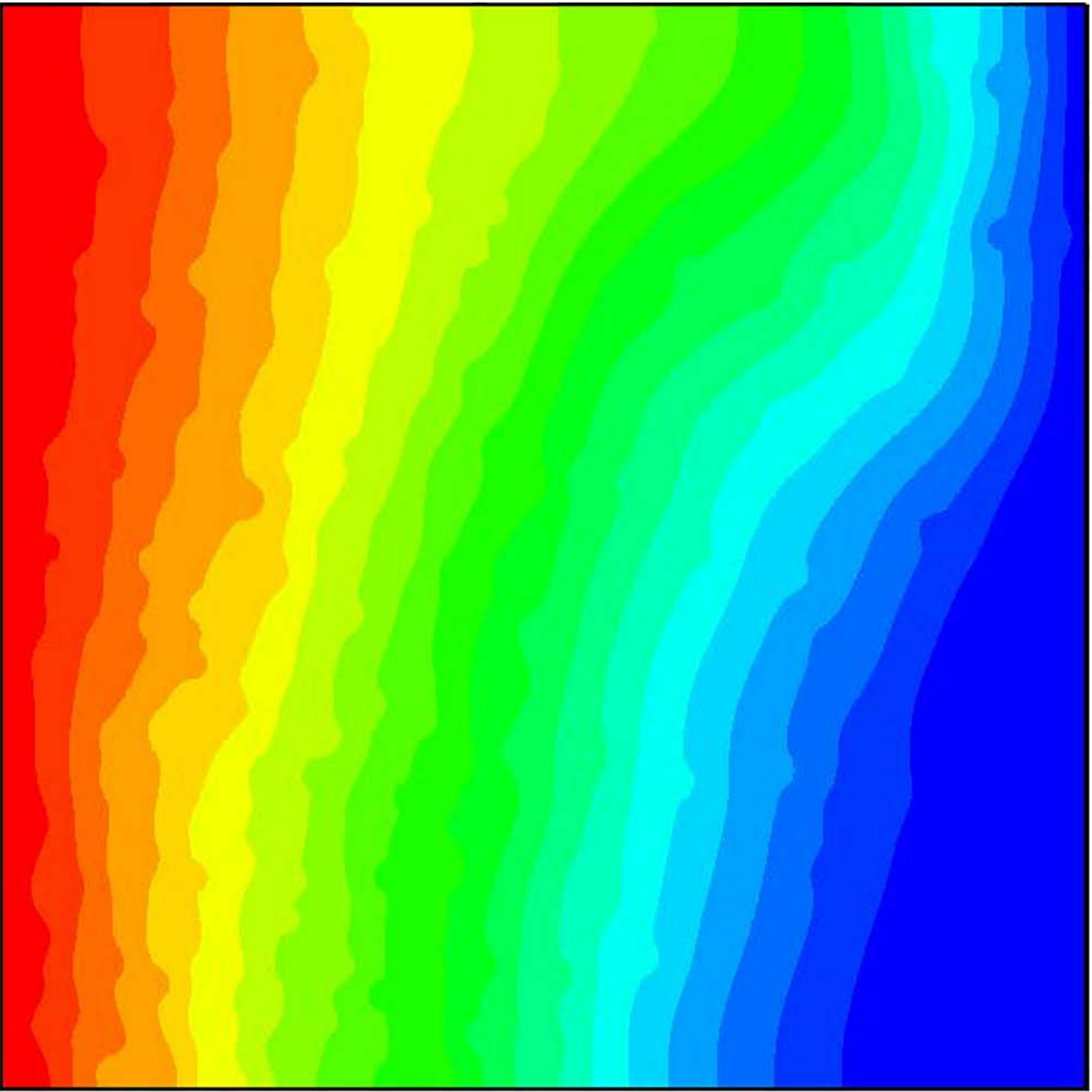}
				\includegraphics[width=\textwidth]{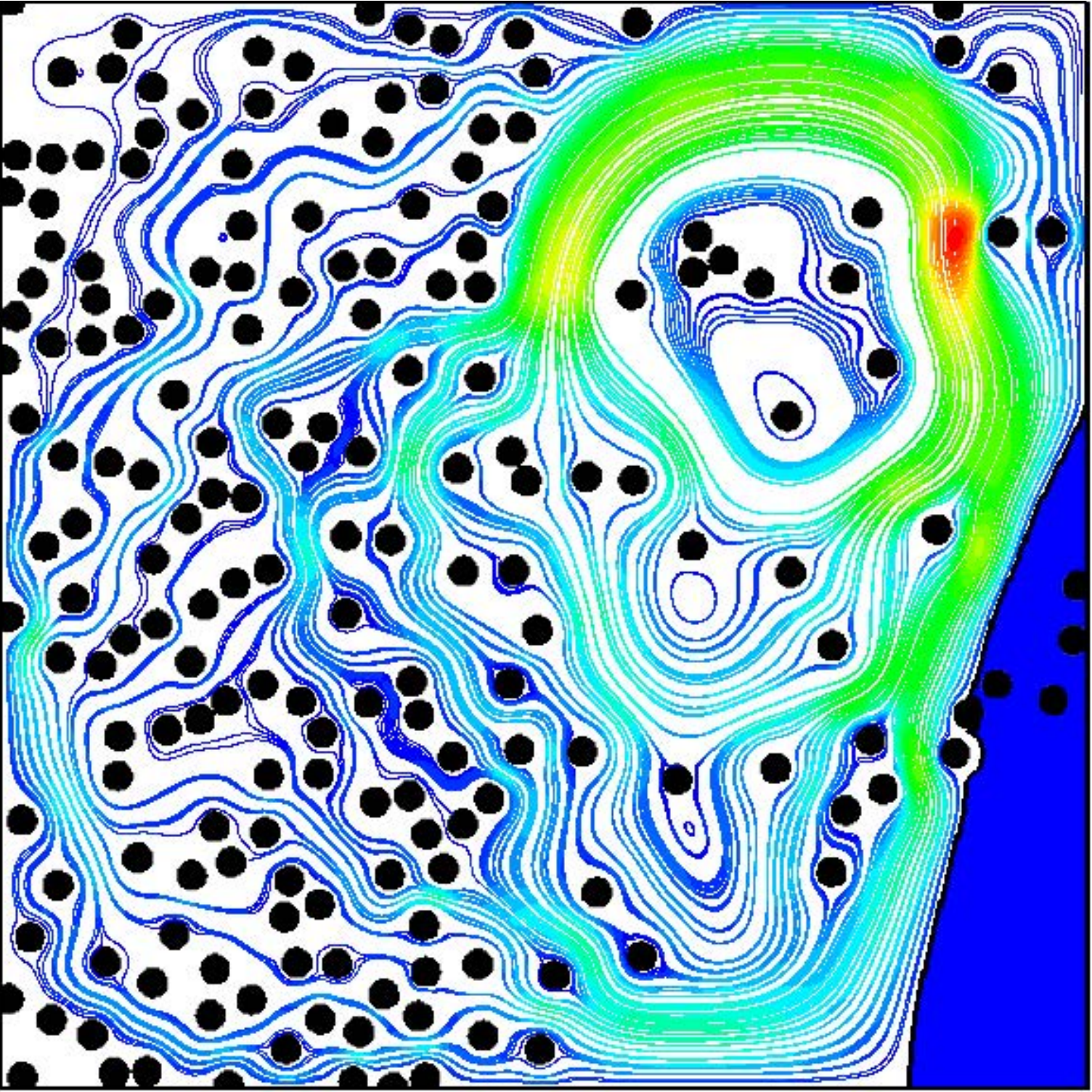}
				\caption*{$ Ra=1.0 \times 10^{5} $}	
			\end{minipage}
		}
		\subfigure{
			\begin{minipage}[b]{0.2\textwidth}
				\includegraphics[width=\textwidth]{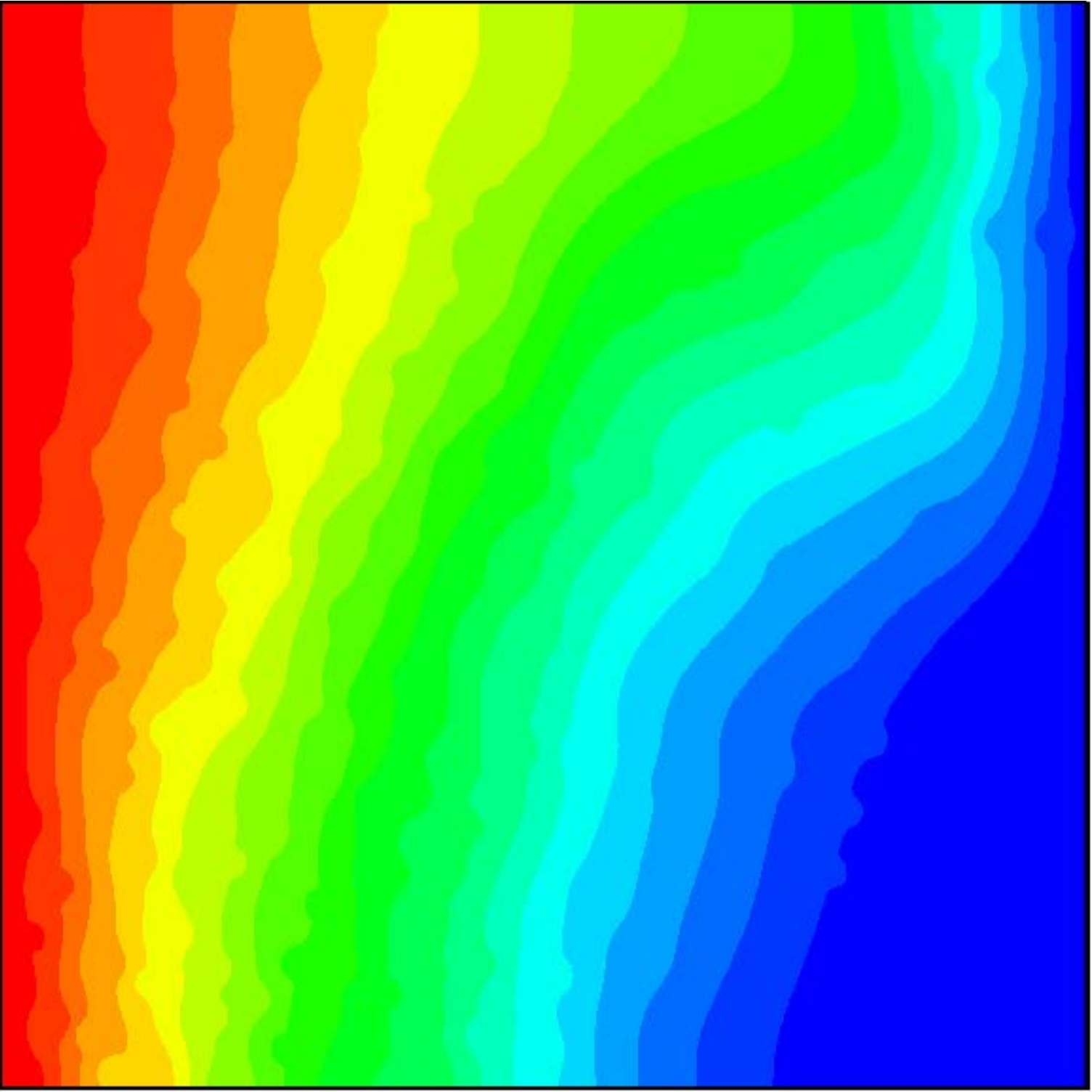}
				\includegraphics[width=\textwidth]{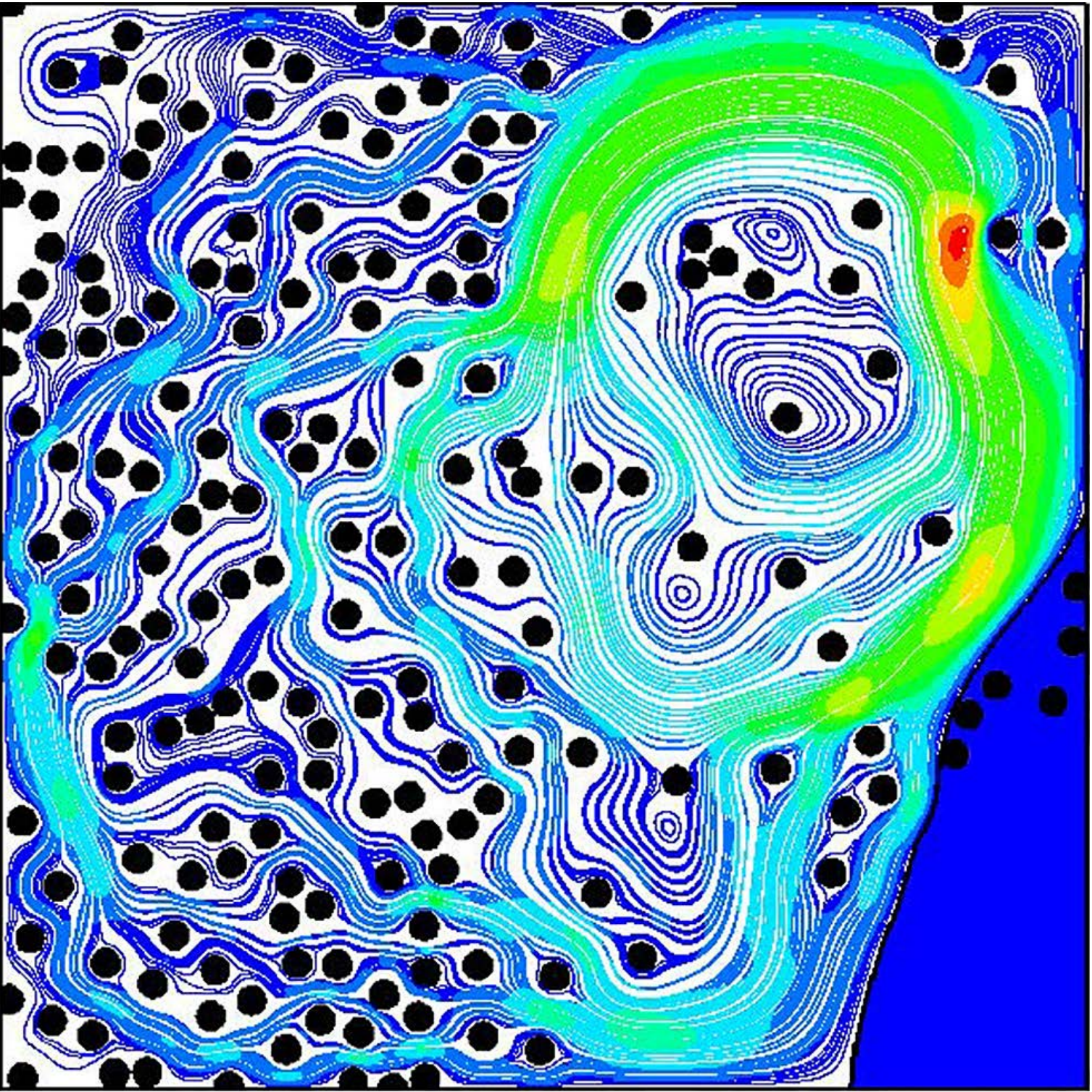}
				\caption*{$ Ra=2.0 \times 10^{5} $}
			\end{minipage}
		}
		\subfigure{
			\begin{minipage}[b]{0.2\textwidth}
				\includegraphics[width=\textwidth]{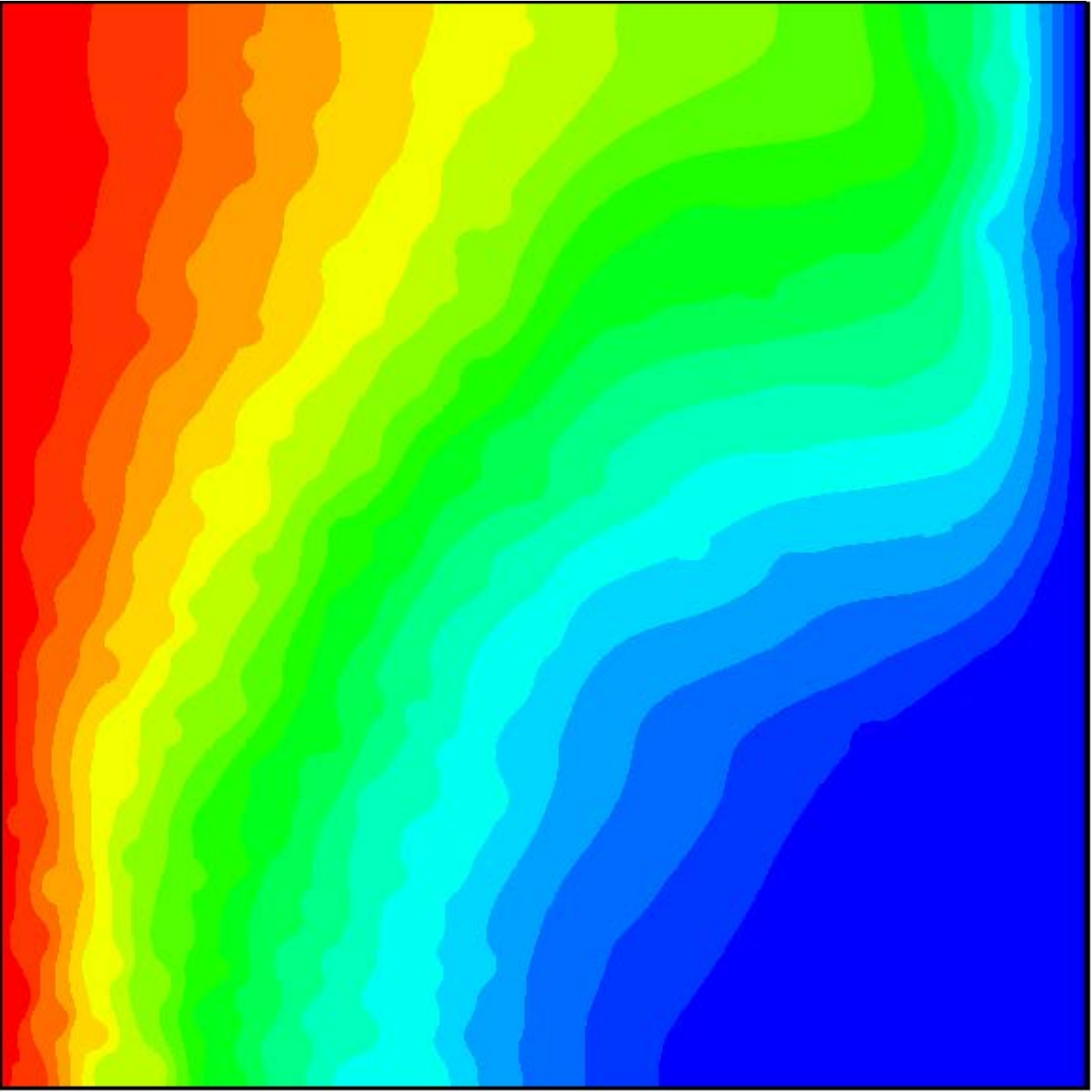}
				\includegraphics[width=\textwidth]{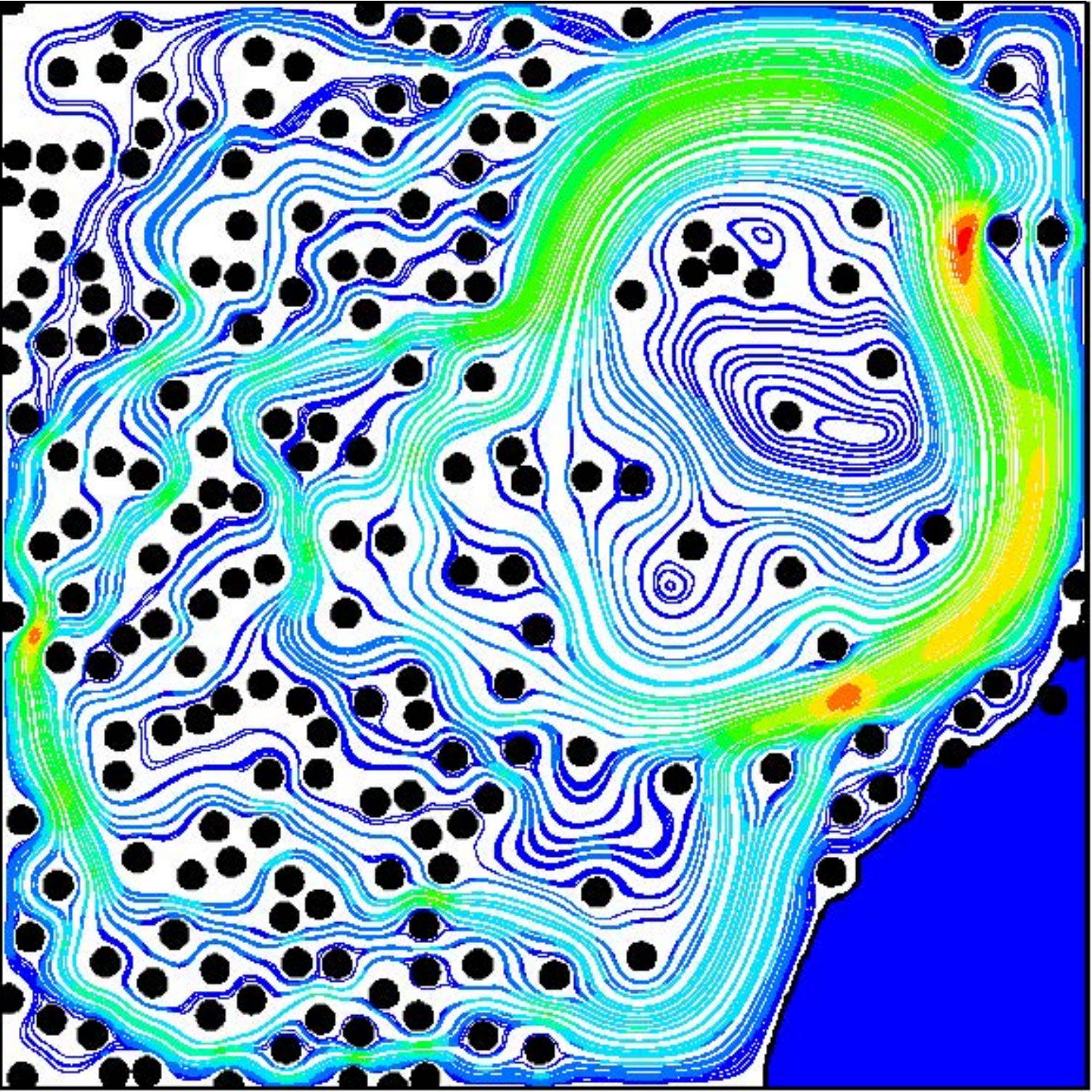}
				\caption*{$ Ra=3.0 \times 10^{5} $}
			\end{minipage}
		}
		\caption{Temperature, streamlines and liquid fraction distributions of turning point for positive gradient of Case A.}
		\label{fig22}
	\end{figure*}

	\noindent  of Case A and Case B. As shown in this figure, a critical point near $ Ra=10^{5} $ is observed, the positive gradient structure is superior to the positive gradient structure when $ Ra $ is less than  this critical point, and when $ Ra $ exceeds this critical point, the negative gradient is more advantageous. As shown in this figure, the positive gradients have similar trends for two cases. Moreover, one can also found that the melting time of the positive gradient structure almost increases to the steady state with the increase of $ Ra $. For further insight into the overall melting situation of the positive gradient under different $ Ra $, Fig. \ref{fig18} illustrates that the dimensionless time dependent the liquid fraction with different $ Ra $ conditions. It can be clearly seen that Rayleigh number has little effect on conduction due to heat conduction is the main heat transfer mechanism, then high Rayleigh number brings stronger natural convection with the increasing liquid phase PCM. To further confirm this observation, we also present the temperature and vertical velocity distributions at the mid height of the domain for positive gradient in Fig. \ref{fig21}. It is clearly observed that the overall downtrend of temperature is gentler than that under $Ra=10^{6}$ condition as for the reduced Rayleigh number and the vertical velocity decrease and the thickness of velocity boundary layer is thickened accordingly with the decrease of Rayleigh number. As the Rayleigh number increases, the natural convection effect is strengthened, and thus causing the intense flow in the liquid PCM, which leads to greater changes in the temperature near the cavity wall. However, it is worth noting that a turning point is observed and the slope of liquid fraction changes dramatically when the time nears $ Fo = 3 $ point. Subsequently, the liquid fraction of low Rayleigh number increases with a higher rate compared to the other cases while the melting rate decreases in the high Rayleigh number. Fig. \ref{fig22} shows the temperature distributions, liquid fraction distributions and streamlines of positive at turning point for different Ra conditions. As it can be seen, the remaining part gradually concentrate towards the lower right corner and heat is concentrated in the upper partas $ Ra $ increases. For negative gradient, the enhancement of natural convection shortens the total melting time for Case A, while the trend of Case B first rises due to the above reasons, and then declines due to the enhancement of natural convection.
	
\begin{figure}[H]
		\centering
		\subfigure[$Ra=10^{4}$]{ \label{fig24}
			\includegraphics[width=0.4\textwidth]{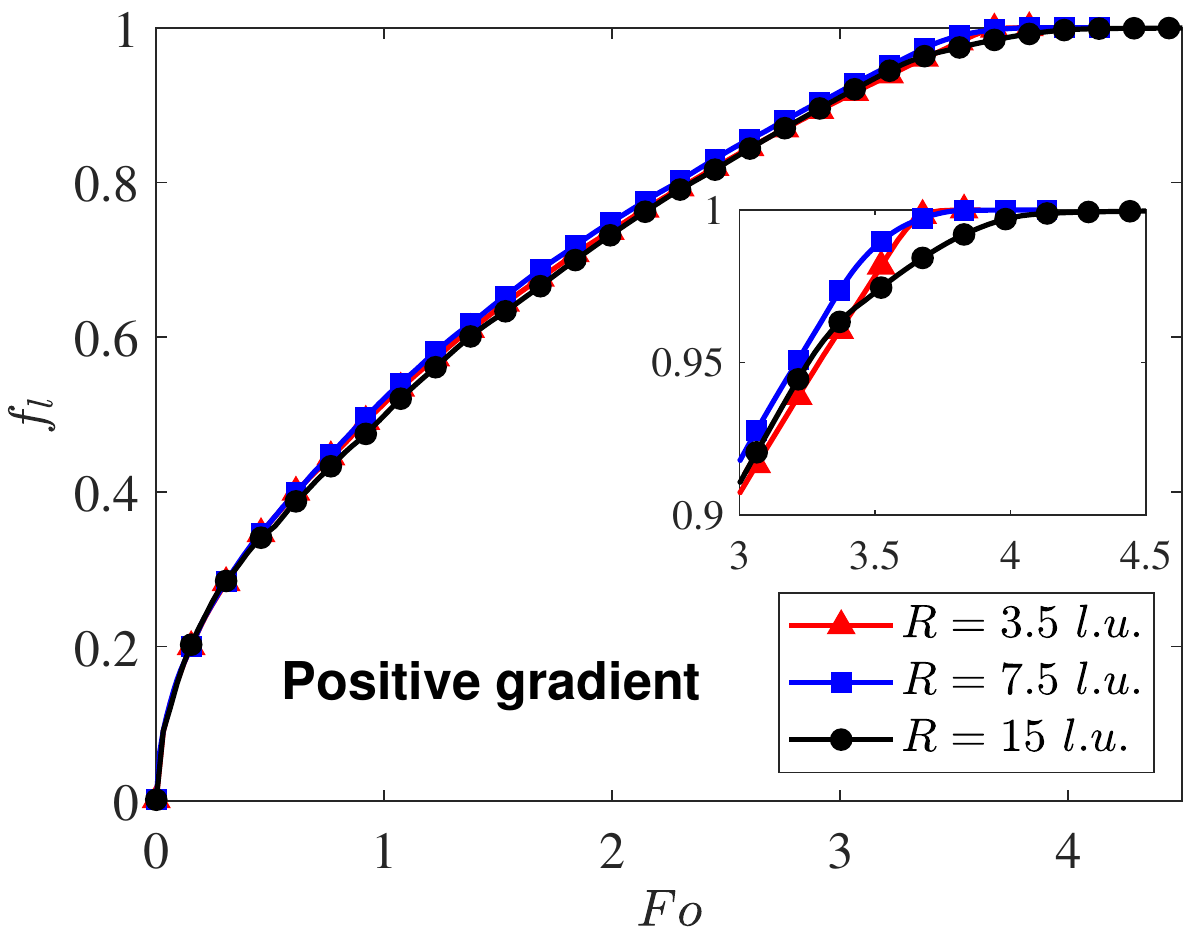}
			\quad
			\includegraphics[width=0.4\textwidth]{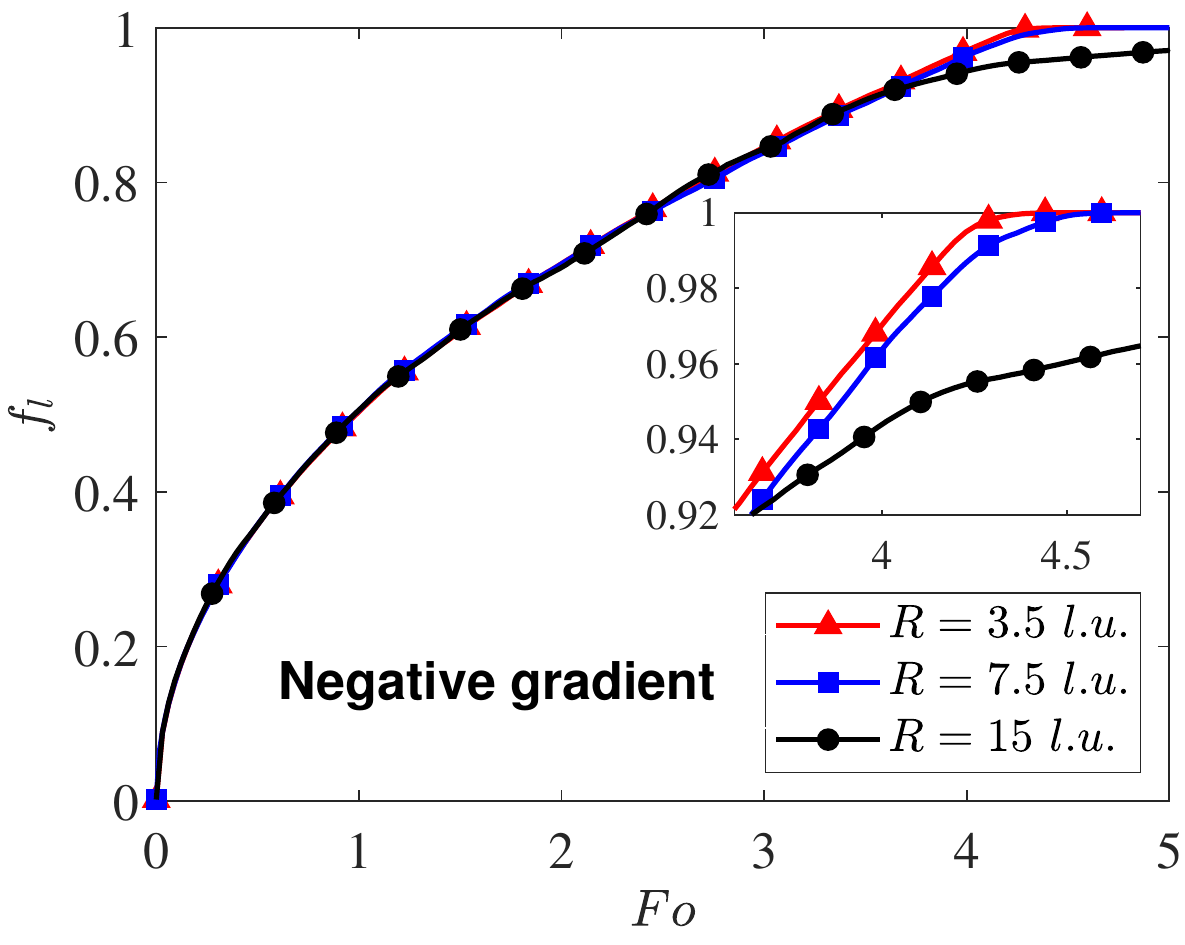}}
		
		\subfigure[$Ra=10^{6}$]{ \label{fig25}
			\includegraphics[width=0.4\textwidth]{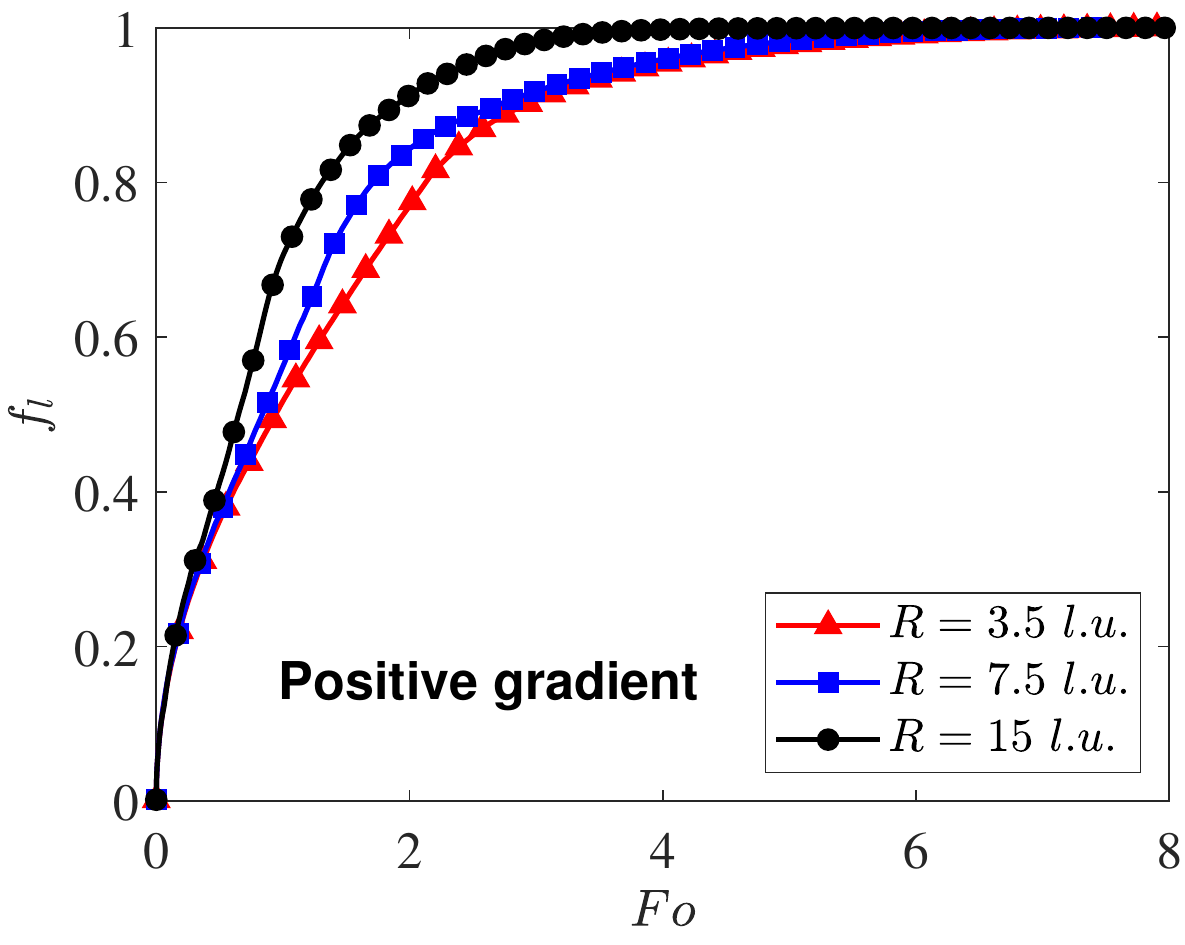}
			\quad
			\includegraphics[width=0.4\textwidth]{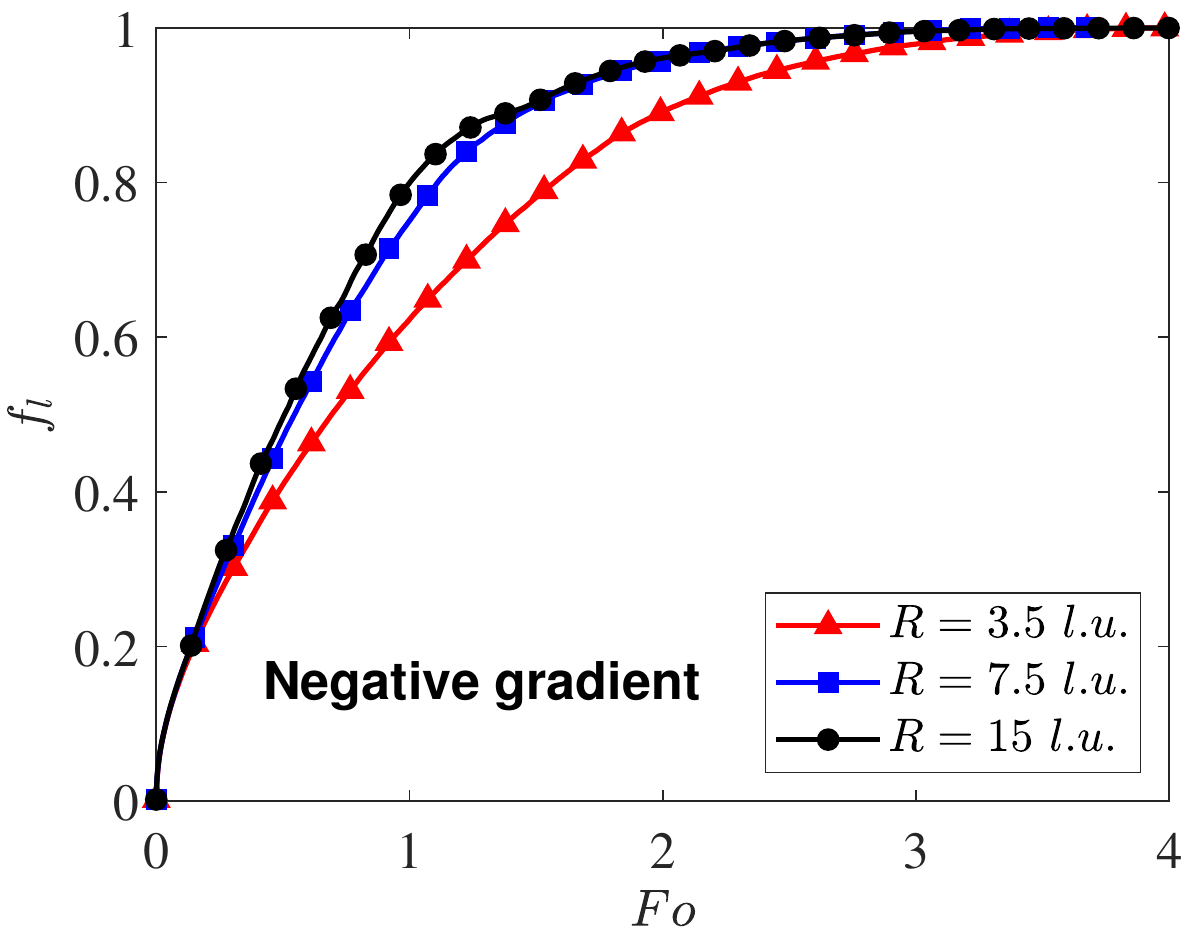}} 
		
		\caption{Effect of particle diameters on the melting evolution of the PCM for Case A.} 
		\label{fig26}
	\end{figure}

\begin{figure}[H]
		\centering
		\subfigure[$Ra=10^{4}$]{ \label{fig27}
			\includegraphics[width=0.4\textwidth]{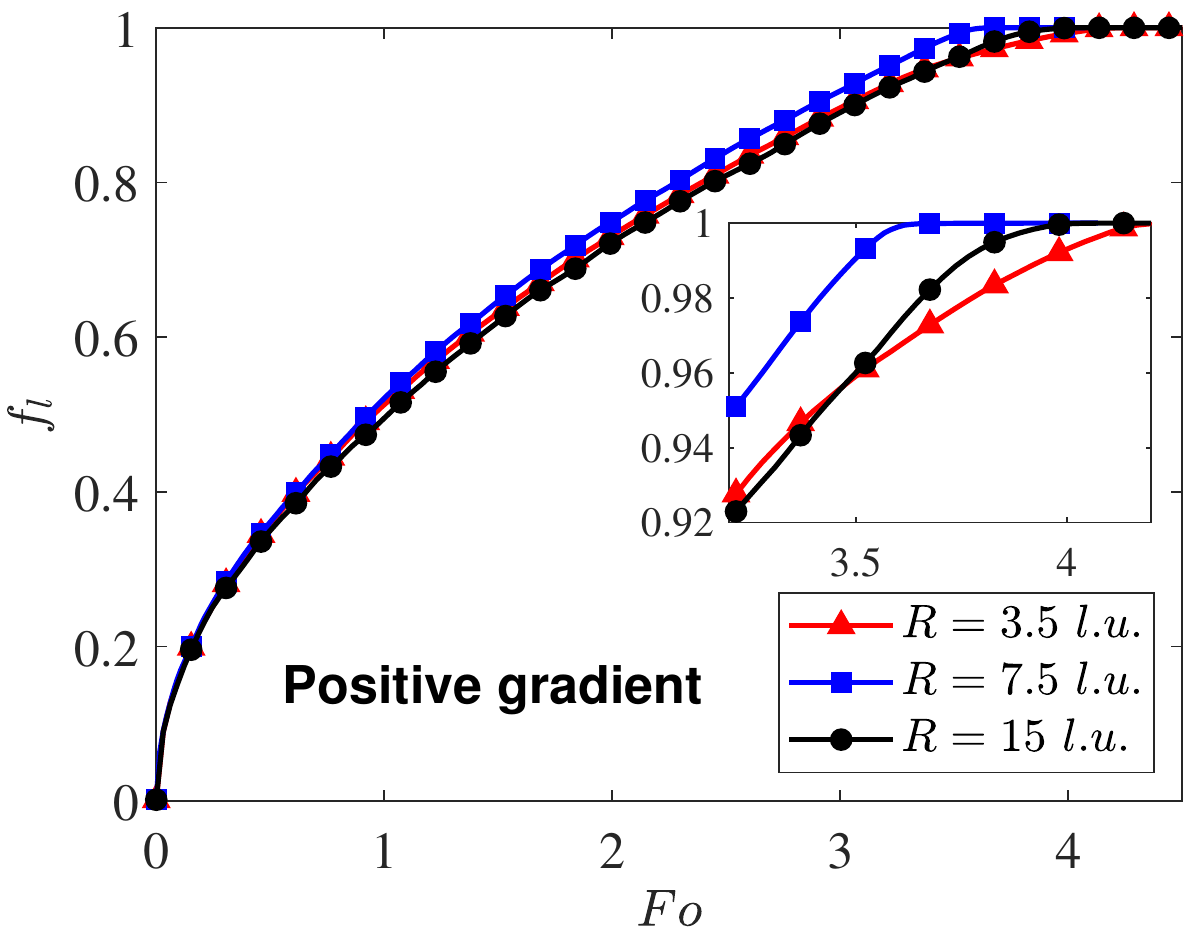}
			\includegraphics[width=0.4\textwidth]{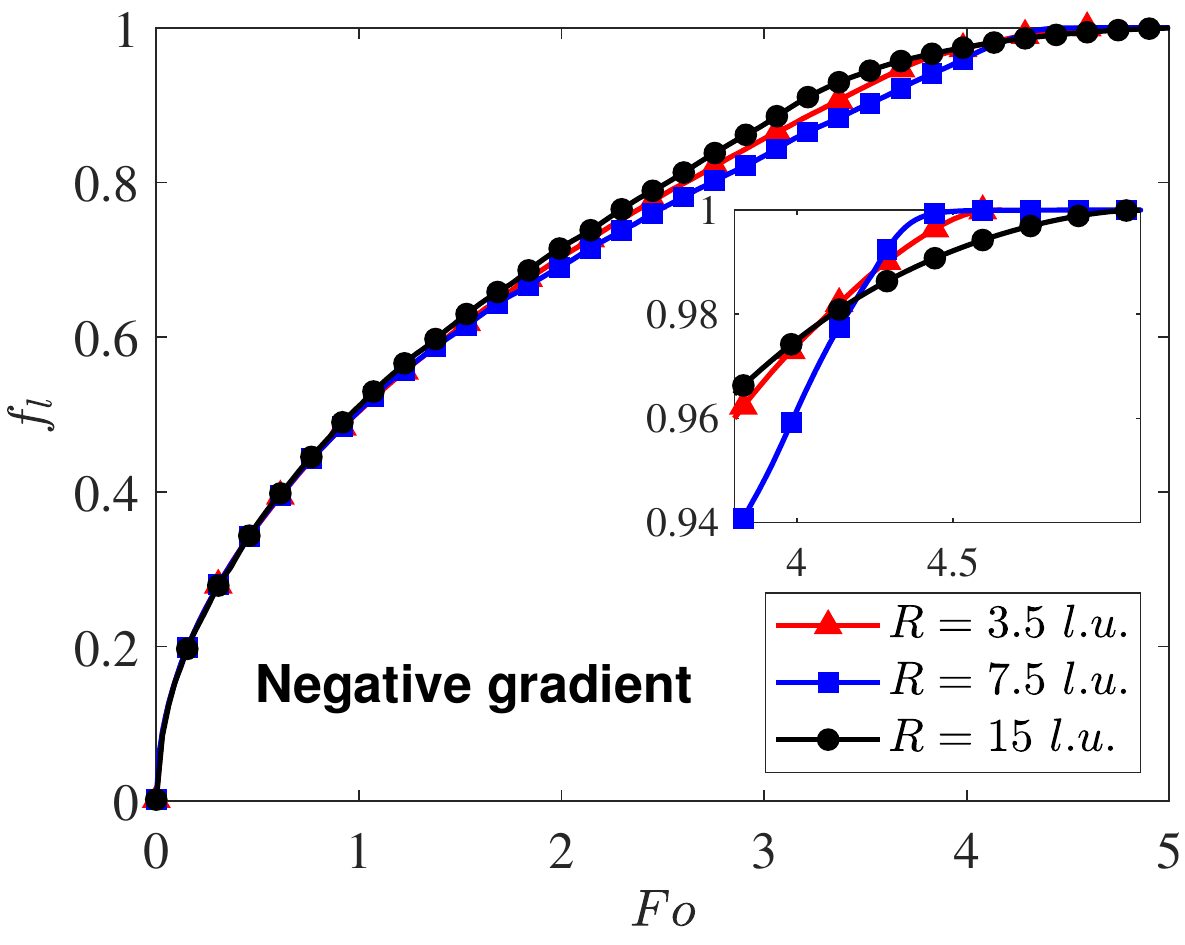}}
		
		\subfigure[$Ra=10^{6}$]{ \label{fig28}
			\includegraphics[width=0.4\textwidth]{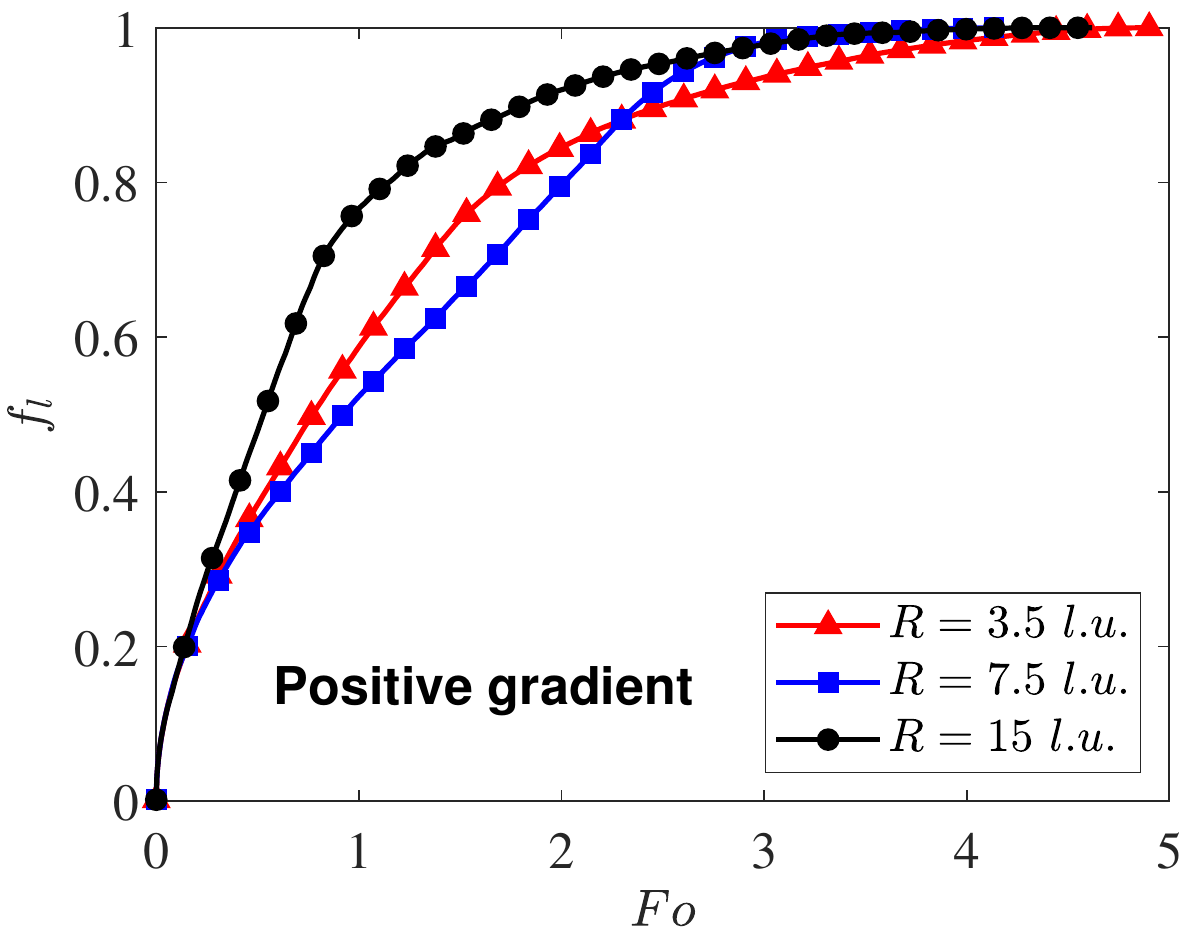}
			\includegraphics[width=0.4\textwidth]{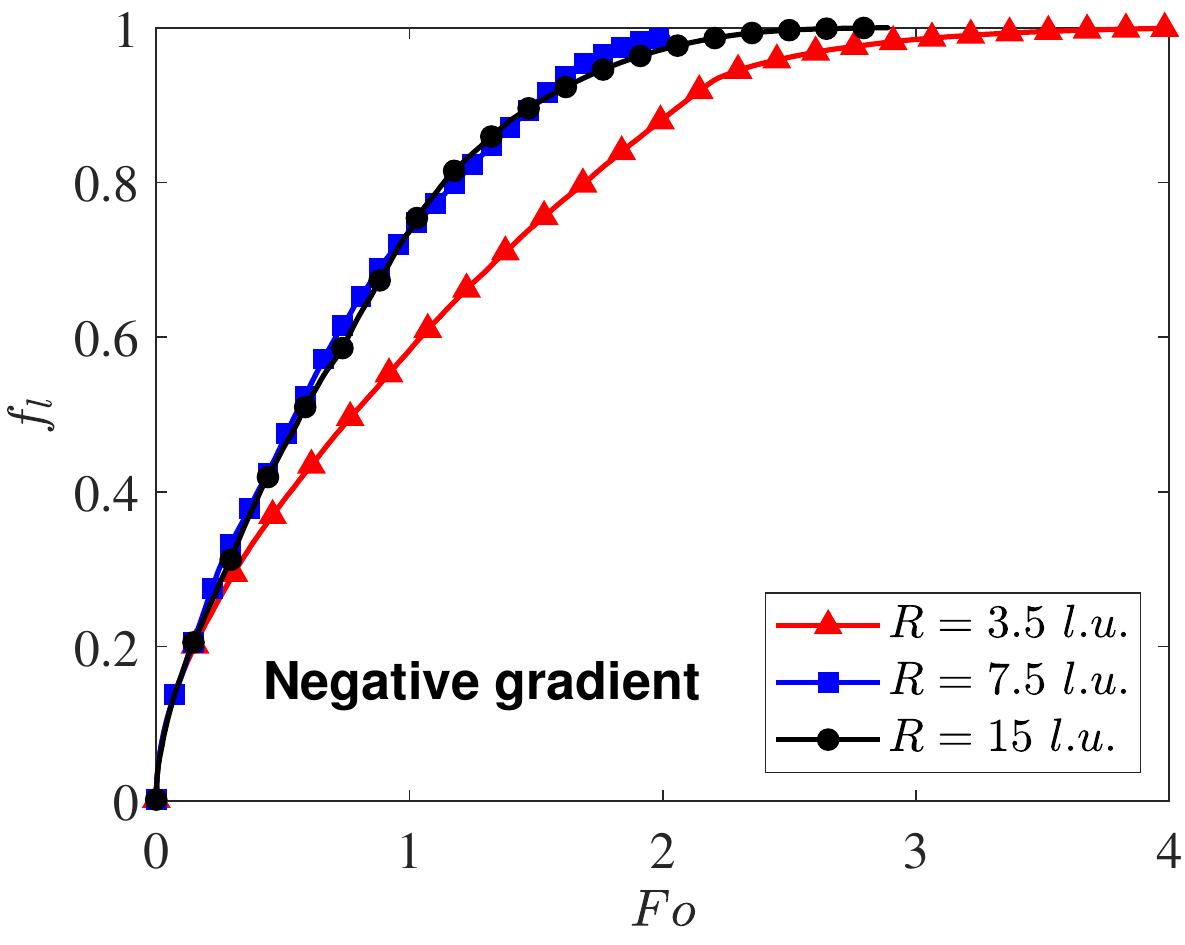}} 
		
		\caption{Effect of particle diameters on the melting evolution of the PCM for Case B.} 
		\label{fig29}
	\end{figure}

	Finally, the influence of the particle diameters $ Ra $ is explored, and the corresponding numerical results are presented in Fig. \ref{fig26} and Fig. \ref{fig29}. It can be seen from Fig. \ref{fig24} that the liquid fraction cannot be distinguished at the beginning of melting process, which indicates that the effect of particle diameters is not such significant on conduction. About $ Fo=3.5 $ later, the influence of particle diameters appears gradually,  which is caused by the decrease of the particle size leads to a growth in the internal surfaces for heat transfer in porous media. However, as Rayleigh number increases further, the situation begins to change and one can observed from Fig. \ref{fig25} and Fig. \ref{fig28} that owing to the increasing of the particle size, the toal melting time of PCM decreases, which is caused by the enhancement of convection. The reason why this phenomenon occurs is that smaller particle diameters indeed enhances the heat transfer by providing larger heat transfer surfaces, but as Rayleigh number increases, natural convection tends to play an increasingly stronger role and smaller particle diameters would lead to lower permeability which suppresses natural convection and then weakens the overall heat transfer. Further, one can deduce that the variation of total melting time depend little on the particle diameters since the

	\noindent  curves in Fig. \ref{fig26} are almost irregular.

	\section{Conclusion}
	\label{my_section5}
	In this paper, the total enthalpy-based lattice Boltzmann model is adopted to simulate the solid–liquid phase change in a square cavity equipped filled with a gradient porous structure. In order to improve computational efficiency, the proposed algorithm is programmed in parallel by using CUDA. The influence of the gradient porosity, Stefan number, gradient direction and Rayleigh number on the heat transfer in the composite are investigated in detail. 
	
	According to the present numerical results, it turns out that porous media with gradient porosity have different effects on the solid-liquid phase change process. The positive gradient porosity shows a further reduction for melting time when the Rayleigh number is small. As the Rayleigh numbers continues to increase and when it exceeds a certain critical value, positive gradient showed further enhancement for melting heat transfer, while the negative gradient deteriorated the heat transfer performance. The melting time of the positive gradient increases as the number of Rayleigh increases, while total melting time of the negative gradient increases as the number of Rayleigh increases when the Rayleigh numbers is greater than  $1.0  \times  10^{5} $ . Furthermore, whether positive gradient or negative gradient, the decreasing of the partical size leads to a growth in effective thermal conductivity, while can create an obstacle to natural convection.

	\section*{Conflict of interest}
	We declare that we have no financial and personal relationships with other people or organizations that can inappropriately influence our work, there is no professional or other personal interest of any nature or kind in any product, service and/or company that could be construed as influencing the position presented in, or the review of, the manuscript entitled.

	\section*{Acknowledgements}
	This work is financially supported by the National Natural Science Foundation of China (Grant No. 12002320), and the Fundamental Research Funds for the Central Universities (Grant Nos. CUG180618 and CUGGC05). We would like to thank Yin Jiang for his program to generate porous media which greatly helped the work.

	%\end{thebibliography}


\begin{thebibliography}{1}
		
		%%%%%%%%%%%%%%%%%%%%%%%%%%%%%%%%introduction_1
		%\bibitem{renewable energy resources}
		%O. Ellabban, H. Abu-Rub, F. Blaabjerg, Renewable energy resources: Current status, future prospects %and their enabling technology, Renew. Sust. Energ. Rev. 39 (2014) 748–764
		
		%\bibitem{geo}
		%Kh.R. Dione, H. Louahlia, M. Marion, J.L. Berçaits, Evaporation heat transfer and pressure drop for geothermal heat pumps working with refrigerants R134a and R407C, Int. Commun. Heat Mass Transf. 93 (2018) 1-1
		
		%\bibitem{tidal}
		%E. Osalusi, J. Side, R. Harris, Structure of turbulent flow in EMEC's tidal energy test site, Int. %Commun. Heat Mass Transf. 36 (2009) 422–431
		
		%\bibitem{solar}
		%A. Jamar, Z.A.A. Mjid, W.H. Azmi, M. Norhafana, A.A. Razak, A review of water heating system for solar energy applications, Int. Commun. Heat Mass Transf. 76 (2016) 178–187
		
		%\bibitem{wind}
		%T. Nosoko, K. Ameku, H. Minakuchi, Double-sided wet fabric evaporator utilizing wind and solar energy efficiently — One-dimensional transient simulations, Int. Commun. Heat Mass Transf. 38 (2011) 723–729
		
		%%%%%%%%%%%%%%%%%%%%%%%%%%%%%%%%introduction_1
		\bibitem{TES1}
		Z. Liao, C. Xu, Y. Ren, F. Gao, X. Ju, X. Du, A novel effective thermal conductivity correlation of the PCM melting in spherical PCM encapsulation for the packed bed TES system, Appl. Therm. Eng. 135 (2018) 116-122.
		
		\bibitem{TES2}
		A. Arteconi, N.J. Hewitt, F. Polonara, Domestic demand-side management (DSM): Role of heat pumps and thermal energy storage (TES) systems, Appl. Therm. Eng. (2013) 155-165.
		
		
		\bibitem{LHTES}
		F. Agyenima, N. Hewitta, P. Eamesb, M. Smyth, A review of materials, heat transfer and phase change problem formulation for latent heat thermal energy storage systems (LHTESS), Renew. Sust. Energ. Rev. 14 (2010) 615–62.
		
		
		\bibitem{LHTES2}
		A. Sarı, A. Karaipekli, Thermal conductivity and latent heat thermal energy storage characteristics of paraffin/expanded graphite composite as phase change material, Appl. Therm. Eng. 27 (2007) 1271–1277.
		
		
		\bibitem{low thermal conductivity}
		M. Liu, W. Saman, F. Bruno, Review on storage materials and thermal performance enhancement techniques for high temperature phase change thermal storage systems, Renew. Sust. Energ. Rev. 16 (2012) 2118– 2132.
		
		%%%%%%%%%%%%%%%%%%%%%%%%%%%%%%%%introduction_2
		\bibitem{particles1}
		T. Xiong, L. Zheng, K.W.Shah, Nano-enhanced phase change materials (NePCMs): A review of numerical simulations, Appl. Therm. Eng. 178 (2020) 115492.
		
		\bibitem{particles2}
		M.H. K. Darvanjooghi, M.N. Esfahany, Experimental investigation of the effect of nanoparticle size on thermal conductivity of in-situ prepared silica–ethanol nanofluid, Int. Commun. Heat Mass Transf. 77 (2016) 148–154.
		
		\bibitem{fin1}
		S.Zhang, L.Pu, L. Xu, R. Liu, Y. Li, Melting performance analysis of phase change materials in different finned thermal energy storage, Appl. Therm. Eng. 176 (2020) 115425.
		
		\bibitem{fin2}
		B, Kok, Examining effects of special heat transfer fins designed for the melting process of PCM and Nano-PCM, Appl. Therm. Eng. 170 (2020) 114989.
		
		
		\bibitem{fin3}
		H. Xu, N. Wang, C. Zhang, Z. Qu, M. Cao, Optimization on the melting performance of triplex-layer PCMs in a horizontal finned shell and tube thermal energy storage unit, Appl. Therm. Eng. 176 (2020) 115409.
		
		\bibitem{multiple PCM1}
		A.H. Mosaffa, L. Garousi Farshi, C.A. Infante Ferreira, M.A. Rosen, Energy and exergy evaluation of a multiple-PCM thermal storage unit for free cooling applications, Renew. Energy 68 (2014) 452e458.
		
		
		\bibitem{multiple PCM2}
		G. Peiro, J. Gasia, L. Miro, L.F. Cabeza, Experimental evaluation at pilot plant scale of multiple PCMs (cascaded) vs. single PCM configuration for thermal energy storage, Renew. Energy 83 (2015) 729e736.
		
		
		
		
		%%%%%%%%%%%%%%%%%%%%%%%%%%%%%%%%introduction_4
		
		\bibitem{chenIJHMT2014}
		Z. Chen, D. Gao, J. Shi, Experimental and numerical study on melting of phase change materials
		in metal foams at pore scale, Int. J. Heat Mass Transf. 72 (2014) 646–655.
		
		\bibitem{yangAE2016}
		J.L. Yang, L. J. Yang, C. Xu, X.Z. Du, Experimental study on enhancement of thermal energy storage with phase-change material, Appl. Energy 169 (2016) 164–17.
		
		\bibitem{zhaoIJHMT2016}
		Y. Zhao, C.Y. Zhao, Z.G. Xu, H.J. Xu, Modeling metal foam enhanced phase change heat transfer in thermal energy storage by using phase field method, Int. J. Heat Mass Transf. 99 (2016) 170–181.
		
		\bibitem{taoATE2016}
		Y.B. Tao, Y. You, Y.L. He, Lattice Boltzmann simulation on phase change heat transfer in metal foams/paraffin composite phase change material, Appl. Therm. Eng. 93 (2016) 476–485.
		
		\bibitem{zhuATE2016}
		F. Zhu, C. Zhang, X.L. Gong, Numerical analysis and comparison of the thermal performance enhancement methods for metal foam/phase change material composite, Appl. Therm. Eng. 109 (2016) 373–383.
		
		\bibitem{yaoIJTS2018}
		Y.P. Yao, H.Y. Wu, Z.Y. Liu, Z.S. Gao, Pore-scale visualization and measurement of paraffin melting in high porosity open-cell copper foam, Int. J. Therm. Sci. 123 (2018) 73-85.
		
		\bibitem{yangICHMT2021}
		X.H. Yang, X.Y. Wang, Z. Liu, Z.X. Guo, L.W. Jin, C. Yang, International Communications in Heat and Mass Transfer, Int. Commun. Heat Mass Transf. 122 (2021) 105127.
		
		
		%%%%%%%%%%%%%%%%%%%%%%%%%%%%%%%%introduction_5
		
		
		\bibitem{yangIJHMT2015}
		J.L. Yang, L.J. Yang, C. Xu, X.Z. Du, Numerical analysis on thermal behavior of solid–liquid phase change within copper foam with varying porosity, Int. J. Heat Mass Transf. 84 (2015) 1008–1018.
		
		\bibitem{yangIJHMT2016}
		X.H. Yang, W.B. Wang, C. Yang, L.W. Jina,  T.J. Lu, Solidification of fluid saturated in open-cell metallic foams with graded morphologies, Int. J. Heat Mass Transf. 98 (2016) 60–69.
		
		\bibitem{zhuATE2017}
		F. Zhu, C. Zhang, X.L. Gong, Numerical analysis on the energy storage efficiency of phase change material embedded in finned metal foam with graded porosity, Appl. Therm. Eng. 123 (2017) 256–265.
		
		
		\bibitem{zhangATE2017}
		Z.Q. Zhang, X.D. He, Three-dimensional numerical study on solid-liquid phase change within open-celled aluminum foam with porosity gradient, Appl. Therm. Eng. 113 (2017) 298–308.
		
		\bibitem{yangAE2020}
		X.H. Yang, P. Wei, X.Y. Wang, Y.L. He, Gradient design of pore parameters on the melting process in a thermal energy storage unit filled with open-cell metal foam, Appl. Energy 268(2020) 115019.
		
		\bibitem{ghahremannezhadATE2020}
		A. Ghahremannezhad, H.J. Xu, M.R. Salimpour, P. Wang, K. Vafai, Thermal performance analysis of phase change materials (PCMs) embedded in gradient porous metal foams, Appl. Therm. Eng. 179 (2020) 115731.
		
		\bibitem{huATE2020}
		C.Z. Hu, H.Y. Li, D.W. Tang, J. Zhu, K.M. Wang, X.F. Hu, M.L. Bai, Pore-scale investigation on the heat-storage characteristics of phase change material in graded copper foam, Appl. Therm. Eng. 178 (2020) 115609.
		
		\bibitem{marriIJHMT2021}
		G.K. Marri, C. Balaji, Experimental and numerical investigations on the effect of porosity and PPI gradients of metal foams on the thermal performance of a composite phase change material heat sink, Int. J. Heat Mass Transf. 164 (2021) 120454.
		
		
		%%%%%%%%%%%%%%%%%%%%%%%%%%%%%%%%introduction_6
		
		%%%%%%%%%%%%%%%%%%%%%%%%%%%%%%%%introduction_7
		\bibitem{boltzmann1}
		C.L. Lu, H.N. Wang, S.Y. Wang, W.J. Cai, L.H. Zhao, H.L. Lu, Effect of heating modes on melting performance of a solid–liquid phase change using lattice Boltzmann model, Int. Commun. Heat Mass Transf. 108 ( 2019 ) 104330.
		
		\bibitem{boltzmann2}
		B.M.  Chen, L.Q. Song, K.K. Gao, F. Liu, Two zone model for mushy region of solid–liquid phase change based on Lattice Boltzmann method, Int. Commun. Heat Mass Transf.   98 (2018) 1-12.
		
		\bibitem{boltzmann3}
		R. Haghani-Hassan-Abadi, M.H. Rahimian, A lattice Boltzmann method for simulation of condensation on liquid-impregnated surfaces, Int. Commun. Heat Mass Transf. 103 (2019) 7-16.
		
		\bibitem{LBM_media1}
		J. Qin, Z.Y. Xu, Z.G. Xu, Pore-scale investigation on flow boiling heat transfer mechanisms in gradient open-cell metal foams by LBM,  Int. Commun. Heat Mass Transf. 119 (2020) 104974.
		
		\bibitem{LBM_media2}
		R. Mabrouk, H. Naji, H. Dhahri, S. Hammouda, Z. Younsi, Numerical investigation of porosity effect on a PCM’s thermal performance in a porous rectangular channel via thermal lattice Boltzmann method,  Int. Commun. Heat Mass Transf. 119 (2020) 104992.
		
		%%%%%%%2222222
		\bibitem{zhouCMS2014}
		F. Zhou, G.X. Cheng, Lattice Boltzmann model for predicting effective thermal conductivity of composite with randomly distributed particles: Considering effect of interactions between particles and matrix, Comput. Mater. Sci. 92 (2014) 157-165.
		
		%\bibitem{Nusselt}
		%B.S. Kim, D.S. Lee, M.Y. Ha, H.S. Yoon, A numerical study of natural convection in a square enclosure with a circular cylinder at different vertical locations, Int. J. Heat Mass Transf. 51 (2008) 1888–1906.
		
		\bibitem{LBGK}
		Z.L. Guo, B.C. Shi, N.C. Wang, Lattice BGK Model for Incompressible Navier–Stokes Equation, J. Comput. Phys. 165, 288–306 (2000).
		
		\bibitem{Guo_F}
		Z.L. Guo, C.G. Zheng, B.C. Shi, Discrete lattice effects on the forcing term in the lattice Boltzmann method, Phys. Rev. E 65 (2002) 046308.
		
		\bibitem{luIJTS2019}
		J.H. Lu, H.Y. Lei, C.S. Dai, An optimal two-relaxation-time lattice Boltzmann equation for solid-liquid phase change: The elimination of unphysical numerical diffusion, Int. J. Therm. Sci. 135 ( 2019 ) 17-29.
		
		\bibitem{GuoPf2002}
		Z.L. Guo, C.G. Zheng, B.C. Shi, An extrapolation method for boundary conditions in lattice Boltzmann method, Phys. Fluids 14 (6) (2002).
		2007–2010
		
		
		\bibitem{huangIJHMT2013}
		R.Z. Huang, H.Y. Wu, P. Cheng, A new lattice Boltzmann model for solid–liquid phase change, Int. J. Heat Mass Transf. 59 (2013) 295–301.
		
		\bibitem{guai}
		F. Zhu, C. Zhang, X.L. Gong, Numerical analysis and comparison of the thermal performance enhancement methods for metal foam/phase change material composite, Appl. Therm. Eng. 109 (2016) 373–383.
		
	\end{thebibliography}
\end{document}